\documentclass[12pt,authoryear]{article}
\newcommand{\blind}{1}
\usepackage[
left=.75in,
right=.75in,
top=.75in,
bottom=.75in
]{geometry}

\usepackage{makecell}
\usepackage{amssymb,amsthm}
\usepackage{amsmath}
\usepackage{natbib}
\usepackage{bibunits}
\usepackage{pifont}

\usepackage{graphics,subcaption, caption, physics}
\usepackage{setspace,enumitem}
\setlist[enumerate]{
	label=(\alph*),
	leftmargin=*,
	itemsep=0pt,
	parsep=0pt,
	topsep=0pt,
	partopsep=0pt,
	labelsep=0.5em
}
\usepackage[dvipsnames]{xcolor}
\usepackage{tikz}
\usetikzlibrary{calc}
\usetikzlibrary{positioning} 
\usetikzlibrary{arrows.meta}
\usetikzlibrary{shadows}
\usepackage{algorithm}   
\usepackage{algpseudocode} 

\usepackage{enumitem}
\usepackage{booktabs}
\usepackage{multirow}
\usepackage{setspace}
\usepackage{colortbl}
\usepackage{longtable,array}
\definecolor{lightgray}{gray}{0.95}
\newcolumntype{C}[1]{>{\centering\arraybackslash}p{#1}}

\newtheorem{theorem}{Theorem}
\newtheorem{remark}{Remark}[section]
\newtheorem{definition}{Definition}[section]
\newtheorem{proposition}{Proposition}
\newtheorem{assumption}{Assumption}[section]
\newtheorem{corollary}{Corollary}[section]
\newtheorem{example}{Example}[section]
\newtheorem{lemma}{Lemma}[section]

\usepackage{chngcntr}
\counterwithin{table}{section}
\counterwithin{algorithm}{section}
\counterwithin{figure}{section}
\numberwithin{equation}{section}

\usepackage[colorlinks=true, linkcolor=blue, urlcolor=blue, citecolor=blue]{hyperref}
\usepackage{xcolor}

\newcommand{\indep}{\perp\!\!\!\perp}

\allowdisplaybreaks[4]
\renewcommand{\arraystretch}{1.2}
\begin{document}
	
	\begin{bibunit}[plainnat]
		\if1\blind
		{
			\title{Estimating Average Treatment Effects with Projected Instrumental Variables}
			\author{Shuyuan Chen, Yifan Cui\thanks{Correspondence to \href{mailto:cuiyf@zju.edu.cn}{cuiyf@zju.edu.cn}.}}
			\date{Zhejiang University    \vspace{-1cm}} 
			\maketitle
		} \fi

		\if0\blind
		{	
			\bigskip
			\bigskip
			\bigskip
			\begin{center}
				{\LARGE\bf Estimating Average Treatment Effects with Projected Instrumental Variables}
			\end{center}
			\medskip
		} \fi
		
		\begin{abstract}
			In this paper, we develop a projected instrumental variable framework that relaxes conventional assumptions for causal identification with instrumental variables. Building on the interpretation of conditional expectation as an orthogonal projection in a Hilbert space, the framework replaces the conditional expectation operator with a general projection operator. We provide an alternative inverse problem formulation based on the Riesz representation. Drawing on semiparametric efficiency theory, we further develop debiased and adaptive machine learning methods for estimation and inference. The proposed framework connects invalid instrumental variables to instrumented difference-in-differences. We also examine invalid projected instrumental variables and discuss their relationship to existing methods. Simulation studies and real data analysis demonstrate that the proposed methods perform favorably compared with existing approaches.
		\end{abstract}
		\noindent {{\bf keywords:}
			Average treatment effect, Bridge function, Instrumented difference-in-differences, Invalid instrumental variable, Projection operator, Riesz representation
		}

		\section{Introduction}
		\subsection{Background}
		
		In many biomedical and econometric applications, unmeasured confounding remains a major obstacle to causal inference. Instrumental variable (IV) methods offer a principled approach to addressing this challenge by exploiting auxiliary variables to identify causal effects in the presence of unmeasured confounding \citep{imbens1994identification,angrist1996identification}.
		
		Despite their widespread use, classical IV methods rely on assumptions that can be difficult to justify in practice. A valid IV must satisfy three core assumptions, possibly conditional on observed covariates: (a) relevance, requiring an association between the IV and the treatment; (b) independence, requiring the IV to be independent of unmeasured confounders; and (c) the exclusion restriction, requiring the IV to affect the outcome only through the treatment. Weak instrument relevance can lead to unstable estimation and unreliable inference, whereas violations of independence or the exclusion restriction can invalidate identification and bias causal effect estimates. 
		
		These challenges have motivated a growing body of work on causal identification and estimation with invalid instruments. A recent review by \citet{kang2025identification} summarizes advances in causal identification and inference with instruments that violate the independence assumption or the exclusion restriction. Such methods are particularly relevant in applications where instrument validity is difficult to justify, including personalized pricing \citep{miao2023personalized}, human-machine interaction \citep{fu2026offline}, and biomedical research \citep{bowden2015mendelian,zhao2020statistical}.
		
		One line of work develops methods for estimation and inference on linear causal effects under assumptions about the composition of the candidate instrument set, without prior knowledge of which instruments are valid. Prominent examples include the majority rule \citep{han2008detecting,kang2016instrumental,windmeijer2019on} and the plurality rule \citep{guo2018confidence,windmeijer2021confidence}. The former requires more than half of the instruments to be valid, whereas the latter permits a smaller valid subset under additional conditions on the invalid instruments.
		
		Another strand of work exploits heteroscedasticity for identification, including the approach of \citet{lewbel2012using} and the G-estimation under no interaction with unmeasured selection (GENIUS) \citep{tchetgen2021GENIUS,sun2022selective,ye2024geniusmawiirobustmendelianrandomization}. Rather than requiring a sufficiently large subset of valid instruments, GENIUS allows violations of the exclusion restriction under alternative structural assumptions. In particular, identification combines variation in the conditional treatment variance across instrument levels with a restriction requiring the conditional covariance between the treatment and the outcome residual to remain invariant across those levels.
		
		More closely related to our work are semiparametric efficient G-estimators that identify effects in partially linear models without requiring prior knowledge of which instruments are valid \citep{sun2023semiparametric,zhang2025mr}. Relaxations of proxy validity assumptions have also been studied in proximal causal inference \citep{yu2025fortified}. In a complementary direction, the instrumented difference-in-differences (IVDID) framework \citep{ye2023instrumented,vo2024structural,zhao2025semiparametric} exploits temporal variation to identify average treatment effects under assumptions that allow for violations of the exclusion restriction.

		Finally, our work connects to Riesz learning, which provides tools for estimating Riesz representers and constructing orthogonal bias corrections \citep{chernozhukov2021automatic,chernozhukov2022debiased}, and to nonparametric instrumental variable methods that study identification and estimation through conditional moment restrictions \citep{newey2003instrumental}. It also connects to proximal causal inference, which identifies causal effects through bridge functions satisfying conditional moment restrictions \citep{miao2018identifying,cui2024semiparametric}. Together, these connections motivate our operator-based perspective on causal identification and estimation.
		
		\subsection{Our contributions} 
		In this paper, we develop a projected IV (PIV) framework that relaxes conventional identification assumptions for IVs. The key idea of the PIV approach is to interpret conditional expectations as projection operators on a suitable Hilbert space and replace the conditional expectation operator with a more general projection operator. Notably, under appropriate identification conditions, this framework can identify causal effects even when none of the candidate instruments satisfies the classical IV validity assumptions.
		
		Our main contributions are fivefold.
		First, we propose the PIV framework from a novel operator-based perspective to identify the average treatment effect (ATE). Identification results for other estimands such as the counterfactual means and average treatment effect on the treated are given in the Appendix. Second, we provide an inverse problem interpretation of the proposed framework based on the Riesz representation. Third, we develop debiased and adaptive estimators using semiparametric efficiency theory. Fourth, we establish connections between invalid IV and IVDID through our PIV framework. Fifth, we examine how one can utilize invalid PIV for identification.
		
		We outline the remainder of the paper as follows. In Section~\ref{sec: projected IV}, we introduce the notions of PIV and projected regular weighting functions (PRWF). 
		In Section~\ref{sec: an inverse problem view}, we interpret the proposed PIV condition from the perspectives of Riesz learning and inverse problems.
		In Section~\ref{sec: semiparametric analysis}, we develop two estimators within the debiased machine learning framework. 
		The first estimator directly employs a prespecified PRWF, while the second adaptively selects an optimal weighting function. We establish that both estimators are consistent and asymptotically normal, provided that the $L_2$-norm convergence rate of the nuisance function estimators is faster than $n^{-1/4}$. 
		Section~\ref{sec: connection to invalid IV} develops a novel application of the proposed PIV framework to invalid IV problems, whereas Section~\ref{sec: connection to exsiting literature} clarifies its connections to the existing invalid IV literature.
		Finally, Section~\ref{sec: simulation} presents simulation studies that illustrate the finite-sample performance of the proposed methods under a range of data-generating processes. Section~\ref{sec: real data analysis} provides a real-data application, and Section~\ref{sec: discussion on invalid PIV} discusses extensions to invalid PIVs.
		
		\subsection{Notation}
		We introduce the notation used throughout the paper. Let $d_Z$, $d_X$, and $d_U$ be positive integers. The outcome is denoted by $Y \in \mathcal{Y} \subseteq \mathbb{R}$, the binary treatment by $A \in\mathcal{A} = \{0,1\}$, the IV by $Z :=\{Z_{[j]}\}_{j=1}^{d_Z}\in \mathcal{Z} \subseteq \mathbb{R}^{d_Z}$, the observed confounders by $X \in \mathcal{X} \subseteq \mathbb{R}^{d_X}$, and the unobserved confounders by $U \in \mathcal{U} \subseteq \mathbb{R}^{d_U}$, where each $Z_{[j]}$ may be discrete or continuous.
		
		Let $Y(a)$ denote the potential outcome that would be observed if $A = a$, and let $O:= \{X, Z, A, Y\}$ denote the observed data vector. The observed sample consists of $n$ independent and identically distributed copies $\{O_i\}_{i=1}^n$ of $O$. Throughout the paper, we assume consistency, $Y=Y(A)$, and $\mathbb{E}[Y(a)^2]<\infty$ for each treatment level under consideration.
		
		Let $L_2(O)$ denote the Hilbert space of measurable, square-integrable functions of $O$. Equip this space with the inner product $\langle f,g\rangle := \mathbb{E}[f(O)g(O)]$ and the norm $\|f\|_2 := \{\mathbb{E}[f(O)^2]\}^{1/2}$. For any closed linear subspace $\mathcal{S}$ of $L_2(Z,X)$, let $\mathcal{S}^{\perp}$ denote its orthogonal complement in $L_2(Z,X)$.

		\section{Projected IV framework}\label{sec: projected IV}
		\subsection{Toy Examples}\label{subsec: toy example}
		We first introduce two toy examples to illustrate the main idea. Suppose that $Z$ is a univariate continuous and bounded IV, and $X = \emptyset$. In these toy examples only, we allow the treatment $A$ to be continuous. For the first example, consider the following model where the IV exclusion restriction is violated
		\begin{align*}
			Y = \beta_1^o + \beta_a^o A + \beta_z^o Z + \epsilon, \qquad \mathbb{E}[\epsilon \mid Z] = 0.
		\end{align*}
		The unmeasured confounder $\epsilon$ can be correlated with the endogenous treatment variable $A$, and our primary interest lies in identifying $\beta_a^o$. 
		Under the standard IV framework, $\beta_a^o$ is not identifiable when the exclusion restriction is violated, i.e., when $\beta_z^o \neq 0$. However, because the support of $Z$ is sufficiently rich, the redundant information contained in different regions of the support can still be exploited. 
		
		Specifically, define $\mathcal{S}_{\mathrm{lm}} := \overline{\{s \in L_2(Z) : s(Z) = \theta_1 Z + \theta_0,\; \theta_0, \theta_1 \in \mathbb{R}\}}$, and for any $f\in L_2(O)$, define $\mathbb{S}_{\mathrm{lm}}[f] := \arg\min_{s\in\mathcal{S}_{\mathrm{lm}}} \mathbb{E}[\{f(O) - s(Z)\}^2].$ Then, for any $\pi \in L_2(Z)$, it is easy to verify that
		\begin{align*}
			\mathbb{E}\bigl[\{\pi(Z) - \mathbb{S}_{\mathrm{lm}}[\pi](Z)\}(Y - \beta_a^o A)\bigr] = \beta_z^o \mathbb{E}\bigl[\{\pi(Z) - \mathbb{S}_{\mathrm{lm}}[\pi](Z)\} Z\bigr] + \beta_1^o \mathbb{E}\bigl[\{\pi(Z) - \mathbb{S}_{\mathrm{lm}}[\pi](Z)\} \bigr] = 0,
		\end{align*}
		where the last equality follows from the fact that $\pi - \mathbb{S}_{\mathrm{lm}}[\pi]$ is orthogonal to all the elements in $\mathcal{S}_{\mathrm{lm}}$. Thus, if there exists a $\pi \in L_2(Z)$ such that $\mathbb{E}\bigl[\{\pi(Z) - \mathbb{S}_{\mathrm{lm}}[\pi](Z)\} A\bigr] \neq 0$, we can identify $\beta_a^o$ by
		\begin{align}\label{eq: linear identification}
			\beta_a^o = \mathbb{E}\bigl[\{\pi(Z) - \mathbb{S}_{\mathrm{lm}}[\pi](Z)\} Y\bigr]/\mathbb{E}\bigl[\{\pi(Z) - \mathbb{S}_{\mathrm{lm}}[\pi](Z)\} A\bigr].
		\end{align}
		Notably, when $Z$ is binary, $\pi(Z) - \mathbb{S}_{\mathrm{lm}}[\pi](Z) \equiv 0$, so the identification strategy fails. In general, $Z$ must take at least three distinct values for identification to be possible.
		
\begin{figure}[t]
			\centering
			\includegraphics[width=\textwidth]{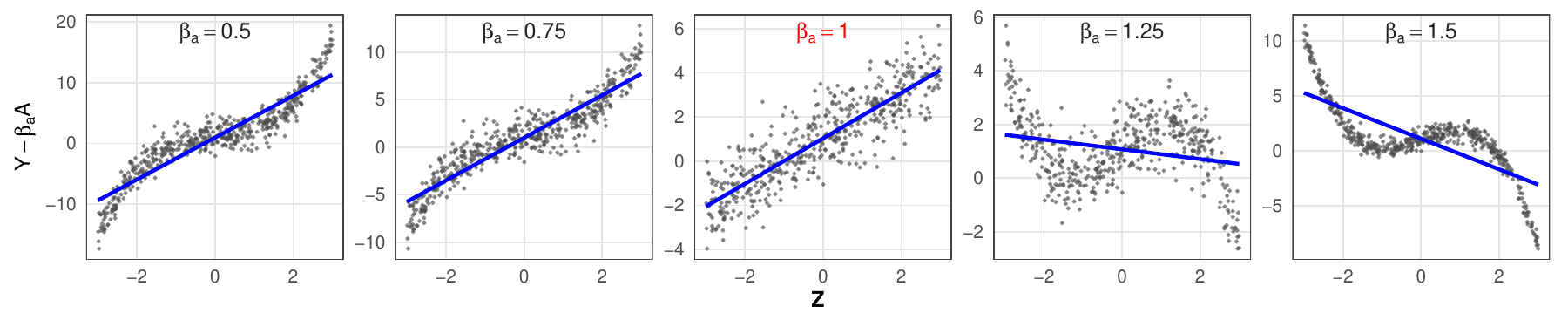}
			\caption{Plots of $Y-A\beta_a$ versus $Z$ under 5 different choices of $\beta_a$ (Example 1).
			}\label{fig: toy_example linear}
			\vspace{1em}
			\includegraphics[width=\textwidth]{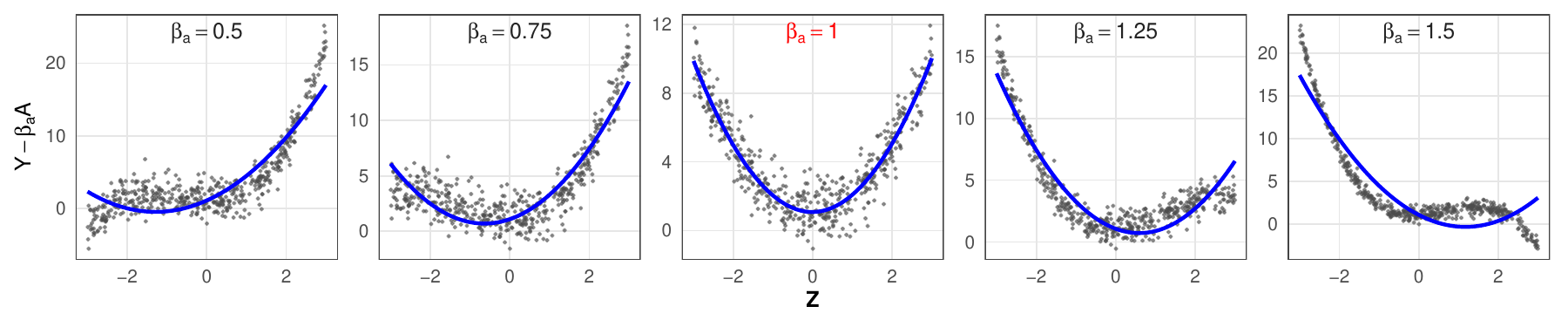}
			\caption{Plots of $Y-A\beta_a$ versus $Z$ under 5 different choices of $\beta_a$ (Example 2). 
			}\label{fig: toy_example quadratic}
		\end{figure}

		In Figure~\ref{fig: toy_example linear}, we illustrate the relationship between $Y-A\beta_a$ and the univariate IV $Z$ through a simple graphical example, where $\epsilon\sim\mathcal{N}(0,1)$, $Z\sim \mathrm{Uniform}(-3,3)$, $A=Z^3+\epsilon-0.5Z$, and $\beta_1^o=\beta_a^o=\beta_z^o=1$. Only the third panel, corresponding to the correctly specified value $\beta_a=1$, successfully removes the nonlinear dependence induced by $A$ and isolates the remaining linear effect of $Z$. By contrast, incorrect choices of $\beta_a$ leave substantial nonlinear residual patterns with respect to $Z$. This illustrates the key intuition underlying identification of $\beta_a^o$.
		
		As a second example, suppose $Y=\beta_1^o+\beta_a^o A+\beta_{z1}^o Z+\beta_{z2}^o Z^2+\epsilon$, with $\mathbb{E}[\epsilon\mid Z]=0$. Define $\mathcal{S}_{\mathrm{qm}}:=\overline{\{s\in L_2(Z): s(Z)=\theta_2 Z^2+\theta_1 Z+\theta_0,\ \theta_0,\theta_1,\theta_2\in\mathbb{R}\}}$, and $\mathbb{S}_{\mathrm{qm}}[f]:=\arg\min_{s\in\mathcal{S}_{\mathrm{qm}}}\mathbb{E}[\{f(O)-s(Z)\}^2]$. Provided that the resulting denominator is nonzero, $\beta_a^o$ is identified by replacing $\mathbb{S}_{\mathrm{lm}}$ with $\mathbb{S}_{\mathrm{qm}}$ in Equation~\eqref{eq: linear identification}. Figure~\ref{fig: toy_example quadratic} shows the relationship between $Y-\beta_a A$ and $Z$ for different values of $\beta_a$. Only the third panel isolates the remaining quadratic effect of $Z$ on $Y$.
		
		When $X$ is nonempty and $d_Z>1$, we instead consider 
		$\mathcal{S}_{\mathrm{lm}}$ as the linear span of $Z$ with $X$-varying coefficients, and define $\mathbb{S}_{\mathrm{lm}}[f]:=\arg\min_{s\in\mathcal{S}_{\mathrm{lm}}}\mathbb{E}\!\left[\left\{f(O)-s(Z,X)\right\}^2\right].$
		Under this formulation, $\beta_a^o$ can be identified analogously by projecting $\pi \in L_2(Z,X)$ onto $\mathcal{S}_{\mathrm{lm}}$.

		\subsection{The projection operator}
        To generalize the idea presented in Section~\ref{subsec: toy example}, we introduce a proper Hilbert subspace $\mathcal{S}$. Intuitively, $\mathcal{S}$ represents the allowable structure of $\mathbb{E}[Y(0)\mid Z,X]$. We formalize this notion below. By convention, we define $L_p(\emptyset):=\mathbb{R}$ for $p\in (1,\infty]$.
		\begin{definition}\label{defn: proper_hilbert_subspace}
			We say that $\mathcal{S}$ is a proper Hilbert subspace of $L_2(Z,X)$ if:
			\begin{enumerate}
				\item $L_2(X) \subseteq \mathcal{S} \subseteq L_2(Z,X)$, and $\mathcal{S}$ is a closed linear subspace of $L_2(Z,X)$;
				\item For any $f_1,f_2 \in \mathcal{S}$ and any $c_1,c_2 \in L_\infty(X)$, we have $c_1 f_1 + c_2 f_2 \in \mathcal{S};$
				\item $\mathcal{S}$ is equipped with the inherited inner product $\langle f_1,f_2\rangle=\mathbb{E}[f_1(Z,X)f_2(Z,X)].$
			\end{enumerate}
		\end{definition}
		Here, ``proper'' denotes the properties in Definition~\ref{defn: proper_hilbert_subspace}; it does not mean a strict subset.
		Obviously, $\mathcal{S}_{\mathrm{lm}}$ and $\mathcal{S}_{\mathrm{qm}}$ are both proper Hilbert subspaces of $L_2(Z,X)$ when $X=\emptyset$.
		Also, $L_2(X)$ itself is the smallest proper Hilbert subspace of $L_2(Z,X)$. 
		Next, define $\mathbb{S}: L_2(O)\to \mathcal{S}$ as a projection operator that maps any function $f\in L_2(O)$ onto the proper Hilbert subspace $\mathcal{S}$:
		\begin{align*}
			\mathbb{S}[f]
			:=\arg\min_{s\in\mathcal{S}}\mathbb{E}\!\left[\left\{f(O)-s(Z,X)\right\}^2\right].
		\end{align*}
		The projection operator $\mathbb{S}$ has the following properties, which are important for identification.
		\begin{proposition}\label{prop: proper Hilbert subspace}
			Let $\mathcal{S}$ be a proper Hilbert subspace of $L_2(Z,X)$.
			\begin{enumerate}
				\item For every $f\in L_2(O)$, $\mathbb{S}[f]$ is the unique function in $\mathcal{S}$ such that, for all $s\in\mathcal{S}$, $$\mathbb{E}[\{f(O)-\mathbb{S}[f](Z,X)\}s(Z,X)\mid X]=0.$$
				\item For all $f_1,f_2 \in L_2(O)$ and any $c_1,c_2 \in L_\infty(X)$, $\mathbb{S}[c_1f_1+c_2f_2]=c_1\mathbb{S}[f_1]+ c_2\mathbb{S}[f_2]$.
				\item For any other proper Hilbert subspace $\mathcal{S}'$ of $L_2(Z,X)$ with $\mathcal{S} \subseteq \mathcal{S}'$ and the corresponding projection operator $\mathbb{S}'$, $\mathbb{S}[f] = \mathbb{S}[\mathbb{S}'[f]]$ for all $f \in L_2(O)$.
			\end{enumerate}
		\end{proposition}

		Proposition~\ref{prop: proper Hilbert subspace}(a) establishes the uniqueness of $\mathbb{S}[f]$ and the orthogonality condition characterizing $\mathbb{S}[f]$ as the projection of $f$ onto $\mathcal{S}$, conditional on $X$. 
		Proposition~\ref{prop: proper Hilbert subspace}(b) demonstrates that $\mathbb{S}$ is a linear projection operator. 
		Proposition~\ref{prop: proper Hilbert subspace}(c) reflects the nesting property of orthogonal projections: projecting onto a smaller subspace is equivalent to first projecting onto a larger subspace and then projecting again onto the smaller one.
		
		In particular, when taking $\mathcal{S} = L_2(X)$, we have $\mathbb{S}[f](Z,X) = \mathbb{E}[f(O)\mid X]$, which corresponds to the conditional expectation operator. When we take $\mathcal{S} = L_2(Z,X)$, we obtain $\mathbb{S}[f](Z,X) = \mathbb{E}[f(O)\mid Z,X]$.

		\subsection{Projected regular weighting function}
		Let $\mathcal{S}$ be a proper Hilbert subspace of $L_2(Z, X)$. We propose the key concept of an $\mathcal{S}$-PRWF, which controls variance and ensures the stability of the functionals proposed later for identification.
		\begin{definition}[$\mathcal{S}$-PRWF]\label{defn: regular weighting function}
			We say that $\pi \in L_2(Z, X)$ is an $\mathcal{S}$-projected regular weighting function (PRWF) if there exists a constant $\epsilon_0 \in (0,1]$ such that
			$\left|\mathbb{E}\left[\left\{\pi(Z,X) - \mathbb{S}[\pi](Z,X)\right\} A \mid X\right]\right| \ge \epsilon_0$, and $\mathbb{E}[\{\pi(Z,X)-\mathbb{S}[\pi](Z,X)\}^2\mid X]\leq 1/\epsilon_0$ almost surely.
		\end{definition}
		Note that $\mathbb{E}[\{\pi(Z,X)-\mathbb{S}[\pi](Z,X)\}^2\mid X]\leq 1/\epsilon_0$ is satisfied for a uniformly bounded $\pi$. Therefore, the existence of an $\mathcal{S}$-PRWF depends critically on the IV relevance structure. Intuitively, when $Z \indep A \mid X$, it is clear that no $\mathcal{S}$-PRWF can exist. Next, we introduce the notion of $\mathcal{S}$-relevance, which formalizes the relevance structure between the IVs and the treatment.
		\begin{definition}[$\mathcal{S}$-relevance]\label{defn: IV relevance}
			Define $\pi_*^o(Z,X) := \mathbb{E}[A \mid Z,X]$. We say that $Z$ is $\mathcal{S}$-relevant to $A$ if there exists an $\epsilon_0\in (0,1]$ such that $ \mathbb{E}\left[\left\{\pi_*^o(Z,X) - \mathbb{S}[\pi_*^o](Z,X)\right\}^2 \middle| X\right] \ge \epsilon_0$ almost surely.
		\end{definition}
		The definition of $\mathcal{S}$-relevance measures IV strength via the $L_2$-distance between $\pi_*^o(Z, X)$ and its projection onto $\mathcal{S}$. As shown in the next proposition, this notion of IV relevance becomes increasingly stringent as the subspace $\mathcal{S}$ grows larger.
		
		\begin{proposition}\label{prop: IV relevance monotonicity}
			Let $\mathcal{S}$, $\mathcal{S}'$ be two proper Hilbert subspaces of $L_2(Z,X)$, where $\mathcal{S}\subseteq\mathcal{S}'$. Then $\mathcal{S}'$-relevance implies $\mathcal{S}$-relevance.	
		\end{proposition}
		
		For the special case $\mathcal{S}=L_2(X)$, $\mathcal{S}$-relevance implies that there exists a positive $\epsilon_0$, such that
		$$\mathbb{E}\!\left[ \{\mathbb{E}[A \mid Z,X] - \mathbb{E}[A \mid X]\}^2 \mid X \right]= \mathrm{Var}\!\left\{ \mathbb{E}[A \mid Z,X] \mid X \right\}\geq \epsilon_0.$$
		This condition requires that the conditional treatment probability varies with $Z$ given $X$, recovering the IV relevance concept in \citet{chen2025identificationdebiasedlearningcausal}. 
		Next, we formalize the relation between the existence of $\mathcal{S}$-PRWF and the $\mathcal{S}$-relevance condition.
		\begin{proposition}[Existence of PRWF]\label{prop: existence of PRWF}
			The following three statements are equivalent: (a) There exists an $\mathcal{S}$-PRWF; (b) $\pi_*^o$ is an $\mathcal{S}$-PRWF; (c) $Z$ is $\mathcal{S}$-relevant to $A$.
		\end{proposition}
		Proposition~\ref{prop: existence of PRWF} shows that an $\mathcal{S}$-PRWF exists if and only if $Z$ contains sufficient information about $A$. Moreover, whenever such a function exists, the weighting function $\pi_*^o$ is an $\mathcal{S}$-PRWF, thereby justifying its role as the canonical optimal weighting function.

		\subsection{Identification with PIV}
		In this subsection, we introduce another key concept of a PIV, which utilizes a proper Hilbert subspace $\mathcal{S}$.
		\begin{definition}[$\mathcal{S}$-PIV]\label{defn: projected IV}
			We say that $Z$ is an $\mathcal{S}$-PIV if (a) $\mathbb{E}[Y(0)\mid Z,X] \in\mathcal{S}$; (b) For any $\pi\in L_2(Z,X)$, $\mathrm{Cov}\{A\{\pi(Z,X) - \mathbb{S}[\pi](Z,X)\},Y(1)-Y(0)\mid X\}=0$ almost surely.
		\end{definition}
		
		Definition~\ref{defn: projected IV}(a) requires that the structure of $\mathcal{S}$ adequately captures the dependence of $Y(0)$ on $Z$. In particular, when $\mathcal{S}=L_2(X)$, the condition requires $\mathbb{E}[Y(0)\mid Z,X]$ to be invariant with respect to $Z$, which is a direct consequence of the IV independence condition when $Z$ is valid.
		
		Definition~\ref{defn: projected IV}(b) serves as an orthogonality restriction, analogous to the no unmeasured common effect modifier assumption \citep{cui2021semiparametric}.
		Specifically, Definition~\ref{defn: projected IV}(b) requires either that the individual treatment effect $Y(1)-Y(0)$ be exogenous or that $A\{\pi-\mathbb{S}[\pi]\}$ be conditionally uncorrelated with the component of $Y(1)-Y(0)$ attributable to unmeasured confounding.

		\begin{example}[Linear model with constant treatment effect]
			When $X=\emptyset$, consider $\mathcal{S}=\mathcal{S}_{\mathrm{lm}}$, as defined in Section~\ref{subsec: toy example}. The linear model \citep{kang2016instrumental,guo2018confidence,sun2023semiparametric} assumes that $Y(a)=\beta^o a+\theta_0^o+\sum_{j=1}^{d_Z}\theta_j^o Z_{[j]}+\epsilon$, where $\mathbb{E}[\epsilon\mid Z]=0.$
			In particular, $\mathbb{E}[Y(0)\mid Z]=\theta_0^o+\sum_{j=1}^{d_Z}\theta_j^o Z_{[j]} \in\mathcal{S}_{\mathrm{lm}},$ so that Definition~\ref{defn: projected IV}(a) is satisfied. In addition, $Y(1)-Y(0)$ is the constant $\beta^o$, so that Definition~\ref{defn: projected IV}(b) is satisfied.
		\end{example}
		To understand the meaning of $A\{\pi-\mathbb{S}[\pi]\}$ being conditionally uncorrelated with $Y(1)-Y(0)$, we provide an alternative description of Definition~\ref{defn: projected IV}(b) when $Z$ is exogenous and present two concrete examples. Let $A(z)$ denote the potential treatment value when the instrument is set to $Z=z$. 
		\begin{proposition}\label{prop: NUCEM equivalent representation}
			Suppose that $\{A(z),Y(1)-Y(0)\}\indep Z\mid X$, and define $g(z,x):=\mathrm{Cov}\{A(z),Y(1)-Y(0)\mid X=x\}$. Then, Definition~\ref{defn: projected IV}(b) is equivalent to requiring that $g\in\mathcal{S}$.
		\end{proposition}
		\begin{example}[Mixture model with $\mathcal{S}=L_2(X)$]
			Suppose that $Z\indep (U,G)\mid X$, $A(z) = (1-G) f_0(z,X) + G f_1(U,X)$, and $Y(a) = e(a,U,X)$. Here, the ranges of $f_0$, $f_1$ are $\mathcal{A}$, and $G\in\{0,1\}$ is a group indicator, satisfying $G\indep U\mid X$. One can verify that $g(z_0,X)-g(z_0',X)\equiv 0$, since $A(z_0)-A(z_0')=(1-G)\{f_0(z_0,X)-f_0(z_0',X)\}$ is independent of $Y(1)-Y(0)=e(1,U,X)-e(0,U,X)$ conditional on $X$. By Proposition~\ref{prop: NUCEM equivalent representation}, Definition~\ref{defn: projected IV}(b) is satisfied for $\mathcal{S}=L_2(X)$.
		\end{example}
		\begin{example}[Mixture model with $\mathcal{S}=\overline{\sum_{j=1,2}L_2(Z_{[j]},X)}$]
			Suppose that $Z\indep (U,G)\mid X$,  $A(z_{[1]},z_{[2]}) = I(G=0) f_0(z_{[1]},z_{[2]},X) + I(G=1) f_1(U,z_{[1]},X) + I(G=2) f_2(U,z_{[2]},X),$ and $Y(a) = e(a,U,X)$. Here, the ranges of $f_0$, $f_1$, $f_2$ are $\mathcal{A}$, and $G\in\{0,1,2\}$ is a group indicator satisfying $G\indep U\mid X$. 
			Then, one can readily verify that $g(z_{[1]},z_{[2]},X)-g(z_{[1]}',z_{[2]},X) - g(z_{[1]},z_{[2]}',X) + g(z_{[1]}',z_{[2]}',X)\equiv 0$, indicating that $g\in\overline{\sum_{j=1,2}L_2(Z_{[j]},X)}$.
			By Proposition~\ref{prop: NUCEM equivalent representation}, Definition~\ref{defn: projected IV}(b) is satisfied for $\mathcal{S}=\overline{\sum_{j=1,2}L_2(Z_{[j]},X)}$.
		\end{example}
		In contrast to $\mathcal{S}$-relevance, an $\mathcal{S}$-PIV is always an $\mathcal{S}'$-PIV for any two proper Hilbert subspaces with $\mathcal{S} \subseteq \mathcal{S}'$. Consequently, choosing a larger $\mathcal{S}$ is more likely to violate the $\mathcal{S}$-relevance condition, leading to increased variance and instability, whereas choosing a smaller $\mathcal{S}$ is more likely to violate the $\mathcal{S}$-PIV condition, thereby invalidating the identifying formula. This highlights a bias-variance trade-off induced by the choice of $\mathcal{S}$.

		We now present identification with PIV in the following theorem.
		\begin{theorem}[Identification]\label{thm: identification}
			Suppose that $Z$ is an $\mathcal{S}$-PIV and $\pi$ is an $\mathcal{S}$-PRWF. Then $\mathbb{E}[Y(1)-Y(0)]=\psi_{\mathcal{S},\pi}^o$, where $\psi_{\mathcal{S},\pi}^o$ is the following inverse probability weighting (IPW) functional:
			\begin{align*}
				\psi_{\mathcal{S},\pi}^o:=\mathbb{E}\left[\dfrac{\pi(Z,X) - \mathbb{S}[\pi](Z,X)}{\mathbb{E}\left[\left\{\pi(Z,X) - \mathbb{S}[\pi](Z,X)\right\}A|X\right]}Y\right].
			\end{align*}
		\end{theorem}
		
		\begin{remark}
			If $Z$ is an $\mathcal{S}$-PIV and $\mathcal{S}=L_2(X)$, then Theorem~\ref{thm: identification} reduces to 
			\begin{align*}
				\mathbb{E}[Y(1)-Y(0)] = \mathbb{E}\left[
				\dfrac{\pi(Z,X) - \mathbb{E}[\pi(Z,X)\mid X]}{\mathrm{Cov}\left\{\pi(Z,X),\,A|X\right\}}Y
				\right],
			\end{align*}
			a result established in Theorem~2.3 of \citet{chen2025identificationdebiasedlearningcausal} for valid IVs.     
		\end{remark}

		\section{An inverse problem view}\label{sec: an inverse problem view}
        In this section, we develop an inverse problem interpretation of the PIV framework through treatment bridge functions \citep{cui2024semiparametric} and outcome bridge functions \citep{miao2018identifying}, highlighting its connection to Riesz learning \citep{chernozhukov2021automatic,chernozhukov2022locally}. Specifically, we find a treatment bridge function $q\in \mathcal{S}^{\perp}$ that satisfies 
		\begin{equation}
			\begin{aligned}\label{eq: treatment bridge}
				\mathbb{E}\left[q(Z,X)A\middle| X\right]\equiv 1\text{ almost surely}; \quad \mathbb{E}[q(Z,X)^2\mid X]\in L_{\infty}(X).
			\end{aligned}
		\end{equation} 
		Notably, Equation~\eqref{eq: treatment bridge} was previously established by \citet{chernozhukov2021automatic,chernozhukov2022locally} for the special case \(\mathcal{S}=L_2(X)\).
		We next provide a necessary and sufficient condition for the existence of a solution to Equation~\eqref{eq: treatment bridge}.
		\begin{proposition}[Existence of treatment bridge]\label{prop: treatment bridge existence}
			There exists a solution \(q\in \mathcal{S}^{\perp}\) to Equation~\eqref{eq: treatment bridge} 
			if and only if \(Z\) is \(\mathcal{S}\)-relevant to \(A\). In addition, for any $\mathcal{S}$-PRWF $\pi$, $q_{\mathcal{S},\pi}(Z,X):=\frac{\pi(Z,X) - \mathbb{S}[\pi](Z,X)}{\mathbb{E}\left[\left\{\pi(Z,X) - \mathbb{S}[\pi](Z,X)\right\}A|X\right]}$ satisfies Equation~\eqref{eq: treatment bridge}.
		\end{proposition}
		Proposition~\ref{prop: treatment bridge existence} reveals a close connection between $\mathcal{S}$-relevance and the solvability of Equation~\eqref{eq: treatment bridge}. In particular, by Proposition~\ref{prop: existence of PRWF}, $q_{\mathcal{S},\pi_*^o}$ provides a natural candidate solution to the treatment bridge equation.
		Now, we can identify the ATE using a treatment bridge function.
		\begin{theorem}[Treatment bridge identification]\label{thm: identification treatment bridge}
			Suppose that $Z$ is an $\mathcal{S}$-PIV. Then, for any $q\in\mathcal{S}^{\perp}$ satisfying Equation~\eqref{eq: treatment bridge}, $\mathbb{E}\left[q(Z,X)Y\right] = \mathbb{E}[Y(1)-Y(0)]$. 
		\end{theorem}
		\begin{remark}
			Equation~\eqref{eq: treatment bridge} may admit multiple solutions, each of which can be used for identification. Selecting an efficient \(q(Z,X)\), or combining multiple treatment bridge solutions to improve the efficiency of the proposed estimator, is an interesting direction for future research but lies beyond the scope of this paper.
		\end{remark}
		Alternatively, we can also identify the ATE by constructing outcome bridge functions. 
		We consider the problem of finding an outcome bridge function $h^o\in L_2(X)$, such that
		\begin{align}\label{eq: outcome bridge}
			\mathbb E\big[d(Z,X)\{Y - Ah^o(X)\}\big] = 0, \qquad \forall\, d \in \mathcal{S}^{\perp}. 
		\end{align}
		To establish the uniqueness of the solution to Equation~\eqref{eq: outcome bridge}, we introduce the definition of $\mathcal{S}$-completeness.
		\begin{definition}[$\mathcal{S}$-Completeness]\label{defn: completeness}
			We say that $\mathcal{S}$-completeness holds if for any $g \in L_2(X)$, $g(X) \equiv 0$ if and only if for all $d \in \mathcal{S}^{\perp}$,
			$\mathbb{E}\!\left[d(Z,X) A\, g(X)\right] = 0$.
		\end{definition}
		Note that the $\mathcal{S}$-completeness condition is a weaker form of the $\mathcal{S}$-relevance condition.
		Now, we can establish identification of the ATE using the solution to Equation~\eqref{eq: outcome bridge} and $\mathcal{S}$-completeness.
		\begin{theorem}[Outcome bridge identification]\label{thm: identification outcome bridge} 
			(a) If there exists a solution $q\in \mathcal{S}^{\perp}$ to Equation~\eqref{eq: treatment bridge}, then $\mathcal{S}$-completeness holds.
			(b) If Equation~\eqref{eq: outcome bridge} admits a solution and $\mathcal{S}$-completeness holds, then the solution is unique. (c) If $Z$ is an $\mathcal{S}$-PIV for $A$, then $h^o(X)=\mathbb{E}\!\left[Y(1)-Y(0)\mid X\right]$ is a solution to Equation~\eqref{eq: outcome bridge}.
		\end{theorem}

		\begin{figure}[t]
			\centering
			\resizebox{0.9\textwidth}{!}{
				\begin{tikzpicture}[every node/.style={draw},>=Stealth, node distance=5cm, thick]
					\node (relevance) at (-15.5,-4) {$\mathcal{S}$-relevance};
					\node (bridge) at (-10,4) [align=center] {Existence of solution \\$q\in\mathcal{S}^{\perp}$ to Equation~\eqref{eq: treatment bridge}};
					\node (completeness) at (-10,-0) {$\mathcal{S}$-completeness};
					\node (PRWF) at (-15.5,4) {Existence of $\mathcal{S}$-PRWF};
					
					\draw[<->] (bridge) to node[midway, sloped, above, draw=none]
					{Proposition~\ref{prop: treatment bridge existence}} (relevance);
					\draw[<->] (bridge) to (PRWF);
					\draw[<->] (PRWF) to node[midway, sloped, above, draw=none]
					{Proposition~\ref{prop: existence of PRWF}}
					(relevance);
					\draw[->] (bridge) to node[midway, sloped, above, draw=none]
					{Theorem~\ref{thm: identification outcome bridge}(a)}(completeness);
					
					\node (identification) at (-2,-4) [align=center] {$h^o(X)=\mathbb{E}[Y(1)-Y(0)|X]$\\ is a solution to Equation~\eqref{eq: outcome bridge} for $\mathcal{S}$};
					\node (uniqueness) at (-10,-4) [align=center] {At most one solution \\to Equation~\eqref{eq: outcome bridge} for $\mathcal{S}$};
					\node[draw, fill=black!10] (id) at (-6, 0) [align=center] {Identification of ATE};
					\node (PIV) at (-2, 4) [align=center] {$\mathcal{S}$-PIV};
					
					\draw[->] (completeness) to node[midway, sloped, above, draw=none]
					{Theorem~\ref{thm: identification outcome bridge}(b)} (uniqueness);
					\draw[->] (PIV) to node[midway, sloped, above, draw=none]
					{Theorem~\ref{thm: identification outcome bridge}(c)} (identification);
					\draw[-] (identification) to (uniqueness);
					
					\draw[-] (bridge) to (PIV);
					\draw[->] (-6,4) to node[midway, sloped, above, draw=none]
					{Theorem~\ref{thm: identification treatment bridge}}(id);
					\draw[->] (-6,-4) to (id);
			\end{tikzpicture}}
			\caption{A flowchart of identification.}
			\label{fig: completeness relation}
		\end{figure}
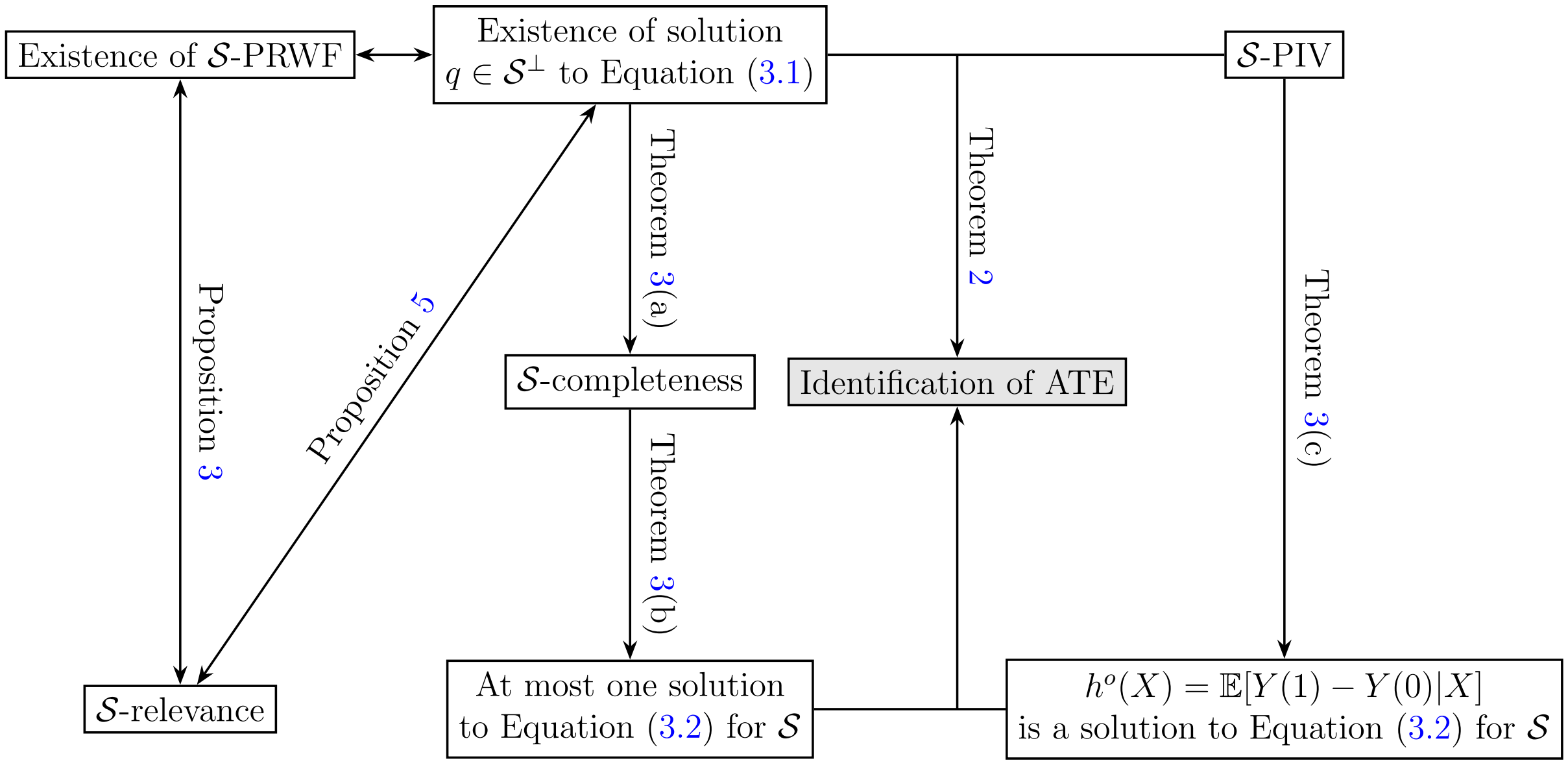	
		Theorem~\ref{thm: identification outcome bridge} shows that we can first find a solution $h^o(X)$ to Equation~\eqref{eq: outcome bridge}, and then identify the ATE by $\mathbb{E}[Y(1)-Y(0)]=\mathbb{E}[h^o(X)]$ if this solution is unique.
		Figure~\ref{fig: completeness relation} provides an overview of the key ideas and identification results developed in this section.
		
		\section{Semiparametric analysis}\label{sec: semiparametric analysis}
		\subsection{Debiased estimator with a fixed PRWF}
		In this subsection, we present semiparametric analysis for $\psi_{\mathcal{S},\pi}^o$ when a prespecified proper Hilbert subspace $\mathcal{S}$ and a corresponding $\mathcal{S}$-PRWF $\pi$ are given. In particular, define a nuisance vector as $\alpha_{\mathcal{S},\pi}^o=[\delta_{\mathcal{S}}^o, \eta_{\mathcal{S}}^o,\rho_{\mathcal{S},\pi}^o,\kappa_{\mathcal{S},\pi}^o,\mu_{\mathcal{S},\pi}^o]$, where
		\begin{align*}
			&\delta_{\mathcal{S}}^o(Z,X) := \mathbb{S}[A](Z,X), \qquad
			\eta_{\mathcal{S}}^o(Z,X) := \mathbb{S}[Y](Z,X),\qquad
			\rho_{\mathcal{S},\pi}^o(Z,X) := \mathbb{S}[\pi](Z,X), \\
			&\kappa_{\mathcal{S},\pi}^o(X) :=\mathbb{E}[\{\pi(Z,X)-\mathbb{S}[\pi](Z,X)\}\{A-\mathbb{S}[A](Z,X)\}\mid X],\\
			&\mu_{\mathcal{S},\pi}^o(X) := \dfrac{\mathbb{E}[\{\pi(Z,X)-\mathbb{S}[\pi](Z,X)\}\{Y-\mathbb{S}[Y](Z,X)\}\mid X]}{\mathbb{E}[\{\pi(Z,X)-\mathbb{S}[\pi](Z,X)\}\{A-\mathbb{S}[A](Z,X)\}\mid X]}.
		\end{align*}
		By the definitions of nuisance functions, it can be readily verified that 
		\[
		\psi_{\mathcal{S},\pi}^o
		= \mathbb{E}\!\left[\frac{\pi(Z,X) - \rho_{\mathcal{S},\pi}^o(Z,X)}{\kappa_{\mathcal{S},\pi}^o(X)}\,Y\right]
		= \mathbb{E}\!\left[\mu_{\mathcal{S},\pi}^o(X)\right].
		\]
		The former representation resembles an IPW functional, whereas the latter is analogous to an outcome regression (OR) functional.
		Next, we derive the efficient influence function (EIF) of $\psi_{\mathcal{S},\pi}^o$.
		\begin{theorem}[EIF]\label{thm: EIF ATE}
			Suppose that $\pi$ is an $\mathcal{S}$-PRWF. Under the differentiability conditions in Lemma~\ref{lem: semiparametric calculation} (see the Appendix for details), the EIF of $\psi_{\mathcal{S},\pi}^o$ in the nonparametric model is
			\begin{align*}
				&\varphi_{\mathcal{S},\pi}(O;\psi_{\mathcal{S},\pi},\alpha_{\mathcal{S},\pi})=\dfrac{\pi(Z,X) - \rho_{\mathcal{S},\pi}(Z,X)}{\kappa_{\mathcal{S},\pi}(X)}\left\{Y - \eta_{\mathcal{S}}(Z,X)\right\}\\
				&+\mu_{\mathcal{S},\pi}(X)\left(1 - \dfrac{\left\{\pi(Z,X) -\rho_{\mathcal{S},\pi}(Z,X)\right\}\left\{A-\delta_{\mathcal{S}}(Z,X)\right\}}{\kappa_{\mathcal{S},\pi}(X)}\right)-\psi_{\mathcal{S},\pi}.
			\end{align*}
		\end{theorem}
		\begin{remark}
			The efficiency analysis in this section is carried out relative to the nonparametric model. The semiparametric theory developed by \citet{dong2025marginalcausaleffectestimation} for the case $\mathcal{S}=L_2(X)$ could potentially be adapted to extend the semiparametric efficiency analysis of the proposed framework.
		\end{remark}
		Since all $\mathcal{S}$-PRWFs can be adopted to identify the ATE, we are interested in selecting an optimal weighting function. We define $\Psi^o(X) := \mathbb{E}[Y(1)-Y(0)\mid X]$, and demonstrate that $\pi_*^o$ is a good candidate. 
		\begin{proposition}[Lower efficiency bound]\label{prop: lower efficiency bound}
			Suppose that $Z$ is an $\mathcal{S}$-PIV that is  $\mathcal{S}$-relevant to $A$, and
			\begin{align}\label{eq: homoskedastic}
				\mathbb{E}\left[\left\{Y-A\Psi^o(X)-\mathbb{S}[Y-A\Psi^o](Z,X)\right\}^2\middle| Z,X\right]\indep Z\mid X.
			\end{align}
			Then $\mathbb{E}\left[\varphi_{\mathcal{S},\pi}(O;\psi_{\mathcal{S},\pi}^o,\alpha_{\mathcal{S},\pi}^o)^2\right]$ achieves its lower bound when taking $\pi=\pi_*^o$, provided that the influence functions being compared have finite second moments.
		\end{proposition}
		\begin{remark}
			The weighting functions that achieve the lower bound in Proposition~\ref{prop: lower efficiency bound} are not unique, since the asymptotic variance $\mathbb{E}\left[\varphi_{\mathcal{S},\pi}(O;\psi_{\mathcal{S},\pi}^o,\alpha_{\mathcal{S},\pi}^o)^2\right]$ induced by $\pi(Z,X)=\pi_*^o(Z,X) f(X)+g(X)$ is the same as that induced by $\pi_*^o(Z,X)$ for all choices such that $\pi$ remains an $\mathcal{S}$-PRWF. 
		\end{remark}
		The conditional independence condition in \eqref{eq: homoskedastic} can be interpreted as a generalized homoskedasticity assumption. Specifically, after removing both the conditional treatment effect component $A\Psi^o(X)$ and the projected component $\mathbb{S}[Y-A\Psi^o](Z, X)$, the remaining residual variation is assumed not to depend on $Z$ conditional on $X$. Under this condition, $\pi_*^o(Z,X)$ minimizes the asymptotic variance within the stated class of fixed-weight functionals. This comparison does not establish the efficiency bound in the smaller causal model defined by the PIV restrictions.

		\subsection{Adaptive estimator with optimal PRWF}\label{subsec: adaptive weighting function}
		In practice, it is unclear how to optimally select an $\mathcal{S}$-PRWF $\pi$ when $Z$ is $\mathcal{S}$-relevant to $A$. Propositions~\ref{prop: existence of PRWF} and \ref{prop: lower efficiency bound} motivate the choice of $\pi_*^o$. However, its true form is generally unknown, and directly utilizing a plug-in estimator $\hat{\pi}_*$ may introduce non-negligible estimation bias. A principled approach to address this challenge is to treat $\pi_*^o$ as a nuisance function and estimate it adaptively from the data. 
		In particular, suppose that $Z$ is $\mathcal{S}$-relevant to $A$. The parameter of interest is defined as
		\begin{align*}
			\psi_{\mathcal{S},*}^o:=\mathbb{E}\left[\frac{\mathbb{E}[A|Z,X] - \mathbb{S}[A](Z,X)}{\mathbb{E}\left[\{\mathbb{E}[A|Z,X] - \mathbb{S}[A](Z,X)\}^2|X\right]}Y\right].
		\end{align*}
		If $Z$ is also an $\mathcal{S}$-PIV, then $\psi_{\mathcal{S},*}^o = \mathbb{E}[Y(1)-Y(0)]$ by setting $\pi=\pi_*^o$ in Theorem~\ref{thm: identification}. 
		
		The unified nuisance function vector $\alpha_{\mathcal{S},*}^o:=[\pi_*^o,\delta_{\mathcal{S}}^o,\xi_{*}^o,\eta_{\mathcal{S}}^o,\kappa_{\mathcal{S},*}^o,\mu_{\mathcal{S},*}^o]$ includes 
		\begin{align*}
			&\pi_*^o(Z,X):=\mathbb{E}[A \mid Z,X],\quad
			\delta_{\mathcal{S}}^o(Z,X):=\mathbb{S}[A](Z,X),\\
			&\xi_{*}^o(Z,X):=\mathbb{E}[Y\mid Z,X],\quad
			\eta_{\mathcal{S}}^o(Z,X) := \mathbb{S}[Y](Z,X),\\
			&\kappa_{\mathcal{S},*}^o(X):=\mathbb{E}[\{\mathbb{E}[A|Z,X]-\mathbb{S}[A](Z,X)\}^2\mid X],\\
			&\mu_{\mathcal{S},*}^o(X) := \dfrac{\mathbb{E}[\{\mathbb{E}[A|Z,X]-\mathbb{S}[A](Z,X)\}\{\mathbb{E}[Y|Z,X]-\mathbb{S}[Y](Z,X)\}\mid X]}{\mathbb{E}[\{\mathbb{E}[A|Z,X]-\mathbb{S}[A](Z,X)\}^2\mid X]}.
		\end{align*}
		We now present a semiparametric analysis of $\psi_{\mathcal{S},*}^o$.
		\begin{theorem}[EIF]\label{thm: ATE EIF adaptive}
			If $Z$ is $\mathcal{S}$-relevant to $A$, then, under the differentiability conditions in Lemma~\ref{lem: semiparametric calculation}, the EIF of $\psi_{\mathcal{S},*}^o$ in the nonparametric model is
			\begin{align*}
				&\varphi_{\mathcal{S},*}(O;\psi_{\mathcal{S},*},\alpha_{\mathcal{S},*})=\frac{1}{\kappa_{\mathcal{S},*}(X)}\left\{(A-\pi_*(Z,X))\xi_{*}(Z,X) - (A-\delta_{\mathcal{S}}(Z,X))\eta_{\mathcal{S}}(Z,X)
				\right\}\\
				&+\frac{\pi_*(Z,X) - \delta_{\mathcal{S}}(Z,X)}{\kappa_{\mathcal{S},*}(X)}Y
				+ \dfrac{\mu_{\mathcal{S},*}(X)}{\kappa_{\mathcal{S},*}(X)}
				\left\{\kappa_{\mathcal{S},*}(X) + (A - \pi_*(Z,X))^2 - (A - \delta_{\mathcal{S}}(Z,X))^2  \right\}-\psi_{\mathcal{S},*}.
			\end{align*}
		\end{theorem}
		
		Compared with Theorem~\ref{thm: EIF ATE}, Theorem~\ref{thm: ATE EIF adaptive} only requires the $\mathcal{S}$-relevance condition, which avoids prespecifying a known $\mathcal{S}$-PRWF. Moreover, the following proposition establishes the efficiency optimality of the adaptive estimator.
		
		\begin{proposition}\label{prop: adaptive estimator achieve lower efficiency bound}
			If $Z$ is an $\mathcal{S}$-PIV for $A$ and is $\mathcal{S}$-relevant to $A$, then $\mathbb{E}[\varphi_{\mathcal{S},*}(O;\psi_{\mathcal{S},*}^o,\alpha_{\mathcal{S},*}^o)^2] = \mathbb{E}[\varphi_{\mathcal{S},\pi_*^o}(O;\psi_{\mathcal{S},\pi_*^o}^o,\alpha_{\mathcal{S},\pi_*^o}^o)^2]$.
		\end{proposition}
		
		Proposition~\ref{prop: adaptive estimator achieve lower efficiency bound} shows that under the restriction of Equation~\eqref{eq: homoskedastic}, the adaptive estimator attains the minimum asymptotic variance characterized in Proposition~\ref{prop: lower efficiency bound}. Because the optimal choice $\pi_*^o$ is generally unknown in practice, we use the adaptive procedure associated with $\psi_{\mathcal{S},*}^o$ as our default approach to estimating the ATE and constructing the corresponding confidence intervals.
		
		\subsection{Asymptotic theory}\label{subsec: asymptotic theory}
		To reduce overfitting bias arising from nuisance function estimation and to relax empirical process conditions, we employ a $K$-fold cross-fitting procedure \citep{schick1986asymptotically, Chernozhukov2018}. 
		Specifically, we partition the data into $K$ folds $\{I_k\}_{k=1}^K$, where $K$ is fixed and $|I_k|=n/K$ (for simplicity, assume that $K$ divides $n$). For each fold $k$, the nuisance functions are estimated using observations in the complement $I_{-k} := \{1,\dots,n\}\setminus I_k$, yielding estimators $\hat\alpha_{\mathcal{S},\pi}^{(n,-k)}$ and $\hat\alpha_{\mathcal{S},*}^{(n,-k)}$. For any integrable function $f$, write $\mathbb{E}_{nk}[f(O)]:=|I_k|^{-1}\sum_{i\in I_k}f(O_i)$.
		The estimators $\hat\psi_{\mathcal{S},\pi}$ and $\hat\psi_{\mathcal{S},*}$ are defined as the solutions to the following estimating equations:
		\begin{align*}
			\dfrac{1}{K}\sum_{k=1}^K\mathbb{E}_{nk}\left[\varphi_{\mathcal{S},\pi}\left(O;\hat\psi_{\mathcal{S},\pi},\hat\alpha_{\mathcal{S},\pi}^{(n,-k)}\right)\right]=0;\quad
			\dfrac{1}{K}\sum_{k=1}^K\mathbb{E}_{nk}\left[\varphi_{\mathcal{S},*}\left(O;\hat\psi_{\mathcal{S},*},\hat\alpha_{\mathcal{S},*}^{(n,-k)}\right)\right]=0.
		\end{align*}
		The estimated asymptotic variances of $\sqrt n(\hat\psi_{\mathcal{S},\pi}-\psi_{\mathcal{S},\pi}^o)$ and $\sqrt n(\hat\psi_{\mathcal{S},*}-\psi_{\mathcal{S},*}^o)$ are given by
		\begin{align*}
			(\hat\sigma_{\mathcal{S},\pi})^2
			&:= \frac{1}{K}\sum_{k=1}^K \mathbb{E}_{nk}\!\left[\varphi_{\mathcal{S},\pi}\!\left(O;\hat\psi_{\mathcal{S},\pi},\hat\alpha_{\mathcal{S},\pi}^{(n,-k)}\right)^2\right];\quad
			(\hat\sigma_{\mathcal{S},*})^2
			:= \frac{1}{K}\sum_{k=1}^K \mathbb{E}_{nk}\!\left[\varphi_{\mathcal{S},*}\!\left(O;\hat\psi_{\mathcal{S},*},\hat\alpha_{\mathcal{S},*}^{(n,-k)}\right)^2\right].
		\end{align*}
		
		To mitigate the bias induced by plugging in estimated nuisance functions, we next establish the mixed bias property \citep{rotnitzky2021characterization} of the proposed cross-fitted estimator. Specifically, we show that its bias decomposes into products of nuisance estimation errors and therefore has a second-order bias structure under mild regularity conditions.
		\begin{proposition}[Mixed bias property]\label{prop: mix bias property}
			Suppose that $\pi$ is an $\mathcal{S}$-PRWF for $A$. For any $\alpha_{\mathcal{S},\pi}=[\delta_{\mathcal{S}}, \eta_{\mathcal{S}},\rho_{\mathcal{S},\pi},\kappa_{\mathcal{S},\pi},\mu_{\mathcal{S},\pi}]$ with $\delta_{\mathcal{S}}, \eta_{\mathcal{S}}, \rho_{\mathcal{S},\pi}\in\mathcal{S}$, $\mathbb{E}[\varphi_{\mathcal{S},\pi}(O;\psi_{\mathcal{S},\pi}^o,\alpha_{\mathcal{S},\pi})]$ equals
			\begin{align*}
				\mathbb{E}\left[\dfrac{1}{\kappa_{\mathcal{S},\pi}(X)}
				\left(\begin{array}{l}
					\left\{\mu_{\mathcal{S},\pi}(X)-\mu_{\mathcal{S},\pi}^o(X)\right\}\left\{\kappa_{\mathcal{S},\pi}(X) - \kappa_{\mathcal{S},\pi}^o(X)\right\}\\
					+\{\rho_{\mathcal{S},\pi}^o(Z,X)-\rho_{\mathcal{S},\pi}(Z,X)\}
					\left\{\eta_{\mathcal{S}}^o(Z,X)-\eta_{\mathcal{S}}(Z,X)\right\}\\
					-\mu_{\mathcal{S},\pi}(X)
					\left\{\rho_{\mathcal{S},\pi}^o(Z,X) -\rho_{\mathcal{S},\pi}(Z,X)\right\}\left\{\delta_{\mathcal{S}}^o(Z,X)-\delta_{\mathcal{S}}(Z,X)\right\}
				\end{array}\right)\right].
			\end{align*}
			Alternatively, under the $\mathcal{S}$-relevance condition, for any $\alpha_{\mathcal{S},*}:=[\pi_*,\delta_{\mathcal{S}},\xi_{*},\eta_{\mathcal{S}},\kappa_{\mathcal{S},*},\mu_{\mathcal{S},*}]$ with $\delta_{\mathcal{S}},\;\eta_{\mathcal{S}}\in \mathcal{S}$,
			$\mathbb{E}[\varphi_{\mathcal{S},*}(O;\psi_{\mathcal{S},*}^o,\alpha_{\mathcal{S},*})]$ equals
			\begin{align*}
				\mathbb{E}\left[
				\frac{1}{\kappa_{\mathcal{S},*}(X)}\left(\begin{array}{l}
					\left\{\mu_{\mathcal{S},*}(X) - \mu_{\mathcal{S},*}^o(X)\right\}
					\left\{\kappa_{\mathcal{S},*}(X) - \kappa_{\mathcal{S},*}^o(X)\right\}\\
					-\{\pi_*(Z,X)-\pi_*^o(Z,X)\}\{\xi_{*}(Z,X)-\xi_{*}^o(Z,X)\} \\
					+ \{\delta_{\mathcal{S}}(Z,X)-\delta_{\mathcal{S}}^o(Z,X)\}\{\eta_{\mathcal{S}}(Z,X)-\eta_{\mathcal{S}}^o(Z,X)\}\\
					+\{(\pi_*(Z,X) - \pi_*^o(Z,X))^2 - (\delta_{\mathcal{S}}(Z,X) - \delta_{\mathcal{S}}^o(Z,X))^2\}\mu_{\mathcal{S},*}(X)
				\end{array}\right)
				\right].
			\end{align*}
		\end{proposition}
		Now, we can establish that $\hat\psi_{\mathcal{S},\pi}$ and $\hat\psi_{\mathcal{S},*}$ are consistent and asymptotically normal, and that $\hat\sigma_{\mathcal{S},\pi}$ and $\hat\sigma_{\mathcal{S},*}$ are consistent.
		\begin{theorem}[Asymptotic normality of $\hat\psi_{\mathcal{S},\pi}$]\label{thm: asymptotic normality}
			Suppose that $Z$ is an $\mathcal{S}$-PIV, $\pi(Z,X)$ is a uniformly bounded $\mathcal{S}$-PRWF for $A$, $Y$ is bounded, $0<(\sigma_{\mathcal{S},\pi}^o)^2\!:=\!\mathbb{E}[\varphi_{\mathcal{S},\pi}(O;\psi_{\mathcal{S},\pi}^o,\alpha_{\mathcal{S},\pi}^o)^2]<\infty$, and $\hat\delta_{\mathcal{S}}^{(n,-k)},\hat\eta_{\mathcal{S}}^{(n,-k)},\hat\rho_{\mathcal{S},\pi}^{(n,-k)}\in\mathcal{S}$ almost surely. Assume further that, for any $k=1,\ldots,K$, $\hat \alpha_{\mathcal{S},\pi}^{(n,-k)}$ is uniformly bounded almost surely, $\hat \kappa_{\mathcal{S},\pi}^{(n,-k)}$ is uniformly bounded away from zero almost surely,
			$\mathbb{E}[\|\hat \alpha_{\mathcal{S},\pi}^{(n,-k)}-\alpha_{\mathcal{S},\pi}^o\|_2^2]=o(1)$, and that
			\begin{align*}
				\left\{\begin{array}{l}
					\|\hat\mu_{\mathcal{S},\pi}^{(n,-k)}-\mu_{\mathcal{S},\pi}^o\|_2\times
					\|\hat\kappa_{\mathcal{S},\pi}^{(n,-k)}-\kappa_{\mathcal{S},\pi}^o\|_2
					+\|\hat\rho_{\mathcal{S},\pi}^{(n,-k)}-\rho_{\mathcal{S},\pi}^o\|_2\times
					\|\hat\delta_{\mathcal{S}}^{(n,-k)}-\delta_{\mathcal{S}}^o\|_2\\
					+\|\hat\rho_{\mathcal{S},\pi}^{(n,-k)}-\rho_{\mathcal{S},\pi}^o\|_2\times
					\|\hat\eta_{\mathcal{S}}^{(n,-k)}-\eta_{\mathcal{S}}^o\|_2\\
				\end{array}\right\}
				=o_p(n^{-1/2}).
			\end{align*}
			Then, $\sqrt{n}(\hat\psi_{\mathcal{S},\pi}-\psi_{\mathcal{S},\pi}^o)/\sigma_{\mathcal{S},\pi}^o\overset{d}{\rightarrow}\mathcal{N}(0,1)$, and $\hat\sigma_{\mathcal{S},\pi}\overset{p}{\rightarrow}\sigma_{\mathcal{S},\pi}^o$.
		\end{theorem}
		
		\begin{theorem}[Asymptotic normality of $\hat\psi_{\mathcal{S},*}$]\label{thm: asymptotic normality adaptive}
			Suppose that $Z$ is an $\mathcal{S}$-PIV that is $\mathcal{S}$-relevant to $A$, $Y$ is bounded, $0<(\sigma_{\mathcal{S},*}^o)^2\!:=\!\mathbb{E}[\varphi_{\mathcal{S},*}(O;\psi_{\mathcal{S},*}^o,\alpha_{\mathcal{S},*}^o)^2]<\infty$, and $\hat\delta_{\mathcal{S}}^{(n,-k)},\hat\eta_{\mathcal{S}}^{(n,-k)}\in\mathcal{S}$ almost surely. In addition, for $k=1,\ldots,K$, assume that $\hat \alpha_{\mathcal{S},*}^{(n,-k)}$ is uniformly bounded almost surely, $\hat \kappa_{\mathcal{S},*}^{(n,-k)}$ is uniformly bounded away from zero almost surely,
			$\mathbb{E}[\|\hat \alpha_{\mathcal{S},*}^{(n,-k)}-\alpha_{\mathcal{S},*}^o\|_2^2]=o(1)$, and that
			\begin{align*}
				\left\{
				\begin{array}{l}
					\|\hat\mu_{\mathcal{S},*}^{(n,-k)}-\mu_{\mathcal{S},*}^o\|_2 \times \|\hat\kappa_{\mathcal{S},*}^{(n,-k)}-\kappa_{\mathcal{S},*}^o\|_2
					+\|\hat\delta_{\mathcal{S}}^{(n,-k)}-\delta_{\mathcal{S}}^o\|_2^2+\|\hat\pi_*^{(n,-k)}-\pi_*^o\|_2^2\\
					+\|\hat\xi_{*}^{(n,-k)}-\xi_{*}^o\|_2\times \|\hat\pi_*^{(n,-k)}-\pi_*^o\|_2
					+\|\hat\eta_{\mathcal{S}}^{(n,-k)}-\eta_{\mathcal{S}}^o\|_2\times \|\hat\delta_{\mathcal{S}}^{(n,-k)}-\delta_{\mathcal{S}}^o\|_2
				\end{array}\right\}
				=o_p(n^{-1/2}).
			\end{align*}
			Then, $\sqrt{n}(\hat\psi_{\mathcal{S},*}-\psi_{\mathcal{S},*}^o)/\sigma_{\mathcal{S},*}^o\overset{d}{\rightarrow}\mathcal{N}(0,1)$, and $(\hat\sigma_{\mathcal{S},*})^2\overset{p}{\rightarrow}(\sigma_{\mathcal{S},*}^o)^2$.
		\end{theorem}
		Notably, in Theorem~\ref{thm: asymptotic normality adaptive}, we only require $\mathcal{S}$-relevance between $Z$ and $A$, and do not need to prespecify an $\mathcal{S}$-PRWF.

		\subsection{Nuisance function estimation}
		In this subsection, we provide guidance on estimating the nuisance functions $\kappa_{\mathcal{S},*}^o$ and $\mu_{\mathcal{S},*}^o$, which depend on other nuisance functions and therefore require estimated pseudo outcomes when fitted by least squares regression. In particular, define $\Gamma_{\mathcal{S},*}^o(X)=\mathbb{E}[\{\mathbb{E}[A\mid Z,X]-\mathbb{S}[A](Z,X)\}\{Y-\mathbb{S}[Y](Z,X)\}\mid X].$
		We estimate $\mu_{\mathcal{S},*}^o$ using the identity $\mu_{\mathcal{S},*}^o(X)=\Gamma_{\mathcal{S},*}^o(X)/\kappa_{\mathcal{S},*}^o(X)$.
		
		The definitions of $\Gamma_{\mathcal{S},*}^o$ and $\kappa_{\mathcal{S},*}^o$ rely on the remaining nuisance functions in $\alpha_{\mathcal{S},*}^o$. Therefore, we utilize the orthogonal statistical learning framework \citep{foster2023orthogonalstatisticallearning} to mitigate the bias induced by using estimated nuisance functions to estimate $\Gamma_{\mathcal{S},*}^o$ and $\kappa_{\mathcal{S},*}^o$. 
		In particular, for each training sample $I_{-k}$, we split the observations into $L$ equally sized folds $\{I_{kl}\}_{l=1}^L$ and estimate $\hat\pi_*^{(n,-kl)}$, $\hat\delta_{\mathcal{S}}^{(n,-kl)}$, $\hat\xi_*^{(n,-kl)}$ and $\hat\eta_{\mathcal{S}}^{(n,-kl)}$ using the observations in $I_{-kl}:=I_{-k}\backslash I_{kl}$. Then, define two augmented squared loss functions
		\begin{align*}
			&\mathcal{L}_{\kappa}(O;\kappa,\delta_{\mathcal{S}},\pi_*):=\frac{1}{2}\{(A-\delta_{\mathcal{S}}(Z,X))^2 - (A-\pi_*(Z,X))^2 - \kappa(X)\}^2;\\
			&\mathcal{L}_{\Gamma}(O;\Gamma,\delta_{\mathcal{S}},\pi_*,\eta_{\mathcal{S}},\xi_*):=\frac{1}{2}\left\{\begin{array}{l}
				(A-\pi_*(Z,X))\xi_*(Z,X) - (A-\delta_{\mathcal{S}}(Z,X))\eta_{\mathcal{S}}(Z,X)\\
				+(\pi_*(Z,X)-\delta_{\mathcal{S}}(Z,X))Y -\Gamma(X)
			\end{array}\right\}^2.
		\end{align*}
		Let $\mathcal{F}_{\kappa}$ and $\mathcal{F}_{\Gamma}$ be two prespecified function classes. For any integrable function $f$, write $\mathbb{E}_{nkl}[f(O)]:=|I_{kl}|^{-1}\sum_{i\in I_{kl}}f(O_i)$. Then, $\hat{\kappa}_{\mathcal{S},*}^{(n,-k)}$ and $\hat{\Gamma}_{\mathcal{S},*}^{(n,-k)}$ can be estimated by
		\begin{align*}
			&\hat{\kappa}_{\mathcal{S},*}^{(n,-k)}:=\underset{\kappa\in\mathcal{F}_{\kappa}}{\arg\min}\left\{ \frac{1}{L}\sum_{l=1}^L\mathbb{E}_{nkl}\left[\mathcal{L}_{\kappa}(O;\kappa,\hat\delta_{\mathcal{S}}^{(n,-kl)},\hat\pi_*^{(n,-kl)})\right]\right\};\\
			&\hat{\Gamma}_{\mathcal{S},*}^{(n,-k)}:=\underset{\Gamma\in\mathcal{F}_{\Gamma}}{\arg\min}\left\{ \frac{1}{L}\sum_{l=1}^L\mathbb{E}_{nkl}\left[\mathcal{L}_{\Gamma}(O;\Gamma,\hat\delta_{\mathcal{S}}^{(n,-kl)},\hat\pi_*^{(n,-kl)},\hat\eta_{\mathcal{S}}^{(n,-kl)},\hat\xi_*^{(n,-kl)})\right]\right\}.
		\end{align*}
		Then, the $L_2$-convergence rates of $\hat{\kappa}_{\mathcal{S},*}^{(n,-k)}$ and $\hat{\Gamma}_{\mathcal{S},*}^{(n,-k)}$ are governed by the critical radii of $\mathcal{F}_{\kappa}$ and $\mathcal{F}_{\Gamma}$, respectively \citep{BartlettBousquetMendelson2005}.
		\begin{theorem}[Convergence rate of nuisance estimators]\label{thm: convergence rate of nuisance estimator}
			Let $\mathcal{F}_{\kappa},\mathcal{F}_{\Gamma}\subseteq L_2(X)$ be two function classes with finite $L_{\infty}$-norms and critical radii $c_{n,\kappa}$ and $c_{n,\Gamma}$, and $\hat\delta_{\mathcal{S}}^{(n,-kl)},\hat\eta_{\mathcal{S}}^{(n,-kl)}\in\mathcal{S}$ almost surely. Assume further that $Y$ and all first-stage nuisance estimators are uniformly bounded by a fixed constant almost surely, and that $\mathcal{F}_{\kappa}$ and $\mathcal{F}_{\Gamma}$ are convex. If $\kappa_{\mathcal{S},*}^o\in\mathcal{F}_{\kappa}$ and $\Gamma_{\mathcal{S},*}^o\in\mathcal{F}_{\Gamma}$, then with probability $1-\epsilon$, for each $k=1,\ldots, K$,
			\begin{align*}
				&\|\hat\kappa_{\mathcal{S},*}^{(n,-k)}-\kappa_{\mathcal{S},*}^o\|_2^2 \lesssim  c_{n,\kappa}^2 +\frac{\log(1/\epsilon)}{n}+ \sum_{l=1}^L\left\{\|\hat\pi_*^{(n,-kl)}-\pi_*^o\|_4^4+\|\hat\delta_{\mathcal{S}}^{(n,-kl)}-\delta_{\mathcal{S}}^o\|_4^4\right\};\\
				&\|\hat\Gamma_{\mathcal{S},*}^{(n,-k)}-\Gamma_{\mathcal{S},*}^o\|_2^2 \lesssim  c_{n,\Gamma}^2 +\frac{\log(1/\epsilon)}{n}+ \sum_{l=1}^L\left\{\begin{array}{l}
					\|\hat\pi_*^{(n,-kl)}-\pi_*^o\|_4^4+\|\hat\delta_{\mathcal{S}}^{(n,-kl)}-\delta_{\mathcal{S}}^o\|_4^4\\
					+\|\hat\xi_{*}^{(n,-kl)}-\xi_{*}^o\|_4^4+\|\hat\eta_{\mathcal{S}}^{(n,-kl)}-\eta_{\mathcal{S}}^o\|_4^4
				\end{array}\right\}.
			\end{align*}
		\end{theorem}
		To achieve the $o_p(n^{-1/4})$ convergence rates of 
		$\|\hat\Gamma_{\mathcal{S},*}^{(n,-k)}-\Gamma_{\mathcal{S},*}^o\|_2$ and 
		$\|\hat\kappa_{\mathcal{S},*}^{(n,-k)}-\kappa_{\mathcal{S},*}^o\|_2$, it suffices to require $o(n^{-1/4})$ convergence rates of the critical radii of $\mathcal{F}_{\kappa}$ and $\mathcal{F}_{\Gamma}$, and $o_p(n^{-1/8})$ convergence rates of $L_4$ norms of $\hat\pi_*^{(n,-kl)}$, $\hat\delta_{\mathcal{S}}^{(n,-kl)}$, $\hat\xi_*^{(n,-kl)}$, and $\hat\eta_{\mathcal{S}}^{(n,-kl)}$.
		The procedure for estimating the nuisance functions in $\alpha_{\mathcal{S},\pi}^o$ can be established similarly and is therefore omitted.

		\section{Application to invalid IV}\label{sec: connection to invalid IV}
		\subsection{Identifying assumptions}
		To illustrate the applicability of our proposed PIV framework, we consider an invalid IV setting. First, we define $\langle d_Z \rangle:=\{1,\dots,d_Z\}$, and let $\nu \subseteq \langle d_Z \rangle$ be any subset with distinct indices. Let $|\nu|$ denote the cardinality of $\nu$. Define $Z_{[\nu]}:=\{Z_{[j]}\}_{j\in\nu}$ and $Z_{[-\nu]}:=\{Z_{[j]}\}_{j\notin\nu}$.
		The following assumptions are stated with respect to a non-empty candidate index set \(\nu_0\subseteq\langle d_Z\rangle\).
		\begin{assumption}[IV independence]\label{as: IV independence}
			$Z_{[\nu_0]}\indep U\mid Z_{[-\nu_0]},X$.
		\end{assumption}
		\begin{assumption}[Strong latent ignorability]\label{as: strong latent ignorability}
			$Y(a)\indep \{A,Z_{[\nu_0]}\}\mid U,Z_{[-\nu_0]},X$ for all $a\in\mathcal{A}$.
		\end{assumption}
		\renewcommand\theassumption{5.2$'$}
		\begin{assumption}[Weak latent ignorability]\label{as: weak latent ignorability}
			$\mathbb{E}[Y(1)-Y(0)| A,U,Z,X]\indep \{A,Z_{[\nu_0]}\}\mid U,Z_{[-\nu_0]},X$.
		\end{assumption}
		\renewcommand{\theassumption}{\thesection.\number\numexpr\value{assumption}-1\relax}
		\begin{assumption}[$Z$-invariant treatment effect]\label{as: Z-invariant treatment effect}
			$\mathbb{E}[Y(1)-Y(0)\mid Z_{[-\nu_0]},X] \indep Z_{[-\nu_0]}\mid X$.
		\end{assumption}
		
		\begin{assumption}\label{as: alternative}
			At least one of the
			following conditions holds:
			\begin{enumerate}
				\item[(a)] \emph{(Additive IV)}
				For some $\gamma\in\{1,\ldots,|\nu_0|\}$, there exist
				$\{b_\nu(U,Z_{[-\nu]},X):
				\nu\subseteq\nu_0,\ |\nu|=\gamma\}$
				and $c(Z,X)$ such that
				$
				\Pr(A=1\mid Z,U,X)
				=
				\sum_{\nu\subseteq\nu_0,|\nu|=\gamma}
				b_\nu(U,Z_{[-\nu]},X)+c(Z,X).
				$
				\item[(b)] \emph{($U$-invariant conditional treatment effect)}
				$\mathbb{E}[Y(1)-Y(0)\mid U,Z_{[-\nu_0]},X] \indep U\mid Z_{[-\nu_0]},X$.
			\end{enumerate}
			If condition~(b) holds, define
			$\gamma_\dagger(\nu_0)=|\nu_0|$.
			Otherwise, define $\gamma_\dagger(\nu_0)$ as the largest
			$\gamma\in\{1,\ldots,|\nu_0|\}$ for which condition~(a) holds.
		\end{assumption}

		\begin{remark}
			As shown in Proposition~\ref{prop: uniqueness of largest nu} of Appendix~\ref{append: uniqueness of maximal index}, under Assumption~\ref{as: Z-invariant treatment effect}, $\gamma_\dagger$ is nondecreasing with respect to set inclusion: $\nu_1\subseteq\nu_2$ implies $\gamma_\dagger(\nu_1)\leq\gamma_\dagger(\nu_2).$ 
		\end{remark}

		Assumption~\ref{as: IV independence} states that $Z_{[\nu_0]}$ is independent of the unmeasured confounders $U$, conditional on $Z_{[-\nu_0]}$ and $X$.
		Assumption~\ref{as: strong latent ignorability} can be interpreted as a latent IV ignorability condition that accounts for unmeasured confounding effects through $U$. 
		Assumption~\ref{as: Z-invariant treatment effect} imposes that $\mathbb{E}[Y(1)-Y(0)\mid Z_{[-\nu_0]},X]$ does not depend on $Z_{[-\nu_0]}$.

		However, despite the simplicity and interpretability of Assumption~\ref{as: strong latent ignorability}, it can be relaxed to Assumption~\ref{as: weak latent ignorability}, which only requires that $\mathbb{E}[Y(1)-Y(0)\mid A,U,Z,X]$ does not depend on $A, Z_{[\nu_0]}$, and is therefore strictly weaker than Assumption~\ref{as: strong latent ignorability}.
		For example, suppose that Assumption~\ref{as: IV independence} holds for $\nu_0=\langle d_Z \rangle$, and that the potential outcomes are generated, for every $a\in\mathcal{A}$, as 
		$$Y(a)=af(X) + e_0(U,X)+\sum_{j=1}^{d_Z}e_j(U,X)g_j(Z_{[j]},X)+\epsilon,\text{ with $\mathbb{E}[\epsilon|Z,U,X]=0$.}$$
		In this setting, Assumption~\ref{as: weak latent ignorability} holds for $\nu_0=\langle d_Z \rangle$, whereas non-constant direct effects of $Z$ on the potential outcomes can violate Assumption~\ref{as: strong latent ignorability}.
		
		Assumption~\ref{as: alternative} offers two alternative identification routes, since it suffices for either condition to hold. On the one hand, Assumption~\ref{as: alternative}(a) imposes an additive structure of the treatment assignment mechanism, in which conditional on $Z_{[-\nu_0]}$ and $X$, the treatment probability admits a decomposition in which each term involving $U$ depends on at most $|\nu_0|-\gamma_{\dagger}(\nu_0)$ components of $Z_{[\nu_0]}$. This relaxes the additive IV structure in \citet{wang2018bounded,cui2023instrumental,Hartwig2023,michael2024instrumental}. 	
		On the other hand, Assumption~\ref{as: alternative}(b) requires that $\mathbb{E}[Y(1)-Y(0)\mid U,Z_{[-\nu_0]},X]$ be invariant to $U$. When $\nu_0=\langle d_Z\rangle$, this reduces to the no additive $U$-$Y$ interaction condition of \citet{wooldridge2010econometric, wang2018bounded, Hartwig2023}.

\begin{figure}[t]
			\centering
			\begin{subfigure}[c]{0.55\textwidth}
				\centering
				\resizebox{\textwidth}{!}{
					\begin{tikzpicture}[>=Stealth, node distance=1cm, thick]
						\node (Z) [draw] at (-2.5,0) {$Z_{[\nu_0]}$};
						\node (A) [draw] at (0,0) {$A$};
						\node (U) [draw, fill=black!10] at (1.25,2.5) {$U$};
						\node (X) [draw] at (-1.25,2.5) {$Z_{[-\nu_0]}$};
						\node (Y0) [draw,fill=black!10] at (2.5,0) {$Y(0)$};
						\node (Y10)[draw,fill=black!10] at (5,0) {$Y(1)-Y(0)$};
						\node (Y)  [draw,fill=red!10] at (3.75,2.5) {$Y$};
						
						\draw[->] (Z) -- (A);
						\draw[->] (U) -- (A);
						\draw[->] (U) -- (Y0);
						
						\draw[->] (U) -- (Y10);
						\draw[->] (X) -- (Z);
						\draw[->] (X) -- (A);
						\draw[->] (X) -- (Y0);
						\draw[->,dashed] (Z) to[bend left=30] (Y0);
						
						\draw[->, red] (A) -- (Y);
						\draw[->, red] (Y0) -- (Y);
						\draw[->, red] (Y10) -- (Y);
						
				\end{tikzpicture}}
				\caption{Satisfying Assumptions~\ref{as: IV independence}, \ref{as: weak latent ignorability}, and \ref{as: Z-invariant treatment effect}.}
				\centering
				\resizebox{\textwidth}{!}{
					\begin{tikzpicture}[>=Stealth, node distance=1cm, thick]
						\node (Z) [draw] at (-2.5,0) {$Z_{[\nu_0]}$};
						\node (A) [draw] at (0,0) {$A$};
						\node (U) [draw, fill=black!10] at (1.25,2.5) {$U$};
						\node (X) [draw] at (-1.25,2.5) {$Z_{[-\nu_0]}$};
						\node (Y0) [draw,fill=black!10] at (2.5,0) {$Y(0)$};
						\node (Y10)[draw,fill=black!10] at (5,0) {$Y(1)-Y(0)$};
						\node (Y)  [draw,fill=red!10] at (3.75,2.5) {$Y$};
						
						\draw[->] (Z) -- (A);
						\draw[->] (U) -- (A);
						\draw[->] (U) -- (Y0);
						\draw[->] (X) -- (Z);
						\draw[->] (X) -- (A);
						\draw[<->] (X) -- (U);
						\draw[->] (X) -- (Y0);
						\draw[->,dashed] (Z) to[bend left=30] (Y0);
						
						\draw[->, red] (A) -- (Y);
						\draw[->, red] (Y0) -- (Y);
						\draw[->, red] (Y10) -- (Y);
				\end{tikzpicture}}
				\caption{Satisfying Assumptions~\ref{as: IV independence}, \ref{as: weak latent ignorability},  \ref{as: Z-invariant treatment effect} and \ref{as: alternative}(b).}
			\end{subfigure}
			\begin{subfigure}[c]{0.425\textwidth}
				\centering
				\resizebox{\textwidth}{!}{
					\begin{tikzpicture}[>=Stealth, node distance=1cm, thick]
						\node (Z) [draw] at (-2.5,0) {$Z_{[\nu_0]}$};
						\node (A) [draw] at (0,0) {$A$};
						\node (Y) [draw] at (2.5,0) {$Y$};
						\node (U) [draw, fill=black!10] at (1.25,2.5) {$U$};
						\node (X) [draw] at (-1.25,2.5) {$Z_{[-\nu_0]}$};
						
						\draw[->] (Z) -- (A);
						\draw[->] (A) -- (Y);
						\draw[->] (U) -- (A);
						\draw[->] (U) -- (Y);
						\draw[->] (X) -- (Z);
						\draw[->] (X) -- (A);
						\draw[<->] (X) -- (U);
						\draw[->] (X) -- (Y);
						\draw[->,dashed] (Z) to[bend left=30] (Y);
				\end{tikzpicture}}
				\caption{Satisfying Assumption~\ref{as: IV independence}.}
			\end{subfigure}
			\caption{Three causal graphs with $X=\emptyset$. Assumption~\ref{as: strong latent ignorability} holds without dashed arrows but fails with them.}
			\label{fig: DAG}
		\end{figure}
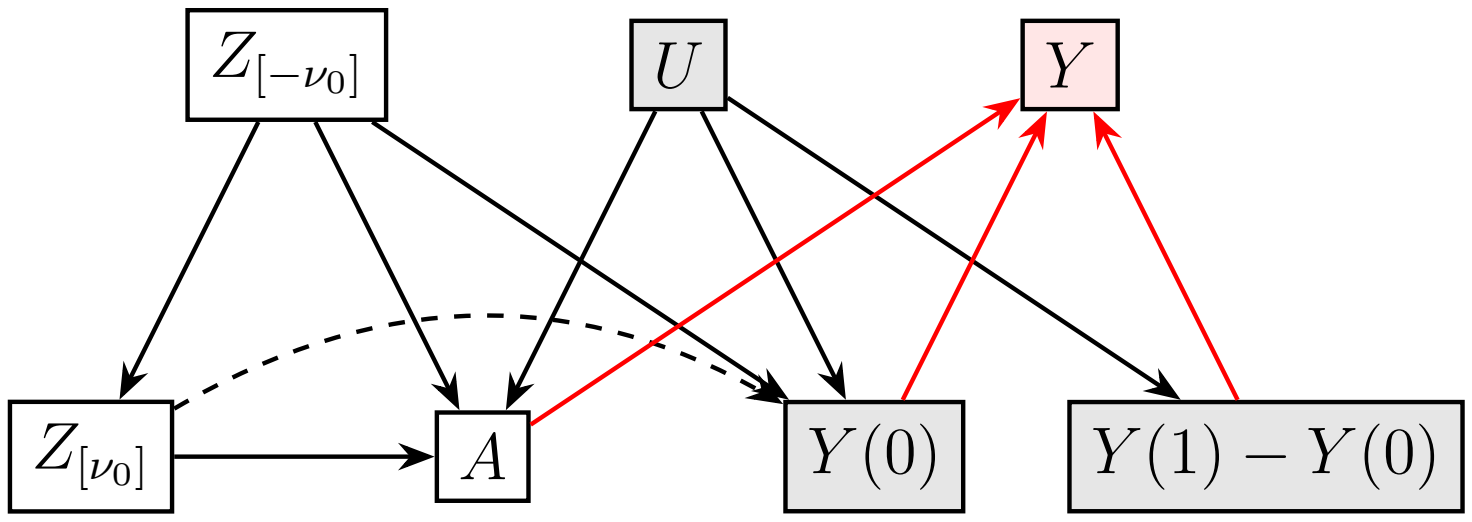
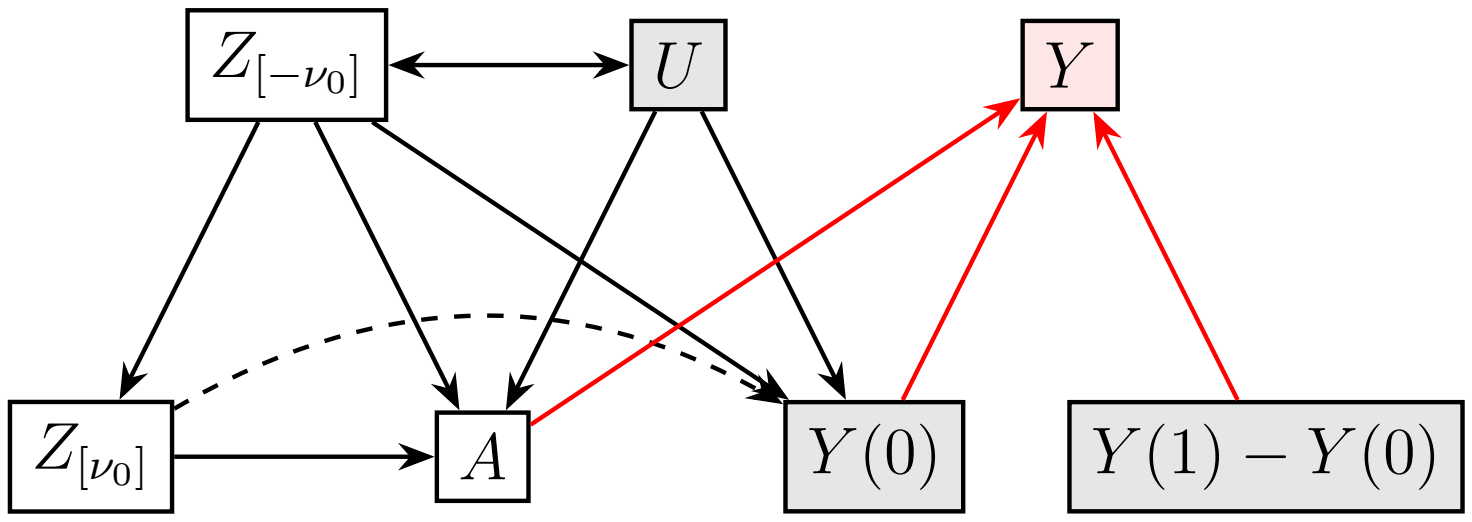
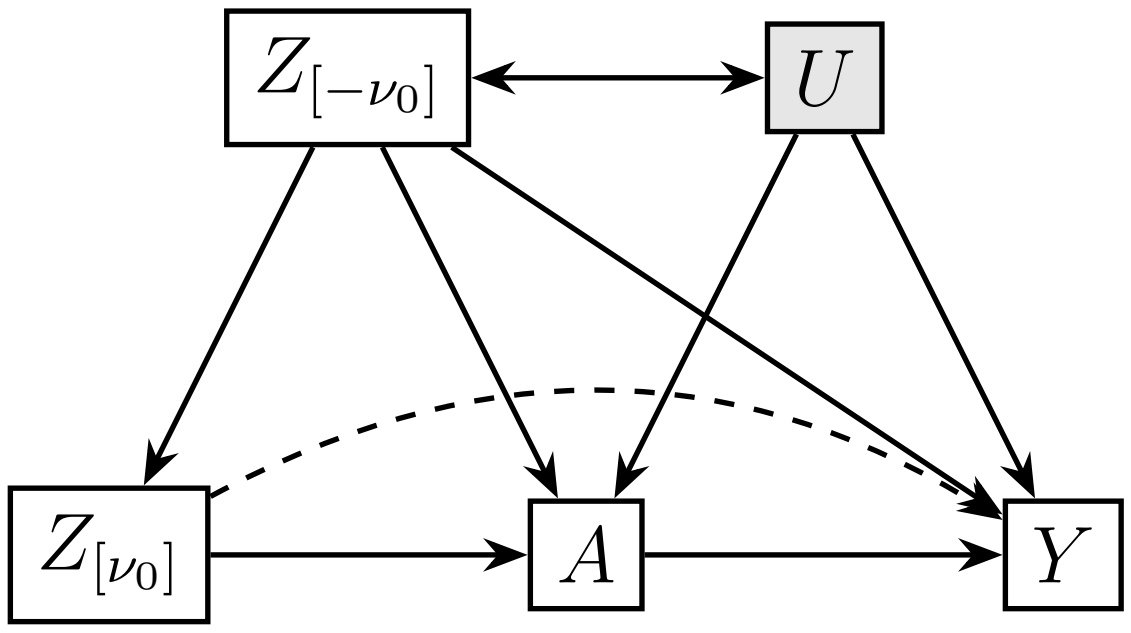

		For illustration, we present three causal graphs in Figure~\ref{fig: DAG}, where $X=\emptyset$ and Assumption~\ref{as: IV independence} is always satisfied. Note that Figure~\ref{fig: DAG}(b) allows an association between $U$ and $Z_{[-\nu_0]}$. 
		Moreover, all three causal graphs allow $Z_{[-\nu_0]}$ to have a direct effect on $Y(0)$, thereby relaxing the classical exclusion restriction.
		When only the solid arrows are present, Assumption~\ref{as: strong latent ignorability} is satisfied. 
		However, in Figures~\ref{fig: DAG}(a) and (b), once the dashed arrows in Figure~\ref{fig: DAG} are introduced,  Assumption~\ref{as: strong latent ignorability} may fail while Assumption~\ref{as: weak latent ignorability} still holds, highlighting the latter's weaker requirements.
		
		In addition, under the causal structure illustrated in Figure~\ref{fig: DAG}(a), Assumption~\ref{as: Z-invariant treatment effect} remains valid while still permitting interactions between the latent confounder $U$ and the treatment effect $Y(1)-Y(0)$. In Figure~\ref{fig: DAG}(b), both Assumptions~\ref{as: Z-invariant treatment effect} and~\ref{as: alternative}(b) are satisfied, allowing interactions between $Z_{[-\nu_0]}$ and $U$ in the treatment assignment mechanism. This demonstrates that our invalid IV setting permits either $U$-$Z_{[-\nu_0]}$ interactions or $U$-$Y(1)-Y(0)$ interactions, but not both simultaneously, under which Assumption~\ref{as: Z-invariant treatment effect} may fail to hold.

		\subsection{Causal identification}
		We now focus on a principled projection space induced by additive model structures.
		Let $\gamma\in\{0,\ldots,d_Z\}$ be a nonnegative integer, and define $\mathcal{S}_{\gamma}:=\overline{\sum_{\nu: |\nu|\geq \gamma} L_2(Z_{[-\nu]},X)}.$ The space $\mathcal{S}_{\gamma}$ consists of all functions that can be represented as additive combinations of components depending on $Z_{[-\nu]}$ and $X$ for subsets $\nu$ satisfying $|\nu|\geq \gamma$, together with their limits in $L_2(Z,X)$.
		It is straightforward to verify that $\mathcal{S}_{\gamma}$ is a proper Hilbert subspace of $L_2(Z, X)$. Moreover, $\mathcal{S}_{\gamma}\subseteq \mathcal{S}_{\gamma'}$ whenever $\gamma\geq \gamma'$.
		Define $\mathbb{S}_{\gamma}: L_2(O)\to \mathcal{S}_{\gamma}$ as a projection operator that maps $f\in L_2(O)$ onto $\mathcal{S}_{\gamma}$.
		When $\gamma=d_Z$, it can be verified that $\mathcal{S}_{d_Z}=L_2(X)$, and hence
		$\mathbb{S}_{d_Z}[f](X)=\mathbb{E}\!\left[f(O)\mid X\right]$.
		Additionally, when $\gamma=0$, it can be verified that $\mathcal{S}_{0}=L_2(Z,X)$, and $\mathbb{S}_0[f](Z,X) = \mathbb{E}[f(O)\mid Z,X]$. 
		
		We next present a key lemma that connects the $\mathcal{S}_{\gamma}$-PIV framework to the invalid IV setting.
		\begin{lemma}\label{lem: connections bewteen PIV and invalid IV}
			(a) Suppose that Assumptions~\ref{as: IV independence} and \ref{as: strong latent ignorability} hold for $\nu_0$, then 
			\begin{align}\label{eq: relation to sun2023}
				\mathbb{E}[Y(0)\mid Z,X]\indep Z_{[\nu_0]} \mid Z_{[-\nu_0]},X.
			\end{align} 
			Consequently, Definition \ref{defn: projected IV}(a) is satisfied for $\mathcal{S}_{\gamma}$ for all $\gamma\leq |\nu_0|$.
			(b) Suppose that Assumptions~\ref{as: IV independence}, \ref{as: weak latent ignorability}, \ref{as: Z-invariant treatment effect}, and \ref{as: alternative} hold for $\nu_0$. Then, Definition \ref{defn: projected IV}(b) is satisfied for $\mathcal{S}_{\gamma}$ for all $\gamma\leq \gamma_\dagger (\nu_0)$. 
		\end{lemma}

		Notably, Equation~\eqref{eq: relation to sun2023} is actually required by \citet{sun2023semiparametric} for some non-empty $\nu_0$. Under Assumption~\ref{as: IV independence}, Assumption~\ref{as: strong latent ignorability} implies this restriction. In turn, this restriction implies Definition~\ref{defn: projected IV}(a) for $\mathcal{S}_{|\nu_0|}$, while the reverse implications do not hold in general.
		
		Moreover, if Definition~\ref{defn: projected IV}(a) holds for $\mathcal{S}_{\gamma_1}$ and Definition~\ref{defn: projected IV}(b) holds for $\mathcal{S}_{\gamma_2}$, then $Z$ is an $\mathcal{S}_{\min\{\gamma_1,\gamma_2\}}$-PIV. Lemma~\ref{lem: connections bewteen PIV and invalid IV} provides sufficient conditions for parts~(a) and~(b) of Definition~\ref{defn: projected IV}, respectively. Combining these observations yields the following sufficient conditions under which $Z$ is an $\mathcal{S}_{\gamma}$-PIV.
		
		\begin{proposition}\label{prop: invalid IV}
			(a) If Assumptions~\ref{as: IV independence}--\ref{as: alternative} hold for $\nu_0$, then, $Z$ is an  $\mathcal{S}_{\gamma}$-PIV for all $\gamma\leq \gamma_\dagger(\nu_0)$. \\
			(b) Let $\gamma_*$ be the maximal number such that  Definition~\ref{defn: projected IV}(a) holds for $\mathcal{S}_{\gamma}$, i.e., $\mathbb{E}[Y(0)\mid Z,X]\in\mathcal{S}_{\gamma}$.
			If Assumptions~\ref{as: IV independence}, \ref{as: weak latent ignorability}, \ref{as: Z-invariant treatment effect}, \ref{as: alternative} hold for $\nu_0$, then $Z$ is an $\mathcal{S}_{\gamma}$-PIV for all $\gamma \leq \min\{\gamma_*,\gamma_\dagger (\nu_0)\}$.
		\end{proposition}
		
		By Lemma~\ref{lem: connections bewteen PIV and invalid IV}(a), we have \(\gamma_*\geq |\nu_0|\), so Proposition~\ref{prop: invalid IV}(a) follows directly from Proposition~\ref{prop: invalid IV}(b). Proposition~\ref{prop: invalid IV}(b) also shows that identification can be achieved even when Assumption~\ref{as: strong latent ignorability} fails, provided Assumption~\ref{as: weak latent ignorability} holds. The admissible range of \(\gamma\) is constrained by both \(\gamma_*\) and \(\gamma_{\dagger}(\nu_0)\).

		To characterize the largest attainable value of $\gamma$ such that the $\mathcal{S}_{\gamma}$-PIV conditions hold, we define an index set $\nu_0\subseteq\langle d_Z\rangle$ to be regular if Assumptions~\ref{as: IV independence}, \ref{as: weak latent ignorability}, \ref{as: Z-invariant treatment effect}, and \ref{as: alternative} hold relative to $\nu_0$. In Appendix~\ref{append: uniqueness of maximal index}, under Assumption~\ref{as: regular condition}, we show that the union of any two regular sets is still regular, indicating that the maximal regular set that contains all regular index sets is unique. 
		We therefore define the unique largest regular index set $\nu_{\max}:=\bigcup_{\nu_0:\,\nu_0\text{ is regular}}\nu_0$, and define $\gamma_{\max}:=\min\{\gamma_*,\gamma_\dagger (\nu_{\max})\}$ as the largest value of $\gamma$ certified by these sufficient conditions.

		\begin{remark}
			Because Assumption~\ref{as: strong latent ignorability} implies Assumption~\ref{as: weak latent ignorability}, $\nu_{\max}$ contains the indices of all IVs satisfying Assumptions~\ref{as: IV independence}--\ref{as: alternative}, and there are no more than $\min\{\gamma_*,|\nu_{\max}|\}$ such IVs. In particular, if $Y(1)-Y(0)$ is constant, the same containment and cardinality conclusions hold for all valid IVs satisfying Assumptions~\ref{as: IV independence} and \ref{as: strong latent ignorability}.
		\end{remark}
		
\begin{figure}[t]
			\centering
			\begin{tikzpicture}[>=stealth, font=\small]
				\draw[->, thick] (-0.4,0) -- (13.7,0)
				node[right] {$\gamma$};
				
				\foreach \p in {0,3.5,8.5,13}{
					\draw[thick] (\p,-0.12) -- (\p,0.12);
				}
				
				\node[below=6pt] at (0,0) {$0$};
				\node[below=6pt] at (3.5,0) {$\gamma_{\min}$};
				\node[below=6pt] at (8.5,0)
				{$\gamma_{\max}=\min\{\gamma_*,\gamma_\dagger (\nu_{\max})\}$};
				\node[below=6pt] at (13,0) {$d_Z$};
				\node at (1.75,0.55) {$\mathcal{S}_{\gamma}$-relevance fails};
				\node at (6,0.55) {Feasible region of $\gamma$};
				\node at (10.75,0.55) {$\mathcal{S}_{\gamma}$-PIV condition may fail};
			\end{tikzpicture}
			\caption{Feasible region of $\gamma$, where $\gamma_{\min}>0$ represents the minimal value of $\gamma$ such that $\mathcal{S}_{\gamma}$-relevance holds.}
			\label{fig: feasible region of gamma}
		\end{figure}
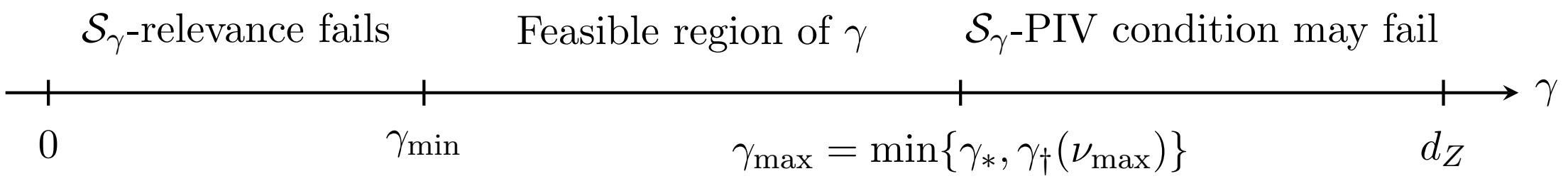

		Figure~\ref{fig: feasible region of gamma} graphically represents the feasible region of $\gamma$ for the identification of ATE, while values of $\gamma$ that are too small or too large may lead to a violation of the $\mathcal{S}_{\gamma}$-relevance or $\mathcal{S}_{\gamma}$-PIV condition, respectively.           
		The following corollary summarizes the identification result and follows directly from Theorem~\ref{thm: identification} and Proposition~\ref{prop: invalid IV}.
		\begin{corollary}
			If $\pi$ is an $\mathcal{S}_{\gamma_{\max}}$-PRWF, then $\mathbb{E}[Y(1)-Y(0)]=\psi_{\mathcal{S}_{\gamma_{\max}},\pi}^o.$
		\end{corollary}

		Notably, when $X=\emptyset$, \citet{sun2023semiparametric} require the existence of at least one $\mathcal{S}_{\gamma_{\max}}$-PRWF in the class $\{\prod_{j\in\nu}Z_{[j]}:|\nu|\geq d_Z-\gamma_{\max}+1\}$. By Proposition~\ref{prop: existence of PRWF}, this requirement is generally stronger than the $\mathcal{S}_{\gamma_{\max}}$-relevance condition; the two conditions are equivalent when all IVs are binary.
		
		For simplicity, we focus on the ATE throughout the main text. In Appendix~\ref{append: identification of other estimands}, we extend our approach to identify mean potential outcomes and the average treatment effect on the treated.
		
		\section{Connections to existing literature}\label{sec: connection to exsiting literature}
		\subsection{IVDID}\label{subsec: connection to IVDID}
		We first suppose that $Z=\{Z_{[1]},Z_{[2]}\}$, where $Z_{[1]}, Z_{[2]} \in \{0,1\}$ are binary variables, and that $Z$ is an $\mathcal{S}_1$-PIV. In this setting, the IVDID functional studied in \citet{ye2023instrumented} is a special case of our functional $\psi_{\mathcal{S},\pi}^o$ with $\mathcal{S}=\mathcal{S}_1$ and $\pi(Z,X)=Z_{[1]}Z_{[2]}$.
		\begin{proposition}\label{prop: IVDID}
			When $Z_{[1]}, Z_{[2]} \in \{0,1\}$ are binary variables, for any $\mathcal{S}_1$-PRWF $\pi$, it holds that
			\begin{align*}
				\psi_{\mathcal{S}_1,\pi}^o=\mathbb{E}\left[
				\dfrac{
					\mathbb{S}_0[Y](1,1,X)
					-\mathbb{S}_0[Y](0,1,X)
					-\mathbb{S}_0[Y](1,0,X)
					+\mathbb{S}_0[Y](0,0,X)
				}{
					\mathbb{S}_0[A](1,1,X)
					-\mathbb{S}_0[A](0,1,X)
					-\mathbb{S}_0[A](1,0,X)
					+\mathbb{S}_0[A](0,0,X)
				}
				\right],
			\end{align*}
			where $\mathbb{S}_0[Y](z_1,z_2,X)=\mathbb{E}[Y| Z_{[1]}=z_1, Z_{[2]}=z_2, X]$ and $\mathbb{S}_0[A](z_1,z_2,X)=\mathbb{E}[A| Z_{[1]}=z_1, Z_{[2]}=z_2, X]$ for $z_1,z_2\in\{0,1\}$.
		\end{proposition}
		Proposition~\ref{prop: IVDID} gives an alternative representation of $\psi_{\mathcal{S}_1,\pi}^o$. If we interpret $Z_{[1]}$ as the group indicator and $Z_{[2]}$ as the time indicator, then $\psi_{\mathcal{S}_1,\pi}^o$ reduces exactly to the IVDID functional studied in \citet{ye2023instrumented}, though the underlying identification framework and nuisance functions might slightly differ from those in their work. For example, 
		Definition~\ref{defn: projected IV}(a) is equivalent to requiring that
		\begin{align*}
			&\mathbb{E}[Y(0)\mid Z_{[1]}=1,Z_{[2]}=1,X]-\mathbb{E}[Y(0)\mid Z_{[1]}=1,Z_{[2]}=0,X]\\
			=&\mathbb{E}[Y(0)\mid Z_{[1]}=0,Z_{[2]}=1,X]-\mathbb{E}[Y(0)\mid Z_{[1]}=0,Z_{[2]}=0,X],
		\end{align*}
        which is a conditional parallel trends restriction on the untreated potential outcome \citep{ye2023instrumented}.

		Moreover, when $A = Z_{[1]}Z_{[2]}$, treatment is received only by units in the treated group ($Z_{[1]}=1$) during the post-treatment period ($Z_{[2]}=1$). The resulting functional further reduces to the classical difference-in-differences functional \citep{abadie2005semiparametric}:
		\begin{align*}
			\psi_{\mathcal{S}_1,\pi}^o
			=
			\mathbb{E}\left[
			\mathbb{S}_0[Y](1,1,X)
			-\mathbb{S}_0[Y](0,1,X)
			-\mathbb{S}_0[Y](1,0,X)
			+\mathbb{S}_0[Y](0,0,X)
			\right].
		\end{align*}
		Consequently, the $\mathcal{S}_{\gamma}$-PIV framework extends the IVDID functional in two important directions. First, it allows both the group index $Z_{[1]}$ and the time index $Z_{[2]}$ to be either discrete or continuous. Second, the PIV framework naturally accommodates multiple group indices, namely $\{Z_{[j]}\}_{j=1}^{d_Z}$ with $d_Z\geq 3$. In particular, when $d_Z=3$, the triple-difference estimator proposed by \citet{olden2022triple} can also be viewed as a special case of our framework.

		\subsection{Partially linear model with invalid IV}\label{sec: connection to partially linear model}
		In this subsection, we suppose that the treatment variable $A\in\mathcal{A}\subseteq\mathbb{R}$ may be continuous or discrete. We also suppose that $Y(a)$ satisfies a partially linear model:
		\begin{equation}
			\begin{aligned}\label{eq: partially linear model}
				&Y(a) = \beta^oa + s^o(Z,X) + \epsilon_y \text{,\qquad where $\mathbb{E}[\epsilon_y|Z,X]=0$,\; and  $s^o\in\mathcal{S}$}.
			\end{aligned}
		\end{equation}
		
		We are interested in estimating $\beta^o$, the partially linear effect, and $Z$ might directly influence $Y$ through $s^o(Z, X)$. In particular, define two functionals of interest:
		\begin{align*}
			&\beta_{\mathcal{S},\pi}^o:= 
			\dfrac{\mathbb{E}[\{\pi(Z,X)-\mathbb{S}[\pi](Z,X)\}\{Y-\mathbb{S}[Y](Z,X)\}]}{\mathbb{E}[\{\pi(Z,X)-\mathbb{S}[\pi](Z,X)\}\{A-\mathbb{S}[A](Z,X)\}]},\\
			&\beta_{\mathcal{S},*}^o:= 
			\dfrac{\mathbb{E}[\{\mathbb{E}[A|Z,X]-\mathbb{S}[A](Z,X)\}\{Y-\mathbb{S}[Y](Z,X)\}]}{\mathbb{E}[\{\mathbb{E}[A|Z,X]-\mathbb{S}[A](Z,X)\}^2]}.
		\end{align*}
		The statistical properties of \(\beta_{\mathcal{S},\pi}^o\) have been studied by \citet{Chernozhukov2018} in the special case \(\mathcal{S}=L_2(X)\), and by \citet{sun2022selective} for \(\mathcal{S}=\mathcal{S}_{\gamma}\).
		We give the identification results as follows.
		\begin{theorem}[Identification]\label{thm: identification partially linear effect}
			Under Equation~\eqref{eq: partially linear model}, suppose that $\pi\in L_2(Z,X)$, and $0<|\mathbb{E}[\{\pi(Z,X)-\mathbb{S}[\pi](Z,X)\}\{A-\mathbb{S}[A](Z,X)\}]|<\infty$. Then $\beta_{\mathcal{S},*}^o=\beta_{\mathcal{S},\pi}^o=\beta^o$.
		\end{theorem}
		Next, we derive the EIFs of $\beta_{\mathcal{S},\pi}^o$ and $\beta_{\mathcal{S},*}^o$. Based on these results, we can conduct statistical inference and construct corresponding confidence intervals for $\beta_{\mathcal{S},\pi}^o$ and $\beta_{\mathcal{S},*}^o$.
		\begin{theorem}[EIFs]\label{thm: EIF partially linear effect}
			If $\tau_{\mathcal{S},\pi}^o:=\mathbb{E}[\{\pi(Z,X)-\rho_{\mathcal{S},\pi}^o(Z,X)\}\{A-\delta_{\mathcal{S}}^o(Z,X)\}]\neq 0$, then, under the differentiability conditions in Lemma~\ref{lem: semiparametric calculation}, the EIF of $\beta_{\mathcal{S},\pi}^o$ is $\phi_{\mathcal{S},\pi}(O;\beta_{\mathcal{S},\pi}^o,\alpha_{\mathcal{S},\pi}^o,\tau_{\mathcal{S},\pi}^o)$, where
			\begin{align*}
				\phi_{\mathcal{S},\pi}(O;\beta_{\mathcal{S},\pi},\alpha_{\mathcal{S},\pi},\tau_{\mathcal{S},\pi})
				=&\frac{1}{\tau_{\mathcal{S},\pi}}
				\{\pi(Z,X)-\rho_{\mathcal{S},\pi}(Z,X)\}\{Y-\eta_{\mathcal{S}}(Z,X)-(A-\delta_{\mathcal{S}}(Z,X))\beta_{\mathcal{S},\pi}\}.
			\end{align*}
			If $\tau_{\mathcal{S},*}^o:=\mathbb{E}[\{\pi_*^o(Z,X)-\delta_{\mathcal{S}}^o(Z,X)\}^2]\neq 0$, then the EIF of $\beta_{\mathcal{S},*}^o$ is $\phi_{\mathcal{S},*}(O;\beta_{\mathcal{S},*}^o,\alpha_{\mathcal{S},*}^o,\tau_{\mathcal{S},*}^o)$, where
			\begin{align*}
				&\phi_{\mathcal{S},*}(O;\beta_{\mathcal{S},*},\alpha_{\mathcal{S},*},\tau_{\mathcal{S},*})
				=\dfrac{1}{\tau_{\mathcal{S},*}}\{\pi_*(Z,X)-\delta_{\mathcal{S}}(Z,X)\}\{Y-\eta_{\mathcal{S}}(Z,X)-(A-\delta_{\mathcal{S}}(Z,X))\beta_{\mathcal{S},*}\}\\
				&+\dfrac{1}{\tau_{\mathcal{S},*}}
				\{A-\pi_*(Z,X)\}\{\xi_*(Z,X)-\eta_{\mathcal{S}}(Z,X)-(\pi_*(Z,X)-\delta_{\mathcal{S}}(Z,X))\beta_{\mathcal{S},*}\}.
			\end{align*}
		\end{theorem}
		Similar to Section~\ref{subsec: asymptotic theory}, we estimate $\beta_{\mathcal{S},\pi}^o$, $\beta_{\mathcal{S},*}^o$, $\tau_{\mathcal{S},\pi}^o$, and $\tau_{\mathcal{S},*}^o$ using the cross-fitting techniques:
		\begin{align*}
			&\dfrac{1}{K}\sum_{k=1}^K\mathbb{E}_{nk}\left[\phi_{\mathcal{S},\pi}\left(O;\hat\beta_{\mathcal{S},\pi},\hat\alpha_{\mathcal{S},\pi}^{(n,-k)},\cdot\right)\right]=0;\quad
			\dfrac{1}{K}\sum_{k=1}^K\mathbb{E}_{nk}\left[\phi_{\mathcal{S},*}\left(O;\hat\beta_{\mathcal{S},*},\hat\alpha_{\mathcal{S},*}^{(n,-k)},\cdot\right)\right]=0;\\
			&\hat\tau_{\mathcal{S},\pi}:=\frac{1}{K}\sum_{k=1}^K\mathbb{E}_{nk}[\{\pi(Z,X)-\hat\rho_{\mathcal{S},\pi}^{(n,-k)}(Z,X)\}\{A-\hat\delta_{\mathcal{S}}^{(n,-k)}(Z,X)\}];\\
			&\hat\tau_{\mathcal{S},*}:=\dfrac{1}{K}\sum_{k=1}^K\mathbb{E}_{nk}[\{A-\hat\delta_{\mathcal{S}}^{(n,-k)}(Z,X)\}^2-\{A-\hat\pi_*^{(n,-k)}(Z,X)\}^2].
		\end{align*}
		Notably, the first two equations do not depend on $\tau_{\mathcal{S},\pi}$ and $\tau_{\mathcal{S},*}$, which are therefore replaced by ``$\cdot$'' for simplicity.
		The variances of $\sqrt{n}\{\hat\beta_{\mathcal{S},\pi}-\beta^o\}$ and $\sqrt{n}\{\hat\beta_{\mathcal{S},*}-\beta^o\}$ can be estimated by
		\begin{align*}
			\frac{1}{K}\sum_{k=1}^K \mathbb{E}_{nk}\!\left[\phi_{\mathcal{S},\pi}\!\left(O;\hat\beta_{\mathcal{S},\pi},\hat\alpha_{\mathcal{S},\pi}^{(n,-k)},\hat\tau_{\mathcal{S},\pi}\right)^2\right],\quad\frac{1}{K}\sum_{k=1}^K \mathbb{E}_{nk}\!\left[\phi_{\mathcal{S},*}\!\left(O;\hat\beta_{\mathcal{S},*},\hat\alpha_{\mathcal{S},*}^{(n,-k)},\hat\tau_{\mathcal{S},*}\right)^2\right],
		\end{align*}
		respectively. Akin to Theorems~\ref{thm: asymptotic normality} and \ref{thm: asymptotic normality adaptive}, $\hat\beta_{\mathcal{S},\pi}$ and $\hat\beta_{\mathcal{S},*}$ are consistent and asymptotically normal, and their variance estimators are consistent. 
		For a discussion of the estimation procedure when multiple $\mathcal{S}$-PRWFs are used to identify $\beta^o$, we refer readers to Appendix~\ref{append: connections to debiased generalized method of moments}.

		\section{Simulation}\label{sec: simulation}
		We compare the proposed $\mathcal{S}_{\gamma}$-PIV estimator with several nonparametric and semiparametric methods, including the G-estimator (GE) \citep{sun2023semiparametric} and the double machine learning method under the assumption of no unmeasured confounding (NUC) \citep{Chernozhukov2018}, where $Z$ and $X$ are both included as measured confounders. For additional comparison, we also include several parametric methods, including sisVIVE \citep{kang2016instrumental} and TSHT \citep{guo2018confidence} estimators. Specifically, the NUC method is implemented using the \texttt{R} package \texttt{DoubleML}, TSHT using \texttt{RobustIV}, and sisVIVE using \texttt{sisVIVE}.

		We consider six DGPs under the invalid IV setting in Section~\ref{sec: connection to invalid IV}. In all six DGPs, $Z=(Z_{[1]},Z_{[2]},Z_{[3]})\in\mathcal{Z}:=\{-1,1\}^2\times\mathbb{R}$. The mechanisms for generating $Z$, $A$, $Y(0)$, and $Y(1)-Y(0)$ are summarized in Table~\ref{tbl: concrete DGPs}.
		We independently generate $U\sim\mathrm{Uniform}(-1,1)$,  $X\sim\mathrm{Uniform}(-0.5,0.5)$, and the instrument disturbances $\epsilon_{[j]}^z\sim\mathrm{Uniform}(-1,1)$ for $j=1,2,3$.
		For the treatment mechanism in DGPs~3 and 4, we generate
		\(
		H\mid X\sim\mathrm{Bernoulli}\!\left\{0.7+0.3\mathrm{expit}(-2X)\right\},
		\; V=\sum_{j=1}^3 Z_{[j]}+\epsilon^a,
		\; \epsilon^a\sim\mathrm{Uniform}(-1,1),
		\)
		where $H$ is conditionally exogenous given $X$, $\mathrm{expit}(t)=(1+e^{-t})^{-1}$. 
		
		\begin{table}[t]
			\centering
			\caption{Data-generating mechanisms for the six simulation settings, with the ATE parameter set to $2$. Here, $\phi_1(s)=\phi_2(s)=2I\{s\ge0\}-1$ and $\phi_3(s)=0.25s$.}
			\label{tbl: concrete DGPs}
			\renewcommand{\arraystretch}{1.75}
			\resizebox{\textwidth}{!}{
				\begin{tabular}{c|c|c|c|c|c}
					\toprule
					DGP & $Z_{[1]}$ & $Z_{[j]}$, $j=2,3$ & $A$ & $Y(0)$ & $Y(1)-Y(0)$ \\
					\midrule
					1 & $\phi_1(X+6U)$ & $\phi_j(X+6\epsilon_{[j]}^z)$ & $I\{(5\mathrm{expit}(5X)+1)\sum_{j=1}^3Z_{[j]}+U\ge0\}$ & $0.05\sum_{j=1}^3Z_{[j]}(0.5+X^2)+U$ & $-4\sin(6X)+2$ \\
					2 & $\phi_1(X+6U)$ & $\phi_j(X+6\epsilon_{[j]}^z)$ & $I\{(5\mathrm{expit}(5X)+1)\sum_{j=1}^3Z_{[j]}+U\ge0\}$ & $0.05\sum_{j=1}^3Z_{[j]}(0.5+X^2)+U$ & $-4X+2$ \\
					3 & $\phi_1(X+6\epsilon_{[1]}^z)$ & $\phi_j(X+6\epsilon_{[j]}^z)$ & $HI\{V\ge0\}+(1-H)I\{2U+Z_{[1]}\ge0\}$ & $0.05\sum_{j=1}^3Z_{[j]}(0.5+X^2)+U$ & $2U-4\sin(6X)+2$ \\
					4 & $\phi_1(X+6\epsilon_{[1]}^z)$ & $\phi_j(X+6\epsilon_{[j]}^z)$ & $HI\{V\ge0\}+(1-H)I\{2U+Z_{[1]}\ge0\}$ & $0.1Z_{[1]}(0.5+X^2)+U$ & $2U-2X+2$ \\
					5 & $\phi_1(X+6\epsilon_{[1]}^z)$ & $\phi_j(X+6\epsilon_{[j]}^z)$ & $I\{(5\mathrm{expit}(5X)+1)\sum_{j=1}^3Z_{[j]}+U\ge0\}$ & $0.05\sum_{j=1}^3Z_{[j]}+U+X$ & $2$ \\
					6 & $\phi_1(X+6\epsilon_{[1]}^z)$ & $\phi_j(X+6\epsilon_{[j]}^z)$ & $I\{(5\mathrm{expit}(5X)+1)\sum_{j=1}^3Z_{[j]}+U\ge0\}$ & $0.1Z_{[1]}+U+X$ & $2$ \\
					\bottomrule
			\end{tabular}}
		\end{table}
		
		\begin{table}[t]
			\centering
			\caption{DGPs satisfying Assumptions~\ref{as: IV independence}, \ref{as: weak latent ignorability}, \ref{as: Z-invariant treatment effect}, and \ref{as: alternative} for $\nu_{\max}$.}
			\label{tbl: DGPs motivation}
				\begin{tabular}{c|cccccc}
					\toprule
					DGP index 
					& 1 & 2 & 3 & 4 & 5 & 6\\
					\midrule
					Valid IV & $\emptyset$ & $\emptyset$ & $\emptyset$ & $\{Z_{[2]},Z_{[3]}\}$ & $\emptyset$ & $\{Z_{[2]},Z_{[3]}\}$\\
					$Z_{[-\nu_{\max}]}$ 
					& $Z_{[1]}$ & $Z_{[1]}$ & $\emptyset$ & $\emptyset$
					& $\emptyset$ & $\emptyset$\\
					Assumption~\ref{as: alternative} for $\nu_{\max}$  &(b)&(b)&(a)&(a)&(b)&(b)\\
					Constant treatment effect & \ding{55} & \ding{55} & \ding{55} & \ding{55} & \ding{51} & \ding{51}\\
					\bottomrule
				\end{tabular}

			\end{table}

			In summary, Assumptions~\ref{as: IV independence}, \ref{as: weak latent ignorability}, \ref{as: Z-invariant treatment effect}, and \ref{as: alternative} hold under all DGPs, with $\nu_{\max}=\{2,3\}$ for DGPs~1 and 2 and $\nu_{\max}=\{1,2,3\}$ for DGPs~3--6. In addition, $\gamma_{\max}=2$ in all six DGPs. Under each DGP, all IVs are mutually independent conditional on $X$. In addition, the assumptions of sisVIVE and TSHT are satisfied under DGP~6, while the partially linear model in Equation~\eqref{eq: partially linear model} holds under DGPs~5 and 6.

			Table~\ref{tbl: DGPs motivation} summarizes the motivation behind the design of these DGPs, each of which can be naturally represented by either Figure~\ref{fig: DAG}(a) or Figure~\ref{fig: DAG}(b). The true ATE is $2$ in all six DGPs. We conduct the cross-fitting procedure with $K=5$ folds and run 500 replications with sample size $n=5,000$. 
			In implementation, we omit the inner cross-fitting step and use the entire training sample $I_{-k}$ for both first-stage nuisance estimation and the subsequent estimation of $\hat\kappa_{\mathcal{S},*}^{(n,-k)}$ and $\hat\Gamma_{\mathcal{S},*}^{(n,-k)}$.
			To estimate the nuisance functions, the additive spline methods are employed to estimate $\mathbb{S}_{\gamma}[f](Z, X)$ for any $f\in L_2(O)$ using the \texttt{R} package \texttt{mgcv} \citep{wood2017generalized}.
			
			\begin{table}[t]
  \centering
  \caption{Comparison of simulation results for $n=5,000$ over 500 replications.}
  \label{tbl: simu 5000}
  \renewcommand{\arraystretch}{1}
  \resizebox{0.8\textwidth}{!}{%
    \begin{tabular}{cc|ccc|ccc|cccc}
      \toprule
      DGP & Metrics & PIV$_1$ & PIV$_2$ & PIV$_3$ & GE$_1$ & GE$_2$ & GE$_3$ & NUC & TSHT & sisVIVE \\
      \midrule
      \multirow{4}{*}{1}
      & BIAS & .0105 & .0127 & .6668 & .0535 & .0490 & .7274 & .3495 & .0665 & .1485 \\
      & SD   & .0601 & .0578 & .0473 & .0557 & .0560 & .0493 & .0533 & .2756 & .0980 \\
      & SE   & .0591 & .0552 & .0446 & .0539 & .0560 & .0472 & .0392 & .0784 & -- \\
      & CR   & .9340 & .9300 & .0000 & .8140 & .8520 & .0000 & .0000 & .8960 & -- \\[10pt]

      \multirow{4}{*}{2}
      & BIAS & .0000 & .0013 & .6712 & .0201 & .0206 & .7336 & .3560 & .1140 & .1227 \\
      & SD   & .0247 & .0238 & .0236 & .0283 & .0284 & .0250 & .0287 & .2510 & .0374 \\
      & SE   & .0248 & .0237 & .0226 & .0281 & .0289 & .0238 & .0164 & .0303 & -- \\
      & CR   & .9400 & .9460 & .0000 & .8960 & .9020 & .0000 & .0000 & .2580 & -- \\[10pt]

      \multirow{4}{*}{3}
      & BIAS & .0043 & .0015 & .1129 & .0944 & .1076 & .1849 & .7848 & .1924 & .1973 \\
      & SD   & .2437 & .2409 & .0694 & .1621 & .1625 & .0702 & .0657 & .0840 & .0846 \\
      & SE   & .2426 & .2299 & .0685 & .1605 & .1606 & .0705 & .0672 & .0828 & -- \\
      & CR   & .9640 & .9520 & .6080 & .9040 & .8980 & .2440 & .0000 & .3580 & -- \\[10pt]

      \multirow{4}{*}{4}
      & BIAS & .0040 & .0010 & .0900 & .0244 & .0250 & .1083 & .8068 & .1310 & .1059 \\
      & SD   & .1300 & .1326 & .0485 & .1304 & .1309 & .0483 & .0522 & .0717 & .0613 \\
      & SE   & .1281 & .1235 & .0478 & .1294 & .1291 & .0482 & .0486 & .0542 & -- \\
      & CR   & .9540 & .9420 & .5120 & .9500 & .9500 & .3560 & .0000 & .3200 & -- \\[10pt]

      \multirow{4}{*}{5}
      & BIAS & .0012 & .0007 & .1489 & .0063 & .0039 & .1603 & .3773 & .2017 & .2017 \\
      & SD   & .0365 & .0361 & .0181 & .0385 & .0382 & .0184 & .0228 & .0208 & .0206 \\
      & SE   & .0368 & .0362 & .0171 & .0391 & .0390 & .0176 & .0115 & .0198 & -- \\
      & CR   & .9560 & .9520 & .0000 & .9560 & .9520 & .0000 & .0000 & .0000 & -- \\[10pt]

      \multirow{4}{*}{6}
      & BIAS & .0012 & .0007 & .1086 & .0063 & .0039 & .1172 & .4007 & .0006 & .0307 \\
      & SD   & .0365 & .0361 & .0182 & .0385 & .0382 & .0185 & .0257 & .0248 & .0277 \\
      & SE   & .0368 & .0362 & .0173 & .0391 & .0390 & .0178 & .0115 & .0250 & -- \\
      & CR   & .9560 & .9520 & .0000 & .9560 & .9520 & .0000 & .0000 & .9480 & -- \\
      \bottomrule
    \end{tabular}
  }
\end{table}

			The results are reported in Table~\ref{tbl: simu 5000}. For notational brevity, the proposed estimators $\hat{\psi}_{\mathcal{S}_j,*}$ are denoted as PIV$_{j}$ for $j=1,2,3$.
			Across all scenarios, we evaluate and compare the absolute value of bias (BIAS), standard deviation (SD), estimated standard error (SE), and empirical coverage rate (CR) of nominal 95\% confidence intervals.
			
			Overall, PIV$_1$ and PIV$_2$ perform stably, with small bias and low variance in moderate regimes. Moreover, their empirical coverage probabilities are close to the nominal 95\% level, and the SDs generally match the SEs, suggesting well-calibrated variance estimation and confidence interval construction.
			By contrast, PIV$_3$ ($\hat\psi_{\mathcal{S}_3,*}$) shows substantial bias due to an overly aggressive choice of $\gamma$, which leads to an overestimation of the validity of the IVs.
			
			The G-estimators (GE) proposed by \citet{sun2023semiparametric} generally perform slightly worse than the PIV estimators. Their implementation is detailed in Appendix~\ref{append: connections to debiased generalized method of moments}. The bias of the GE method becomes particularly pronounced when $\mathbb{E}[\{\pi_{*}^o(Z,X) - \mathbb{S}_\gamma[A](Z,X)\}^2 \mid X]$ and $\mathbb{E}[Y(1)-Y(0)\mid X]$ exhibit strong heterogeneity with respect to $X$, especially in DGPs 1--4. This deterioration arises because the G-estimator effectively imposes a constant treatment effect assumption, which is violated under heterogeneous treatment effects.

			The TSHT and sisVIVE estimators exhibit nonnegligible bias under DGPs~1--5, where both the plurality and majority rules are violated under DGPs~1, 2, 3, 5, and the treatment effect varies with the covariate $X$ under DGPs~1--4. Under DGP~6, the plurality rule holds, and TSHT achieves coverage close to the nominal 95\% level.
			In summary, the results highlight the advantages of the PIV framework in terms of robustness and stability across a wide range of empirical environments.
			
			\begin{figure}
				\centering
				\includegraphics[width=0.95\linewidth]{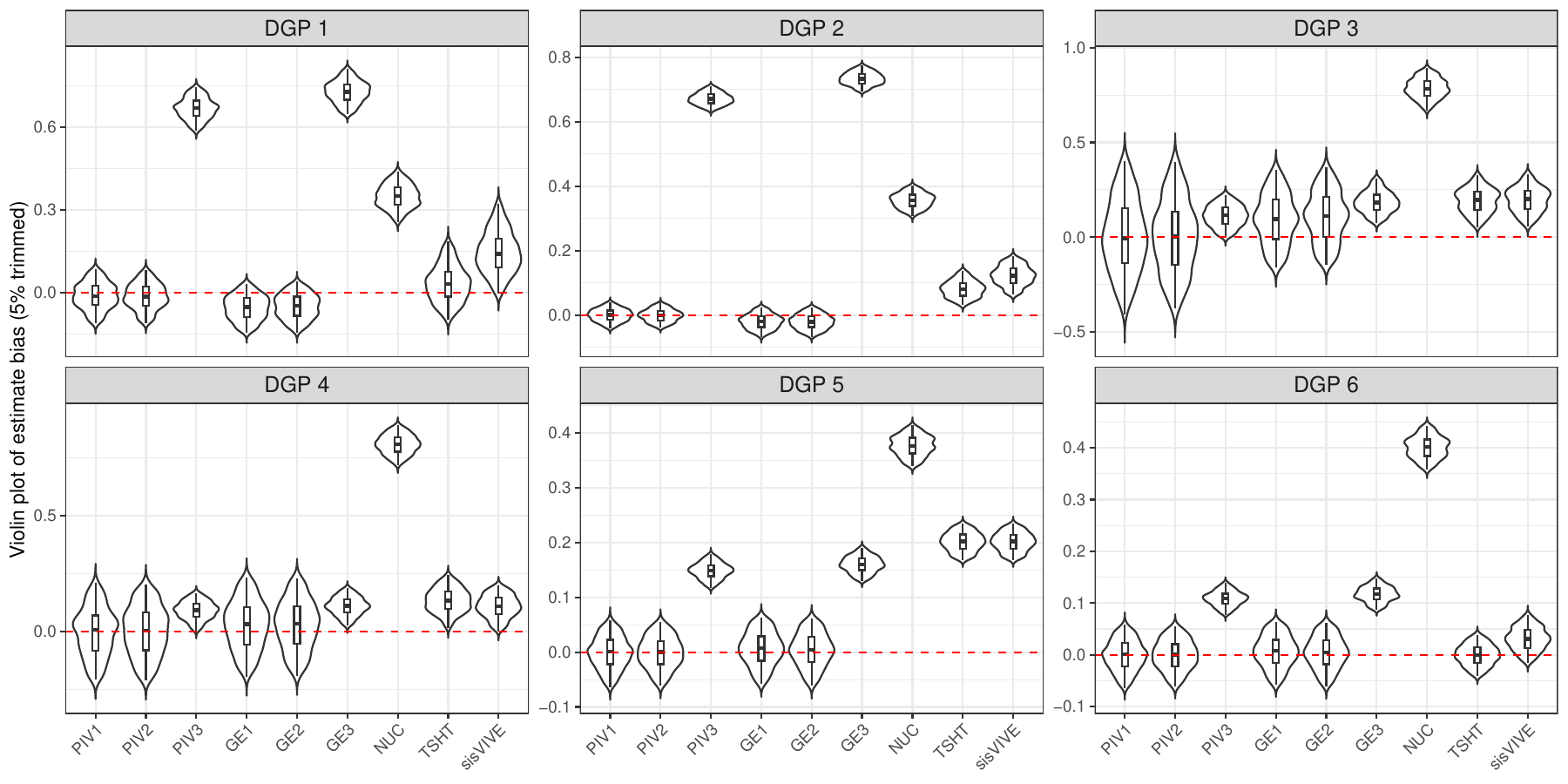}
				\caption{Violin plots of the residual bias in estimated treatment effects under DGPs 1--6 when $n=5000$ for various methods.}
				\label{fig: boxplot 5000}
			\end{figure}
			
			Finally, the violin plots of the residual bias in Figure~\ref{fig: boxplot 5000} show that PIV$_1$ and PIV$_2$ are centered close to zero across all DGPs, indicating small bias and stable performance. In contrast, PIV$_3$, the GE estimators, NUC, TSHT, and sisVIVE exhibit noticeable bias in several settings. 
			
			\section{Real data analysis}\label{sec: real data analysis}
			We apply our PIV method to the Australian Longitudinal Survey (ALS) data \citep{mcrae2019australian}. The ALS is conducted annually from 1984 and collects information on the education, employment, earnings, and demographic characteristics of young Australians. We use the 1984 and 1985 waves as repeated cross-sectional data. The 1984 wave contains 2,403 individuals aged 15--24, while the 1985 wave contains 8,998 individuals aged 16--25. 
			We are interested in estimating the ATE of having more than 12 years of education, compared with 12 years or fewer, on hourly wages.
            
			Following \citet{zhao2025semiparametric}, we construct hourly wages and years of education separately for the 1984 and 1985 waves. For the 1984 wave, the reported pay categories are converted into representative weekly monetary amounts, and the categories of weekly working hours are converted into representative numbers of hours. For the 1985 wave, reported earnings are converted to a weekly amount based on payment frequency. In both waves, hourly wages are calculated by dividing weekly earnings by weekly working hours. We exclude all individuals with missing hourly wages and impute the remaining missing values. The final dataset contains 5,661 individuals.
			
			To construct the years of education, school-leaving grades are converted into years of schooling, while post-school qualifications are assigned between one and four additional years according to their type. We then compare this constructed value with the respondent’s reported years of schooling and retain the larger value.
			Finally, the education level is binarized at 12 years.
			
			We include the survey wave indicator, residence in a capital city, birth in Australia, and attitudes toward the labor market in the PIV vector $Z$, with gender serving as the observed confounder $X$. Residence in a capital city indicates whether the respondent had mostly lived in a capital city, while birth in Australia indicates whether the respondent was born in Australia. Attitudes toward women’s employment and gender roles are measured using a composite index constructed by summing responses to six survey questions, with higher values reflecting more egalitarian attitudes toward gender roles for women’s employment. Unlike \citet{zhao2025semiparametric}, we retain the nonbinary attitude index, as our PIV framework accommodates nonbinary instruments.
			
			Our choice of candidate PIV is motivated by the related IVDID literature \citep{su2013local,zhao2025semiparametric}. These variables may be directly associated with wage levels and therefore may not satisfy the conventional exclusion restriction. However, IVDID allows such direct associations when they remain stable over time. In our PIV framework, this requirement is represented by the condition $\mathbb{E}[Y(0)\mid Z,X]\in\mathcal{S}_{\gamma}$, which allows the conditional mean of the hourly wage to depend on the candidate PIV through the structured function class $\mathcal{S}_{\gamma}$.
			
			\begin{table}[t]
    \centering
    \begin{tabular}{c|ccc ccc c}
        \toprule
        Method & PIV$_2$ & PIV$_3$ & PIV$_4$ & GE$_2$ & GE$_3$ & GE$_4$ & NUC \\
        \midrule
        EST & 2.75 & 2.84 & 6.27 & 0.45 & 0.90 & 7.04 & 1.66 \\
        SE & 1.51 & 1.29 & 1.60 & 2.88 & 1.57 & 0.86 & 0.08 \\
        $p$-value & 0.069 & 0.028 & <0.001 & 0.877 & 0.569 & <0.001 & <0.001 \\
        \bottomrule
    \end{tabular}
    \caption{Estimated effects of education on hourly wages.}
    \label{tbl: ALS_results}
\end{table}

			\begin{figure}
				\centering
				\includegraphics[width=0.7\linewidth]{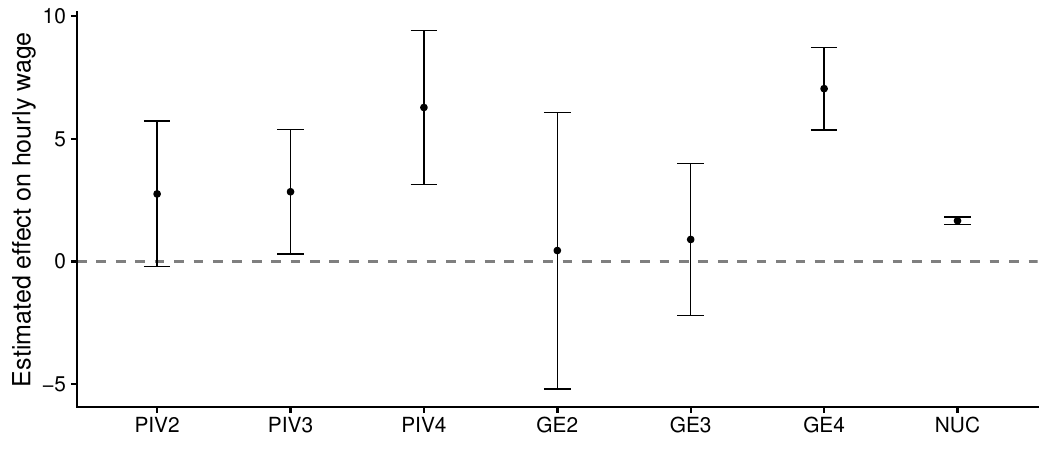}
				\caption{Point estimates and 95\% confidence intervals for the causal effect of having more than 12 years of education on hourly wages.}
				\label{fig:ALS_estimates_CI}
			\end{figure} 
			
			We compare our PIV framework with the G-estimator proposed by \citet{sun2023semiparametric} and a method based on the no unmeasured confounding assumption, denoted by NUC. The NUC method is implemented using the double machine learning approach of \citet{Chernozhukov2018}. Specifically, the survey wave indicator, residence in a capital city, birth in Australia, attitudes toward the labor market, and gender are all treated as measured confounders. The results are summarized in Table~\ref{tbl: ALS_results}.
			
			In general, all methods produce positive point estimates, suggesting that having more than 12 years of education increases hourly wages. PIV$_3$ and PIV$_4$ are statistically significant at the 5\% level, whereas PIV$_2$ is significant only at the 10\% level. 
			The estimated values of $\mathrm{PIV}_2$ and $\mathrm{PIV}_3$ are close to each other, suggesting that $Z$ might be an $\mathcal{S}_{\gamma_{\max}}$-PIV for $A$ with $\gamma_{\max}=3$. The estimated value of $\mathrm{PIV}_4$ differs substantially from that of $\mathrm{PIV}_2$, suggesting that $\mathrm{PIV}_4$ is biased because not all the instruments are valid.
			The GE estimates also exhibit a similar pattern: $\mathrm{GE}_4$ yields a large and statistically significant estimate, whereas $\mathrm{GE}_2$ and $\mathrm{GE}_3$ yield smaller and statistically insignificant estimates.
			
			The NUC method gives a smaller estimate of 1.66 with less uncertainty, although it relies on the absence of unmeasured confounding. Figure~\ref{fig:ALS_estimates_CI} further shows that the 95\% confidence intervals for PIV$_3$, PIV$_4$, GE$_4$, and NUC exclude zero, whereas those for PIV$_2$, GE$_2$, and GE$_3$ include zero.

			\section{Discussion on invalid PIV}\label{sec: discussion on invalid PIV}
			In this paper, we develop a novel projected instrumental variable framework for causal identification. By viewing conditional expectation as an orthogonal projection in a Hilbert space, the framework explores the use of more general projection operators. We also provide an alternative inverse problem formulation based on the Riesz representation and establish a connection between invalid IV and IVDID. Building on semiparametric efficiency theory, we have also developed debiased and adaptive machine learning methods for estimation and inference. 
			
			An important extension of the proposed framework is to accommodate invalid PIVs. One possible approach is to consider alternative formulations of the bridge Equation~\eqref{eq: outcome bridge}. We illustrate this approach with three examples below and leave the exploration of additional settings to future work.
			
			1) Covariance-based moment conditions \citep{lewbel2012using,tchetgen2021GENIUS}. Specifically, let $\mathcal{S}$ be a proper Hilbert subspace and consider finding a function $h^o(X)$ that satisfies
			\begin{align}\label{eq: outcome bridge'}
				\mathbb E\big[d(Z,X)\{Y - Ah^o(X)\}\{A-\mathbb{E}[A\mid Z,X]\}\big] = 0, \qquad \forall\, d \in \mathcal{S}^{\perp}. 
			\end{align}
			Notably, Equation~\eqref{eq: outcome bridge'} is analogous to the form considered in Equation~\eqref{eq: outcome bridge}. When $\mathcal{S}=L_2(X)$, Equation~\eqref{eq: outcome bridge'} is equivalent to finding an $h^o(X)$ such that $\mathrm{Cov}\{\pi(Z,X),\mathrm{Cov}\{Y-Ah^o(X),A\mid Z,X\}\mid X\}\equiv 0$ for all $\pi(Z,X)$.
			In particular, when $Z$ is binary, $h^o(X)$ can be solved in closed-form
			\begin{align*}
				h^o(X)=\dfrac{\mathrm{Cov}\{Z,\mathrm{Cov}\{Y,A\mid Z,X\}\mid X\}}{\mathrm{Cov}\{Z,\mathrm{Cov}\{A,A\mid Z,X\}\mid X\}}.
			\end{align*}
			\citet{sun2022selective} identify the ATE using $\mathbb{E}[h^o(X)]$. Therefore, the identification strategy adopted in \citet{sun2022selective} can be viewed as a special case of solving Equation~\eqref{eq: outcome bridge'}.
			
			2) Proximal causal inference with invalid proxies \citep{yu2025fortified}. 
			Let $\mathcal{S}$ be a proper Hilbert subspace of $L_2(Z,X)$, and find $h^o(W,A,X)$ satisfying 
			\begin{align*}
				\mathbb E\big[d(Z,A,X)\{Y - h^o(W,A,X)\}\big] = 0, \qquad \forall\, d \in \mathcal{S}^{\perp}, 
			\end{align*}
			where the orthogonal complement is taken in $L_2(Z,A,X)$, and $Z,W$ are known as the treatment-inducing proxy and outcome-inducing proxy, respectively.
			When $\mathcal{S}=\mathcal{S}_{\gamma}$, \cite{yu2025fortified} identify the ATE using $\mathbb{E}[h^o(W,1,X)-h^o(W,0,X)]$.
			
			3) Multiple PIVs are available, and some may be invalid. One can propose linear models and combine the projected idea with the majority and plurality rules \citep{han2008detecting,kang2016instrumental,guo2018confidence}. 
			For example, suppose that $X$ is empty, and there exists a collection $\{\phi_1(Z),\ldots, \phi_p(Z)\}$ of linearly independent functions and a proper Hilbert subspace $\mathcal{S}$ such that $\mathcal{S}\cap\operatorname{span}\{\phi_1,\ldots,\phi_p\}=\{0\}$. Define the partially linear subspace $\mathcal{S}_{\mathrm{pl}}:=\{s(Z)+\sum_{i=1}^p\theta_i\phi_i(Z): s\in\mathcal{S},\ \theta\in\mathbb{R}^p\}$, a proper Hilbert subspace of $L_2(Z)$ induced by $\mathcal{S}$.
			Assume the following outcome model
			\begin{align*}
				&Y = \beta^o A + s^o(Z) + \sum_{i=1}^p\theta_i^o\phi_i(Z) +\epsilon, \text{ with $s^o\in\mathcal{S}$ and $\mathbb{E}[\epsilon\mid Z]=0$, and we have}\\
				&\mathbb{S}_{\mathrm{pl}}[Y](Z)=s_Y^o(Z) + \sum_{i=1}^p \gamma_i^o\phi_i(Z),\quad
				\mathbb{S}_{\mathrm{pl}}[A](Z)=s_A^o(Z) + \sum_{i=1}^p \alpha_i^o\phi_i(Z), \text{ where }\\
				&\gamma_i^o = \theta_i^o + \alpha_i^o\beta^o,\text{ and } s_Y^o(Z)=s^o(Z)+\beta^o s_A^o(Z) \text{ with } s_Y^o,s_A^o\in \mathcal{S}.
			\end{align*}
			Without loss of generality, assume that $\alpha_i^o\neq 0$ for $i=1,\ldots,p$, so that all candidate components are relevant to the treatment. Let $\mathcal{V}:=\{i:\theta_i^o=0\}$. The majority rule requires $|\mathcal{V}|>{p}/{2}.$ Therefore, $\beta^o$ can be characterized as the value of $\beta$ that makes the vector $\left\{\gamma_i^o-\alpha_i^o\beta\right\}_{i=1}^p$ as sparse as possible. When $\mathcal{S}=\mathbb{R}$, this reduces to the model considered in \citet{kang2016instrumental}.
			
			\subsection*{Conflict of interest}
			
			None declared. 
			
			\subsection*{Funding}
			
			The authors were partially supported by the National Key R\&D Program of China (2024YFA1015600), the National Natural Science Foundation of China (12471266 and U23A2064). 
			
			\subsection*{Data availability}
			
			The datasets are available at \href{https://doi.org/10.26193/HE3BAO}{https://doi.org/10.26193/HE3BAO}; \href{https://doi.org/10.26193/HMBXJV}{https://doi.org/10.26193/HMBXJV}. 
            The computer code is available at \href{https://github.com/chensy123-sys/Projected-IV}{https://github.com/chensy123-sys/Projected-IV}.
			
			{\setlength{\baselineskip}{0.63\baselineskip}
				\let\newline\space
				\putbib[ref]
			}
			
		\end{bibunit}

		\clearpage
		\newpage
		\begin{bibunit}[plainnat]

			\linespread{1.57} 
			\appendix
			\section*{Appendix}

			\section{Identification of other estimands}\label{append: identification of other estimands}
			All the preceding discussion focuses on identifying the ATE. In this section, we consider several estimands beyond the ATE.
			\subsection{Mean potential outcome}
			We present the identification of the mean potential outcome $\mathbb{E}[Y(a)]$, where $A$ can be multi-categorical.
			\begin{assumption}[$Z$-invariant counterfactual mean]\label{as: Z-invariant treatment effect'}
				$\mathbb{E}[Y(a)\mid Z_{[-\nu_0]},X] \indep Z_{[-\nu_0]}\mid X$.
			\end{assumption}
			
			\begin{assumption}
				\label{as: alternative'}
				At least one of the two conditions holds: 
				\begin{enumerate}
					\item[(a)] \emph{(Additive IV)}
					For some $\gamma\in\{1,\ldots,|\nu_0|\}$, there exist $\{b_{\nu,a}(U,Z_{[-\nu]},X):
					\nu\subseteq\nu_0,\ |\nu|=\gamma\}$ and $c_a(Z,X)$
					such that
					$\Pr(A=a\mid Z,U,X)
					=
					\sum_{\nu\subseteq\nu_0,|\nu|=\gamma}
					b_{\nu,a}(U,Z_{[-\nu]},X)
					+c_a(Z,X).
					$
					
					\item[(b)] \emph{($U$-invariant conditional counterfactual mean)} $\mathbb{E}[Y(a)\mid U,Z_{[-\nu_0]},X] \indep U\mid Z_{[-\nu_0]},X$.
				\end{enumerate}
				If condition~(b) holds, define
				$\gamma_{\dagger,a}(\nu_0)=|\nu_0|$.
				Otherwise, define $\gamma_{\dagger,a}(\nu_0)$ as the largest
				$\gamma\in\{1,\ldots,|\nu_0|\}$ for which condition~(a) holds.
			\end{assumption}
			Assumption~\ref{as: Z-invariant treatment effect'} states that
			$\mathbb{E}[Y(a)\mid Z_{[-\nu_0]},X]$ does not depend on $Z_{[-\nu_0]}$,
			paralleling the role of Assumption~\ref{as: Z-invariant treatment effect}. Notably, under Assumptions~\ref{as: IV independence}, \ref{as: strong latent ignorability},
			and \ref{as: Z-invariant treatment effect'}, one can verify that
			$\mathbb{E}[Y(a)\mid Z,X]$ does not depend on $Z$, thereby ruling out cases in which $Z$
			has a non-negligible effect on $Y(a)$. Under the additional Assumption~\ref{as: alternative'}(b),
			one can further verify that $\mathbb{E}[Y(a)\mid A,U,Z,X]$ does not depend on $A$, $U$, or $Z$,
			thereby degenerating to the NUC setting. Compared to \citet{chen2026double}, the main novelty of Assumption~\ref{as: alternative'}(a) lies in relaxing the restrictive additive IV assumption.
			We can now identify the mean potential outcome as follows.
			\begin{theorem}[Identification of mean potential outcome]\label{thm: identification of MPO}
				If Assumptions~\ref{as: IV independence}, \ref{as: strong latent ignorability}, \ref{as: Z-invariant treatment effect'}, and \ref{as: alternative'} hold for $\nu_0$, and $\pi$ is an $\mathcal{S}_{\gamma}$-PRWF for the binary indicator $I\{A=a\}$ with $\gamma\leq \gamma_{\dagger,a} (\nu_0)$, then 
				\begin{align*}
					\mathbb{E}[Y(a)]=\mathbb{E}\left[\dfrac{\pi(Z,X) - \mathbb{S}_{\gamma}[\pi](Z,X)}{\mathbb{E}\left[\left\{\pi(Z,X) - \mathbb{S}_{\gamma}[\pi](Z,X)\right\}I\{A=a\}|X\right]}I\{A=a\}Y\right].
				\end{align*}
			\end{theorem}
			Following \citet{chen2026double}, this theorem can be extended to identify dose-response curves when the treatment $A$ is continuous.
			
			\subsection{Average treatment effect on the treated}
			In this subsection, we aim to identify the average treatment effect on the treated (ATT), $\mathbb{E}[Y(1)-Y(0)\mid A=1]$. First, we reformulate Assumptions~\ref{as: Z-invariant treatment effect} and \ref{as: alternative} as follows.
			\begin{assumption}[$Z$-invariant treated counterfactual mean]\label{as: Z-invariant treatment effect''}
				$\mathbb{E}[Y(0)| A=1,Z_{[-\nu_0]},X] \indep Z_{[-\nu_0]}| X$.
			\end{assumption}
			\begin{assumption}\label{as: alternative''}
				At least one of the following conditions holds:
				\begin{enumerate}
					\item \emph{(Multiplicative IV)} There exist two measurable functions $b(U,Z_{[-\nu_0]},X)$ and $c(Z,X)$ such that
					$\Pr(A=1\mid Z, U, X) = b(U,Z_{[-\nu_0]}, X) \times c(Z, X).$
					\item \emph{($U$-invariant treated counterfactual mean)} $\mathbb{E}[Y(0)\mid A=1,U,Z_{[-\nu_0]},X] \indep U\mid Z_{[-\nu_0]},X$.
				\end{enumerate}
			\end{assumption}
			We are now ready to establish identification of the ATT.
			\begin{theorem}[Identification of ATT]\label{thm: identification ATT}
				Suppose that Assumptions~\ref{as: IV independence}, \ref{as: strong latent ignorability}, \ref{as: Z-invariant treatment effect''}, and \ref{as: alternative''} hold for $\nu_0$, and $\pi$ is an $\mathcal{S}_{\gamma}$-PRWF with $\gamma\leq |\nu_0|$. Then, 
				\begin{align*}
					\mathbb{E}[Y(1)-Y(0)\mid A=1]=\mathbb{E}\left[Y-\dfrac{\mathbb{E}[\{\pi(Z,X) - \mathbb{S}_{\gamma}[\pi](Z,X)\}(A-1)Y\mid X]}{\mathbb{E}\left[\{\pi(Z,X) - \mathbb{S}_{\gamma}[\pi](Z,X)\}A\middle| X\right]}\middle| A=1\right].
				\end{align*}
			\end{theorem}
			When all the IVs are valid, Theorem~\ref{thm: identification ATT} reduces to the identification strategy proposed in \citet{liu2026multiplicative}. In general, when $|\nu_0|<d_Z$, Assumption~\ref{as: alternative''}(a) relaxes the multiplicative IV assumption in \citet{liu2026multiplicative}.
			
			\section{Connections to debiased generalized method of moments}\label{append: connections to debiased generalized method of moments}
			In this section, we provide further discussion of the partially linear model considered in Section~\ref{sec: connection to partially linear model}.
			Following \citet{sun2023semiparametric} and \citet{zhang2025mr}, we formulate a debiased generalized method of moments (DGMM) estimator \citep{chernozhukov2022locally} within the PIV framework using multiple $\mathcal{S}$-PRWFs.
			
			In particular, under Equation~\eqref{eq: partially linear model}, there exists a solution to Equation~\eqref{eq: outcome bridge} by setting $h^o(X)=\beta^o$. 
			Let $\{\pi_{l}(Z,X)\}_{l=1}^{L}$ be $L$ prespecified weighting functions.
			Define the vector of weighting functions by
			$\Pi(Z,X):=[\pi_1(Z,X),\ldots,\pi_L(Z,X)]^\top\in\mathbb{R}^{L}$.
			For each $l$, define $\rho_{\mathcal{S},\pi_l}^o(Z,X):=\mathbb{S}[\pi_l](Z,X)$, and let
			$\rho_{\mathcal{S},\Pi}^o(Z,X):=[\rho_{\mathcal{S},\pi_1}^o(Z,X),\ldots,\rho_{\mathcal{S},\pi_L}^o(Z,X)]^\top\in\mathbb{R}^{L}$.
			We can identify $\beta^o$ by the estimating equation $$\mathbb{E}\left[(\pi_{l}(Z,X)-\rho_{\mathcal{S},\pi_l}^o(Z,X)) \left\{Y - A \beta^o \right\}\right]=0$$ 
			if there exists at least one $l$ such that $\mathbb{E}\left[(\pi_{l}(Z,X)-\rho_{\mathcal{S},\pi_l}^o(Z,X))A\right]\neq 0$.
			
			Specifically, for $x\in\mathbb{R}^L$, define $\|x\|_{\Sigma}^2 := x^\top \Sigma x$ and $x^{\otimes 2}:=xx^\top\in\mathbb{R}^{L\times L}$, and $$\hat\Sigma_{\mathcal{S}} := \left(\frac{1}{K}\sum_{k=1}^K\mathbb{E}_{nk}\left[\left\{\Pi(Z,X)-\hat\rho_{\mathcal{S},\Pi}^{(n,-k)}(Z,X)\right\}^{\otimes 2}\right]\right)^{-1} \in \mathbb{R}^{L \times L}.$$ 
			Then, the DGMM estimator $\hat\beta_{\mathcal{S}}$ can be constructed as
			\begin{align*}
				&\hat\beta_{\mathcal{S}}:=\underset{\beta\in\mathbb{R}}{\arg\min}\left\|\dfrac{1}{K}\sum_{k=1}^K\mathbb{E}_{nk}\left[\hat\varphi_{\mathcal{S},\Pi}^{(n,-k)}(\beta)\right]\right\|_{\hat\Sigma_{\mathcal{S}}}^2,\text{ where }\\
				&\hat\varphi_{\mathcal{S},\Pi}^{(n,-k)}(\beta):=\left\{\Pi(Z,X)-\hat\rho_{\mathcal{S},\Pi}^{(n,-k)}(Z,X)\right\} \times \left\{Y-\hat\eta_{\mathcal{S}}^{(n,-k)}(Z,X) - \{A - \hat\delta_{\mathcal{S}}^{(n,-k)}(Z,X)\}\beta \right\}\in\mathbb{R}^{L}.
			\end{align*}
			
			Algorithm~\ref{alg: DGMM} provides a computationally efficient procedure for calculating the DGMM estimator $\hat\beta_{\mathcal{S}}$. Compared to the DGMM framework developed by \citet{zhang2025mr}, our approach neither requires mutual independence among the IVs nor restricts $\mathcal{S}$ to the specific form $\mathcal{S}_{\gamma}$.
			For theoretical completeness, we present the statistical properties of $\hat\beta_{\mathcal{S}}$; these results follow directly from standard DGMM theory.
			\begin{algorithm}[t]
				\caption{DGMM estimator}
				\label{alg: DGMM}
				\begin{algorithmic}[1]
					\State \textbf{Input}: Data $\{O_i\}_{i=1}^n$, number of folds $K$, a proper Hilbert space $\mathcal{S}$, and weighting functions $\{\pi_l\}_{l=1}^L$.
					
					\State Partition the sample into $K$ folds $\{I_k\}_{k=1}^K$.
					
					\State For all $k=1,\dots,K$, use training sample $I_{-k}$ to train $\hat\delta_{\mathcal{S}}^{(n,-k)}(Z,X)$, $\hat\eta_{\mathcal{S}}^{(n,-k)}(Z,X)$, and $\hat\rho_{\mathcal{S},\Pi}^{(n,-k)}(Z,X)$.
					
					\State Fit an ordinary least squares (OLS) regression without an intercept:
					\[
					\left\{ A_i - \hat\delta_{\mathcal{S}}^{(n,-k_i)}(Z_i,X_i) \right\}
					\sim
					\left\{ \pi_l(Z_i,X_i)-\hat\rho_{\mathcal{S},\pi_l}^{(n,-k_i)}(Z_i,X_i): l=1,\dots, L\right\},
					\] 
					where $k_i$ is an index such that $i \in I_{k_i}$. Calculate fitted values $\{\hat{A}_i\}_{i=1}^n$.
					
					\State Fit OLS without an intercept:
					$
					\left\{ Y_i - \hat\eta_{\mathcal{S}}^{(n,-k_i)}(Z_i,X_i) \right\}
					\sim
					\left\{ \hat{A}_i \right\}
					$ to obtain $\hat{\beta}_{\mathcal{S}}$.
					
					\State \textbf{Output}: Estimated coefficient $\hat{\beta}_{\mathcal{S}}$.
				\end{algorithmic}
			\end{algorithm}

			\begin{theorem}\label{thm: DGMM}
                Under Equation~\eqref{eq: partially linear model},
				suppose that $L$ is fixed, and $\pi_l(Z,X)$ are uniformly bounded prespecified weighting functions, where $l=1,\ldots,L$.
				Assume that $Y$ and $A$ are bounded, $\hat\rho_{\mathcal{S},\pi_l}^{(n,-k)}$, $\hat\delta_{\mathcal{S}}^{(n,-k)}$, $\hat\eta_{\mathcal{S}}^{(n,-k)}\in\mathcal{S}$ for every $k$ and $l$, and that these nuisance estimators and the corresponding true nuisance functions are uniformly bounded.
				For every $k$, assume further that
				\begin{align*}
					&\mathbb{E}\!\left[\sum_{l=1}^{L}\|\hat\rho_{\mathcal{S},\pi_l}^{(n,-k)}-\rho_{\mathcal{S},\pi_l}^o\|_2^2\right]
					+\mathbb{E}\!\left[\|\hat\delta_{\mathcal{S}}^{(n,-k)}-\delta_{\mathcal{S}}^o\|_2^2\right]
					+\mathbb{E}\!\left[\|\hat\eta_{\mathcal{S}}^{(n,-k)}-\eta_{\mathcal{S}}^o\|_2^2\right]=o(1),\\
					&\sum_{l=1}^{L}\|\hat\rho_{\mathcal{S},\pi_l}^{(n,-k)}-\rho_{\mathcal{S},\pi_l}^o\|_2
					\left\{\|\hat\delta_{\mathcal{S}}^{(n,-k)}-\delta_{\mathcal{S}}^o\|_2
					+\|\hat\eta_{\mathcal{S}}^{(n,-k)}-\eta_{\mathcal{S}}^o\|_2\right\}=o_p(n^{-1/2}).
				\end{align*}
				Define 
				\begin{align*}
					&d_{\mathcal{S}}(Z,X):=\Pi(Z,X)-\rho_{\mathcal{S},\Pi}^o(Z,X),\quad 
					u_{\mathcal{S}}^o(O):=Y-\eta_{\mathcal{S}}^o(Z,X)-\{A-\delta_{\mathcal{S}}^o(Z,X)\}\beta^o,\\
					&q_{\mathcal{S}}:=\mathbb{E}[d_{\mathcal{S}}(Z,X)A],\quad
					\Sigma_{\mathcal{S}}:=\left\{\mathbb{E}[d_{\mathcal{S}}(Z,X)^{\otimes 2}]\right\}^{-1},\quad
					\Omega_{\mathcal{S}}:=\mathbb{E}[\{u_{\mathcal{S}}^o(O)\}^2d_{\mathcal{S}}(Z,X)^{\otimes 2}].
				\end{align*}
				Assume that $\mathbb{E}[d_{\mathcal{S}}(Z,X)^{\otimes 2}]$ is nonsingular,
				$q_{\mathcal{S}}^\top\Sigma_{\mathcal{S}}q_{\mathcal{S}}>0$, and
				$0<q_{\mathcal{S}}^\top\Sigma_{\mathcal{S}}\Omega_{\mathcal{S}}\Sigma_{\mathcal{S}}q_{\mathcal{S}}<\infty$.
				Then $\sqrt{n}(\hat\beta_{\mathcal{S}}-\beta^o)/\sigma_{\mathcal{S}}^o\overset{d}{\to}\mathcal{N}(0,1)$, where $(\sigma_{\mathcal{S}}^o)^2:=(q_{\mathcal{S}}^\top\Sigma_{\mathcal{S}}q_{\mathcal{S}})^{-2}q_{\mathcal{S}}^\top\Sigma_{\mathcal{S}}\Omega_{\mathcal{S}}\Sigma_{\mathcal{S}}q_{\mathcal{S}}.$
			\end{theorem}
			Theorem~\ref{thm: DGMM} establishes the theoretical properties of the DGMM estimator proposed in Algorithm~\ref{alg: DGMM}. In practice, one can implement the two-stage least squares (2SLS) estimator using the \texttt{ivreg} function from the \texttt{AER} package in \texttt{R}.

			\section{Uniqueness of the maximal regular index}\label{append: uniqueness of maximal index}
			In this section, we demonstrate the uniqueness of $\nu_{\max}$ defined in Section~\ref{sec: connection to invalid IV}. In particular, we need the following regularity condition.
			\begin{assumption}\label{as: regular condition}
				For almost every $x\in\mathcal{X}$, the conditional support of $Z$ is a Cartesian product $\mathcal{Z}(x)=\prod_{j=1}^{d_Z}\mathcal{Z}_{[j]}(x)$, and $p_{Z\mid X}(z\mid x)>0$ for every $z\in\mathcal{Z}(x)$. 
			\end{assumption}
			We now present the following proposition.
			\begin{proposition}\label{prop: uniqueness of largest nu}
				Under Assumption~\ref{as: regular condition}, let $\nu_1$ and $\nu_2$ be two index sets that satisfy Assumptions~\ref{as: IV independence}--\ref{as: alternative} (or \ref{as: IV independence}, \ref{as: weak latent ignorability}, \ref{as: Z-invariant treatment effect}, \ref{as: alternative}). Then, $\nu_1\cup \nu_2$ is also an index set that satisfies the corresponding assumptions. In addition, if $\nu_1\subseteq\nu_2$, then $\gamma_{\dagger}(\nu_1)\leq \gamma_{\dagger}(\nu_2)$.
			\end{proposition}
			By Proposition~\ref{prop: uniqueness of largest nu}, if the family of regular index sets is nonempty, then it is closed under finite unions, and therefore has a unique largest element $\nu_{\max}$. In addition, $\gamma_{\dagger}(\nu_{\max})$ is the largest value of $\gamma_{\dagger}(\nu_0)$ over all regular index sets $\nu_0$.
			
			\section{Proof of Propositions}
			
			\subsection{Proof of Proposition~\ref{prop: proper Hilbert subspace}}
			\begin{proof}[Proof of Proposition~\ref{prop: proper Hilbert subspace}(a):]
				By the definition of $\mathbb{S}$, we know that for all $s\in\mathcal{S}$,
				\begin{align*}
					0\equiv& \nabla_{t}\mathbb{E}[\{f(O)-\mathbb{S}[f](Z,X)-ts(Z,X)\}^2]\eval_{t=0}\\
					=&-2\mathbb{E}[\{f(O)-\mathbb{S}[f](Z,X)-ts(Z,X)\}s(Z,X)]\eval_{t=0}\\
					=&-2\mathbb{E}[\{f(O)-\mathbb{S}[f](Z,X)\}s(Z,X)].
				\end{align*}
				Since this equality holds for all $\tilde s(Z,X)=c(X)s(Z,X)$ with $c\in L_{\infty}(X)$ by the definition of a proper Hilbert subspace, we know that for all $s\in\mathcal{S}$,
				\begin{align*}
					\mathbb{E}[\{f(O)-\mathbb{S}[f](Z,X)\}s(Z,X)c(X)]\equiv 0
					\quad\Rightarrow\quad
					 \mathbb{E}[\{f(O)-\mathbb{S}[f](Z,X)\}s(Z,X)\mid X]\equiv 0.
				\end{align*}
				Next, if $s^*\in\mathcal{S}$ is another function such that the equation in Proposition~\ref{prop: proper Hilbert subspace}(a) holds, then for all $s\in\mathcal{S}$,
				\begin{align*}
					&\mathbb{E}\!\left[\left\{f(O)-\mathbb{S}[f](Z,X)\right\}s(Z,X)\mid X\right]=0,\qquad
					\mathbb{E}\!\left[\left\{f(O)-s^*(Z,X)\right\}s(Z,X)\mid X\right]=0.\\
					\Rightarrow\quad&
					\mathbb{E}\!\left[\left\{s^*(Z,X)-\mathbb{S}[f](Z,X)\right\}s(Z,X)\mid X\right]=0.\\
					\Rightarrow\quad&
					\mathbb{E}\!\left[\left\{s^*(Z,X)-\mathbb{S}[f](Z,X)\right\}^2\mid X\right]=0.
					\tag{Setting $s(Z,X)=s^*(Z,X)-\mathbb{S}[f](Z,X)$}
				\end{align*}
				This indicates that $s^*(Z,X)=\mathbb{S}[f](Z,X)$ almost surely, finishing the proof of Proposition~\ref{prop: proper Hilbert subspace}(a).
			\end{proof}
			\begin{proof}[Proof of Proposition~\ref{prop: proper Hilbert subspace}(b):]
				We directly calculate that for all $s\in\mathcal{S}$,
				\begin{align*}
					&\mathbb{E}\left[\{c_1(X)f_1(O)+c_2(X)f_2(O)-
					c_1(X)\mathbb{S}[f_1](Z,X) - c_2(X)\mathbb{S}[f_2](Z,X)
					\}s(Z,X)\middle| X\right]\\
					=&c_1(X)\mathbb{E}\left[\{f_1(O)- \mathbb{S}[f_1](Z,X)\}s(Z,X)\mid X\right]\\
					&+c_2(X)\mathbb{E}\left[\{f_2(O)- \mathbb{S}[f_2](Z,X)\}s(Z,X)\mid X\right]=0.
				\end{align*}
				The final step follows from the property of $\mathbb{S}[f_1]$ and $\mathbb{S}[f_2]$. Therefore, by Proposition~\ref{prop: proper Hilbert subspace}(a),
				$\mathbb{S}[c_1f_1+c_2f_2](Z,X) = 
				c_1(X)\mathbb{S}[f_1](Z,X) + c_2(X)\mathbb{S}[f_2](Z,X).$
			\end{proof}
			\begin{proof}[Proof of Proposition~\ref{prop: proper Hilbert subspace}(c):]
				We directly calculate that for all $s\in\mathcal{S}$,
				\begin{align*}
					&\mathbb{E}[\{f(O)-\mathbb{S}[\mathbb{S}'[f]](Z,X)\}s(Z,X)\mid X]\\
					=&\mathbb{E}[\{f(O)-\mathbb{S}'[f](Z,X)\}s(Z,X)\mid X]
					+\mathbb{E}[\{\mathbb{S}'[f](Z,X)-\mathbb{S}[\mathbb{S}'[f]](Z,X)\}s(Z,X)\mid X]=0.
				\end{align*}
				The final step follows from $s\in\mathcal{S}\subseteq \mathcal{S}'$ and the definition of $\mathbb{S}$. 
			\end{proof}
			
			\subsection{Proof of Proposition~\ref{prop: IV relevance monotonicity}}
			On the one hand,
			\begin{align*}
				&\mathbb{E}\left[\left\{\pi_*^o(Z,X) - \mathbb{S}[\pi_*^o](Z,X)\right\}^2 \mid X\right]\\
				=&\mathbb{E}\left[\left\{\pi_*^o(Z,X) - \mathbb{S}'[\pi_*^o](Z,X)\right\}^2 \mid X\right]+\mathbb{E}\left[\left\{\mathbb{S}'[\pi_*^o](Z,X) - \mathbb{S}[\pi_*^o](Z,X)\right\}^2 \mid X\right]\\
				&+2\mathbb{E}\left[\left\{\pi_*^o(Z,X) - \mathbb{S}'[\pi_*^o](Z,X)\right\} \left\{\mathbb{S}'[\pi_*^o](Z,X) - \mathbb{S}[\pi_*^o](Z,X)\right\}\mid X\right]\\
				=&\mathbb{E}\left[\left\{\pi_*^o(Z,X) - \mathbb{S}'[\pi_*^o](Z,X)\right\}^2 \mid X\right]+\mathbb{E}\left[\left\{\mathbb{S}'[\pi_*^o](Z,X) - \mathbb{S}[\pi_*^o](Z,X)\right\}^2 \mid X\right]\\
				\geq&\mathbb{E}\left[\left\{\pi_*^o(Z,X) - \mathbb{S}'[\pi_*^o](Z,X)\right\}^2 \mid X\right]\geq \epsilon_0.
			\end{align*}
			The second step follows because $\pi_*^o - \mathbb{S}'[\pi_*^o]$ is orthogonal to $\mathcal{S}'$ and $\mathbb{S}'[\pi_*^o] - \mathbb{S}[\pi_*^o]\in\mathcal{S}'$ since $\mathcal{S}\subseteq \mathcal{S}'$.
			This completes the proof of $\mathcal{S}$-relevance.
			
			\subsection{Proof of Proposition~\ref{prop: existence of PRWF}}
			\begin{proof}[Proof of (a) $\Rightarrow$ (b):]
				Recall that $\pi_*^o(Z,X)=\Pr(A=1\mid Z,X)$, which is uniformly bounded.
				If there exists an $\mathcal{S}$-PRWF $\pi\in L_2(Z,X)$, then there exists a positive $\epsilon_0$ such that 
				\begin{align*}
					\epsilon_0\leq& \left|\mathbb{E}\left[\{\pi(Z,X) - \mathbb{S}[\pi](Z,X)\}A\middle|X\right]\right|\\
					=&|\mathbb{E}\left[\{\pi(Z,X) - \mathbb{S}[\pi](Z,X)\}
					\{\pi_*^o(Z,X) - \mathbb{S}[\pi_*^o](Z,X)\}\middle|X\right]|\\
					\leq&\mathbb{E}\left[\{\pi(Z,X) - \mathbb{S}[\pi](Z,X)\}^2\middle|X\right]^{1/2}\times \mathbb{E}\left[
					\{\pi_*^o(Z,X) - \mathbb{S}[\pi_*^o](Z,X)\}^2\middle|X\right]^{1/2}\\
					\leq& \frac{1}{\sqrt{\epsilon_0}}\mathbb{E}\left[
					\{\pi_*^o(Z,X) - \mathbb{S}[\pi_*^o](Z,X)\}^2\middle|X\right]^{1/2}
					=\frac{1}{\sqrt{\epsilon_0}}\mathbb{E}\left[
					\{\pi_*^o(Z,X) - \mathbb{S}[\pi_*^o](Z,X)\}A\middle|X\right]^{1/2}.
				\end{align*}
				The final step follows because $\pi_*^o(Z,X) - \mathbb{S}[\pi_*^o](Z,X)$ is orthogonal to $\mathbb{S}[\pi_*^o](Z,X)$ by Proposition~\ref{prop: proper Hilbert subspace}(a).
				In addition, since $\pi_*^o$ is uniformly bounded, $$\mathbb{E}[\{\pi_*^o(Z,X) - \mathbb{S}[\pi_*^o](Z,X)\}^2\mid X]\leq \mathbb{E}[\pi_*^o(Z,X)^2\mid X]\leq 1.$$
				As a result, $\pi_*^o$ is an $\mathcal{S}$-PRWF.
			\end{proof}
			\begin{proof}[Proof of (b) $\Rightarrow$ (c):]
				If $\pi_*^o$ is an $\mathcal{S}$-PRWF, then there exists a positive $\epsilon$, such that 
				\begin{align*}
					\epsilon\leq &\mathbb{E}\left[
					\{\pi_*^o(Z,X) - \mathbb{S}[\pi_*^o](Z,X)\}A\middle|X\right]\\
					=&\mathbb{E}\left[
					\{\pi_*^o(Z,X) - \mathbb{S}[\pi_*^o](Z,X)\}\pi_*^o(Z,X)\middle|X\right]\\
					=&\mathbb{E}\left[
					\{\pi_*^o(Z,X) - \mathbb{S}[\pi_*^o](Z,X)\}^2\middle|X\right].
				\end{align*}
				Thus, $Z$ is $\mathcal{S}$-relevant to $A$.
			\end{proof}
			\begin{proof}[Proof of (c) $\Rightarrow$ (a):]
				If $Z$ is $\mathcal{S}$-relevant to $A$, then there exists a positive $\epsilon_0>0$ such that
				\begin{align*}
					\epsilon_0\leq &\mathbb{E}\left[
					\{\pi_*^o(Z,X) - \mathbb{S}[\pi_*^o](Z,X)\}^2\middle|X\right]=\mathbb{E}\left[
					\{\pi_*^o(Z,X) - \mathbb{S}[\pi_*^o](Z,X)\}\pi_*^o(Z,X)\middle|X\right]\\
					=&\mathbb{E}\left[
					\{\pi_*^o(Z,X) - \mathbb{S}[\pi_*^o](Z,X)\}A\middle|X\right].
				\end{align*}
				Since $\pi_*^o$ is uniformly bounded, it is an $\mathcal{S}$-PRWF.
			\end{proof}
			This completes the proof of Proposition~\ref{prop: existence of PRWF}.

			\subsection{Proof of Proposition~\ref{prop: NUCEM equivalent representation}}
			For every $\pi\in L_2(Z,X)$,
			\begin{align*}
				&\operatorname{Cov}\!\left\{
				A\{\pi(Z,X)-\mathbb{S}[\pi](Z,X)\},
				Y(1)-Y(0)\mid X
				\right\}\\
				&\quad=
				\mathbb{E}\!\left[
				\{\pi(Z,X)-\mathbb{S}[\pi](Z,X)\}
				A\{Y(1)-Y(0)-\mathbb{E}[Y(1)-Y(0)\mid X]\}
				\mid X
				\right].
			\end{align*}
			As $\pi$ varies over $L_2(Z,X)$,
			$\pi-\mathbb{S}[\pi]$ ranges over $\mathcal{S}^{\perp}$.
			By iterated expectations and conditional orthogonality,
			the above covariance vanishes for every $\pi$ if and only if
			\[
			\mathbb{E}\!\left[
			A\{Y(1)-Y(0)-\mathbb{E}[Y(1)-Y(0)\mid X]\}
			\mid Z,X
			\right]\in\mathcal{S}.
			\]
			Moreover, if $Z\indep\{A(z),Y(1)-Y(0)\}\mid X$ for every $z$,
			consistency gives
			\begin{align*}
				g(z,X)
				&:=\mathbb{E}\!\left[
				A(z)\{Y(1)-Y(0)-\mathbb{E}[Y(1)-Y(0)\mid X]\}
				\mid X
				\right]\\
				&=\mathbb{E}\!\left[
				A\{Y(1)-Y(0)-\mathbb{E}[Y(1)-Y(0)\mid X]\}
				\mid Z=z,X
				\right].
			\end{align*}
			Hence, Definition~\ref{defn: projected IV}(b) is equivalent
			to $g\in\mathcal{S}$.

			\subsection{Proof of Proposition~\ref{prop: treatment bridge existence}}
			If $Z$ is $\mathcal{S}$-relevant to $A$, then by Proposition~\ref{prop: existence of PRWF}, there exist a function $\pi(Z,X)$ and a positive constant $\epsilon_0$ such that $|\mathbb{E}[\{\pi(Z,X) - \mathbb{S}[\pi](Z,X)\}A\mid X]|\geq \epsilon_0.$ Therefore, if we define
			$$
			q_{\mathcal{S},\pi}(Z,X):= \dfrac{\pi(Z,X) - \mathbb{S}[\pi](Z,X)}{\mathbb{E}[\{\pi(Z,X) - \mathbb{S}[\pi](Z,X)\}A\mid X]}\in\mathcal{S}^{\perp},
			$$
			then $q_{\mathcal{S},\pi}$ satisfies Equation~\eqref{eq: treatment bridge}, since
			\begin{align*}
				\mathbb{E}[q_{\mathcal{S},\pi}(Z,X)A\mid X]
				=&\mathbb{E}\left[A\dfrac{\pi(Z,X) - \mathbb{S}[\pi](Z,X)}{\mathbb{E}[\{\pi(Z,X) - \mathbb{S}[\pi](Z,X)\}A\mid X]}\middle| X\right]\equiv 1.
			\end{align*} 
			In addition, one can verify that
			\begin{align*}
				&\mathbb{E}[\{q_{\mathcal{S},\pi}(Z,X)-\mathbb{S}[q_{\mathcal{S},\pi}](Z,X)\}^2\mid X]\\
				=&\dfrac{\mathbb{E}[\{\pi(Z,X)-\mathbb{S}[\pi](Z,X)\}^2\mid X]}{\mathbb{E}[\{\pi(Z,X) - \mathbb{S}[\pi](Z,X)\}A\mid X]^2}\leq \dfrac{1}{\epsilon_0^2}\mathbb{E}[\{\pi(Z,X)-\mathbb{S}[\pi](Z,X)\}^2\mid X].
			\end{align*} 
			Therefore, since $\mathbb{E}[\{\pi(Z,X)-\mathbb{S}[\pi](Z,X)\}^2\mid X]$ is uniformly bounded, $\mathbb{E}[\{q_{\mathcal{S},\pi}(Z,X)-\mathbb{S}[q_{\mathcal{S},\pi}](Z,X)\}^2\mid X]$ is also uniformly bounded.
			
			Conversely, if there exists a solution $q(Z,X)$ to Equation~\eqref{eq: treatment bridge} that satisfies $\mathbb{E}[q(Z,X)^2\mid X]$ is uniformly bounded, then it is apparent that $q(Z,X)$ is itself an $\mathcal{S}$-PRWF, since 
			\begin{align*}
				\mathbb{E}[A\{q(Z,X)-\mathbb{S}[q](Z,X)\}\mid X]
				=\mathbb{E}[Aq(Z,X)\mid X]\equiv 1,
			\end{align*}
			which is uniformly bounded away from zero almost surely.
			By Proposition~\ref{prop: existence of PRWF}, the existence of an \(\mathcal{S}\)-PRWF implies that \(Z\) is \(\mathcal{S}\)-relevant to \(A\).

			\subsection{Proof of Proposition~\ref{prop: lower efficiency bound}}
			From the proof of Theorem~\ref{thm: identification}, if $Z$ is an $\mathcal{S}$-PIV and $\mathcal{S}$-relevant to $A$, then for any $\mathcal{S}$-PRWF $\pi$, $\mu_{\mathcal{S},\pi}^o(X)=\mathbb{E}[Y(1)-Y(0)|X]=\Psi^o(X)$, which does not depend on $\pi$. In addition, it is easy to verify that $\mathbb{E}\left[\varphi_{\mathcal{S},\pi}(O;\psi_{\mathcal{S},\pi}^o,\alpha_{\mathcal{S},\pi}^o)\middle| X\right] = \mu_{\mathcal{S},\pi}^o(X)-\psi_{\mathcal{S},\pi}^o = \Psi^o(X) - \mathbb{E}[Y(1)-Y(0)]$, which does not depend on the choice of $\pi$. 
			
			For notational simplicity, we also define 
			\begin{align*}
				W:=&Y - \eta_{\mathcal{S}}^o(Z,X)-\mu_{\mathcal{S},\pi}^o(X)(A-\delta_{\mathcal{S}}^o(Z,X))\\
				=&Y - \mathbb{S}[Y](Z,X) - \mathbb{E}[Y(1)-Y(0)|X](A-\mathbb{S}[A](Z,X))\\
				=&Y - \mathbb{S}[Y](Z,X) - \Psi^o(X)(A-\mathbb{S}[A](Z,X))\\
				=&Y - A \Psi^o(X) - \mathbb{S}[Y- A\Psi^o](Z,X).
			\end{align*}
			From Equation~\eqref{eq: homoskedastic}, $\mathbb{E}[W^2|Z,X]\indep Z|X$. Next, one can directly verify that
			\begin{align*}
				&\mathrm{Var}\left\{\varphi_{\mathcal{S},\pi}(O;\psi_{\mathcal{S},\pi}^o,\alpha_{\mathcal{S},\pi}^o)\middle| X\right\}\\
				=&\mathrm{Var}\left\{\begin{array}{l}
					\dfrac{\pi(Z,X) - \rho_{\mathcal{S},\pi}^o(Z,X)}{\kappa_{\mathcal{S},\pi}^o(X)}\left\{Y - \eta_{\mathcal{S}}^o(Z,X)\right\}-\psi_{\mathcal{S},\pi}^o\\
					+\mu_{\mathcal{S},\pi}^o(X)\left(1 - \dfrac{\left\{\pi(Z,X) -\rho_{\mathcal{S},\pi}^o(Z,X)\right\}\left\{A-\delta_{\mathcal{S}}^o(Z,X)\right\}}{\kappa_{\mathcal{S},\pi}^o(X)}\right)
				\end{array}\middle| X\right\}\\
				=&\mathrm{Var}\left\{\begin{array}{l}
					\mu_{\mathcal{S},\pi}^o(X)-\psi_{\mathcal{S},\pi}^o
					+\dfrac{\pi(Z,X) - \rho_{\mathcal{S},\pi}^o(Z,X)}{\kappa_{\mathcal{S},\pi}^o(X)}\\
					\times\left\{Y - \eta_{\mathcal{S}}^o(Z,X)-\mu_{\mathcal{S},\pi}^o(X)(A-\delta_{\mathcal{S}}^o(Z,X))\right\}
				\end{array}\middle| X\right\}\\
				=&\mathbb{E}\left[
				\left(\dfrac{\pi(Z,X) - \rho_{\mathcal{S},\pi}^o(Z,X)}{\kappa_{\mathcal{S},\pi}^o(X)}\right)^2
				\times\left\{Y - \eta_{\mathcal{S}}^o(Z,X)-\mu_{\mathcal{S},\pi}^o(X)(A-\delta_{\mathcal{S}}^o(Z,X))\right\}^2
				\middle| X\right]\\
				=&\mathbb{E}\left[
				\left(\dfrac{\pi(Z,X) - \rho_{\mathcal{S},\pi}^o(Z,X)}{\kappa_{\mathcal{S},\pi}^o(X)}\right)^2 W^2
				\middle| X\right]
				=\mathbb{E}\left[
				\left(\dfrac{\pi(Z,X) - \rho_{\mathcal{S},\pi}^o(Z,X)}{\kappa_{\mathcal{S},\pi}^o(X)}\right)^2
				\middle| X\right]
				\mathbb{E}\left[W^2\middle| X\right]\tag{\text{By Equation~\eqref{eq: homoskedastic}} that $\mathbb{E}[W^2|Z,X]\perp Z\mid X$}\\
				=&\dfrac{\mathbb{E}\left[ \left\{\pi(Z,X) - \mathbb{S}[\pi](Z,X)\right\}^2\middle| X\right]}
				{\mathbb{E}\left[ \{\pi(Z,X)-\mathbb{S}[\pi](Z,X)\}\{A-\mathbb{S}[A](Z,X)\}\middle| X\right]^2}\mathbb{E}\left[W^2\middle| X\right]\\
				=&\dfrac{\mathbb{E}\left[ \left\{\pi(Z,X) - \mathbb{S}[\pi](Z,X)\right\}^2\middle| X\right]}
				{\mathbb{E}\left[ \{\pi(Z,X)-\mathbb{S}[\pi](Z,X)\}\{\pi_*^o(Z,X)-\mathbb{S}[\pi_*^o](Z,X)\}\middle| X\right]^2}\mathbb{E}\left[W^2\middle| X\right]\\
				\geq&
				\dfrac{\mathbb{E}\left[ \left\{\pi(Z,X) - \mathbb{S}[\pi](Z,X)\right\}^2\middle| X\right]}
				{\mathbb{E}\left[ \{\pi(Z,X)-\mathbb{S}[\pi](Z,X)\}^2\middle| X\right]
					\mathbb{E}\left[ \{\pi_*^o(Z,X)-\mathbb{S}[\pi_*^o](Z,X)\}^2\middle| X\right]
				}\mathbb{E}\left[W^2\middle| X\right]
				\tag{\text{By the Cauchy-Schwarz inequality}}\\
				=&\dfrac{\mathbb{E}\left[W^2\middle| X\right]}{\mathbb{E}\left[ \{\pi_*^o(Z,X)-\mathbb{S}[\pi_*^o](Z,X)\}^2\middle| X\right]}.
			\end{align*}
			The inequality becomes an equality if and only if $\pi(Z, X)-\mathbb{S}[\pi](Z,X)=f(X)\{\pi_*^o(Z, X)-\mathbb{S}[\pi_*^o](Z,X)\}$ for some nonzero $f\in L_{\infty}(X)$, provided that $\mathbb{E}[W^2\mid X]>0$ almost surely. Finally, we verify that
			\begin{align*}
				&\mathrm{Var}\left\{\varphi_{\mathcal{S},\pi}(O;\psi_{\mathcal{S},\pi}^o,\alpha_{\mathcal{S},\pi}^o)\right\}\\
				=&\mathrm{Var}\left\{\mathbb{E}\left[
				\varphi_{\mathcal{S},\pi}(O;\psi_{\mathcal{S},\pi}^o,\alpha_{\mathcal{S},\pi}^o)\middle|X
				\right]\right\}
				+\mathbb{E}\left[\mathrm{Var}\left\{\varphi_{\mathcal{S},\pi}(O;\psi_{\mathcal{S},\pi}^o,\alpha_{\mathcal{S},\pi}^o)\middle|X\right\}\right]\\
				\geq &\mathbb{E}\left[(\Psi^o(X) - \mathbb{E}[Y(1)-Y(0)])^2\right]+
				\mathbb{E}\left[\dfrac{\mathbb{E}\left[W^2\middle| X\right]}{\mathbb{E}\left[ \{\pi_*^o(Z,X)-\mathbb{S}[\pi_*^o](Z,X)\}^2\middle| X\right]}\right]\\
				=&\mathrm{Var}\left\{\mathbb{E}[Y(1)-Y(0)\mid X]\right\}+
				\mathbb{E}\left[\dfrac{\mathbb{E}\left[W^2\middle| X\right]}{\mathbb{E}\left[ \{\pi_*^o(Z,X)-\mathbb{S}[\pi_*^o](Z,X)\}^2\middle| X\right]}\right].
			\end{align*}
			Again, the inequality becomes an equality if and only if $\pi(Z, X)-\mathbb{S}[\pi](Z,X)=f(X)\{\pi_*^o(Z, X)-\mathbb{S}[\pi_*^o](Z,X)\}$ for some nonzero $f\in L_{\infty}(X)$, provided that $\mathbb{E}[W^2\mid X]>0$ almost surely. This completes the proof of Proposition~\ref{prop: lower efficiency bound}.

			\subsection{Proof of Proposition~\ref{prop: adaptive estimator achieve lower efficiency bound}}
			It suffices to show that the two influence functions coincide almost surely. Since $Z$ is an $\mathcal{S}$-PIV for $A$, we have
			\begin{align*}
				&\psi_{\mathcal{S},\pi_*^o}^o=\psi_{\mathcal{S},*}^o,\qquad
				\rho_{\mathcal{S},\pi_*^o}^o(Z,X)=\delta_{\mathcal{S}}^o(Z,X),\\
				&\kappa_{\mathcal{S},\pi_*^o}^o(X)=\kappa_{\mathcal{S},*}^o(X),\qquad
				\mu_{\mathcal{S},\pi_*^o}^o(X)=\mu_{\mathcal{S},*}^o(X).
			\end{align*}
			Direct substitution into Theorems~\ref{thm: EIF ATE} and~\ref{thm: ATE EIF adaptive} yields
			\begin{align*}
				&\varphi_{\mathcal{S},\pi_*^o}
				(O;\psi_{\mathcal{S},\pi_*^o}^o,\alpha_{\mathcal{S},\pi_*^o}^o)
				-\varphi_{\mathcal{S},*}
				(O;\psi_{\mathcal{S},*}^o,\alpha_{\mathcal{S},*}^o)\\
				={}&
				\frac{\pi_*^o(Z,X)-\delta_{\mathcal{S}}^o(Z,X)}
				{\kappa_{\mathcal{S},*}^o(X)}
				\{Y-\eta_{\mathcal{S}}^o(Z,X)\}
				-\psi_{\mathcal{S},*}^o\\
				&+\mu_{\mathcal{S},*}^o(X)
				\left[
				1-
				\frac{\{\pi_*^o(Z,X)-\delta_{\mathcal{S}}^o(Z,X)\}
					\{A-\delta_{\mathcal{S}}^o(Z,X)\}}
				{\kappa_{\mathcal{S},*}^o(X)}
				\right]\\
				&-\frac{1}{\kappa_{\mathcal{S},*}^o(X)}
				\left[
				\{A-\pi_*^o(Z,X)\}\xi_*^o(Z,X)
				-\{A-\delta_{\mathcal{S}}^o(Z,X)\}\eta_{\mathcal{S}}^o(Z,X)
				\right]\\
				&-\frac{\pi_*^o(Z,X)-\delta_{\mathcal{S}}^o(Z,X)}
				{\kappa_{\mathcal{S},*}^o(X)}Y
				+\psi_{\mathcal{S},*}^o\\
				&-\frac{\mu_{\mathcal{S},*}^o(X)}
				{\kappa_{\mathcal{S},*}^o(X)}
				\left[
				\kappa_{\mathcal{S},*}^o(X)+\{A-\pi_*^o(Z,X)\}^2
				-\{A-\delta_{\mathcal{S}}^o(Z,X)\}^2
				\right]\\
				={}&
				-\frac{\{\pi_*^o(Z,X)-\delta_{\mathcal{S}}^o(Z,X)\}
					\eta_{\mathcal{S}}^o(Z,X)}
				{\kappa_{\mathcal{S},*}^o(X)}\\
				&+\frac{\mu_{\mathcal{S},*}^o(X)}
				{\kappa_{\mathcal{S},*}^o(X)}
				\left[
				\kappa_{\mathcal{S},*}^o(X)
				-\{\pi_*^o(Z,X)-\delta_{\mathcal{S}}^o(Z,X)\}
				\{A-\delta_{\mathcal{S}}^o(Z,X)\}
				\right]\\
				&-\frac{1}{\kappa_{\mathcal{S},*}^o(X)}
				\left[
				\{A-\pi_*^o(Z,X)\}\xi_*^o(Z,X)
				-\{A-\delta_{\mathcal{S}}^o(Z,X)\}\eta_{\mathcal{S}}^o(Z,X)
				\right]\\
				&-\frac{\mu_{\mathcal{S},*}^o(X)}
				{\kappa_{\mathcal{S},*}^o(X)}
				\left[
				\kappa_{\mathcal{S},*}^o(X)+\{A-\pi_*^o(Z,X)\}^2
				-\{A-\delta_{\mathcal{S}}^o(Z,X)\}^2
				\right]\\
				={}&
				\frac{\mu_{\mathcal{S},*}^o(X)}
				{\kappa_{\mathcal{S},*}^o(X)}
				\left[
				\{A-\delta_{\mathcal{S}}^o(Z,X)\}^2
				-\{A-\pi_*^o(Z,X)\}^2
				-\{\pi_*^o(Z,X)-\delta_{\mathcal{S}}^o(Z,X)\}
				\{A-\delta_{\mathcal{S}}^o(Z,X)\}
				\right]\\
				&-\frac{A-\pi_*^o(Z,X)}
				{\kappa_{\mathcal{S},*}^o(X)}
				\{\xi_*^o(Z,X)-\eta_{\mathcal{S}}^o(Z,X)\}\\
				={}&
				\frac{\mu_{\mathcal{S},*}^o(X)}
				{\kappa_{\mathcal{S},*}^o(X)}
				\{\pi_*^o(Z,X)-\delta_{\mathcal{S}}^o(Z,X)\}
				\{A-\pi_*^o(Z,X)\}\\
				&-\frac{A-\pi_*^o(Z,X)}
				{\kappa_{\mathcal{S},*}^o(X)}
				\{\xi_*^o(Z,X)-\eta_{\mathcal{S}}^o(Z,X)\}\\
				={}&\frac{A-\pi_*^o(Z,X)}{\kappa_{\mathcal{S},*}^o(X)}
				\left[
				\{\pi_*^o(Z,X)-\delta_{\mathcal{S}}^o(Z,X)\}
				\mu_{\mathcal{S},*}^o(X)
				-\xi_*^o(Z,X)+\eta_{\mathcal{S}}^o(Z,X)
				\right].
			\end{align*}
			By Theorem~\ref{thm: identification outcome bridge}(c), $\mu_{\mathcal{S},*}^o(X)$ solves Equation~\eqref{eq: outcome bridge}. Hence,
			\begin{align*}
				s^o(Z,X)
				&:=\mathbb{E}\left[Y-A\mu_{\mathcal{S},*}^o(X)\mid Z,X\right]
				=\xi_*^o(Z,X)-\pi_*^o(Z,X)\mu_{\mathcal{S},*}^o(X)
				\in\mathcal{S}.
			\end{align*}
			Since $s^o\in\mathcal{S}$, Proposition~\ref{prop: proper Hilbert subspace}(a)--(b) gives
			\begin{align*}
				s^o(Z,X)
				&=\mathbb{S}\left[Y-A\mu_{\mathcal{S},*}^o(X)\right](Z,X)
				=\eta_{\mathcal{S}}^o(Z,X)
				-\delta_{\mathcal{S}}^o(Z,X)\mu_{\mathcal{S},*}^o(X).
			\end{align*}
			Therefore,
			\[
			\{\pi_*^o(Z,X)-\delta_{\mathcal{S}}^o(Z,X)\}\mu_{\mathcal{S},*}^o(X)
			-\xi_*^o(Z,X)+\eta_{\mathcal{S}}^o(Z,X)=0,
			\]
			which implies
			\[
			\varphi_{\mathcal{S},\pi_*^o}
			(O;\psi_{\mathcal{S},\pi_*^o}^o,\alpha_{\mathcal{S},\pi_*^o}^o)
			=
			\varphi_{\mathcal{S},*}
			(O;\psi_{\mathcal{S},*}^o,\alpha_{\mathcal{S},*}^o)
			\]
			almost surely. Taking second moments on both sides completes the proof of Proposition~\ref{prop: adaptive estimator achieve lower efficiency bound}.
			
			\subsection{Proof of Proposition~\ref{prop: mix bias property}}
			\begin{proof}[Proof of the first equality:]
				We first calculate the conditional expectation of $\varphi_{\mathcal{S},\pi}(O;\psi_{\mathcal{S},\pi}^o,\alpha_{\mathcal{S},\pi})$ as 
				\begin{align}
					&\mathbb{E}[\varphi_{\mathcal{S},\pi}(O;\psi_{\mathcal{S},\pi}^o,\alpha_{\mathcal{S},\pi})\mid X]\notag\\
					=&\dfrac{1}{\kappa_{\mathcal{S},\pi}(X)}
					\mathbb{E}\left[
					\{\pi(Z,X) - \rho_{\mathcal{S},\pi}(Z,X)\}
					\left\{Y - \eta_{\mathcal{S}}(Z,X)\right\}
					\middle|X\right]-\psi_{\mathcal{S},\pi}^o\notag\\
					&+\mu_{\mathcal{S},\pi}(X)\left(1 - \dfrac{1}{\kappa_{\mathcal{S},\pi}(X)}
					\mathbb{E}\left[\left\{\pi(Z,X) -\rho_{\mathcal{S},\pi}(Z,X)\right\}\left\{A-\delta_{\mathcal{S}}(Z,X)\right\}\middle| X\right]\right)\notag\\
					=&\dfrac{1}{\kappa_{\mathcal{S},\pi}(X)}
					\mathbb{E}\left[
					\{\pi(Z,X) - \rho_{\mathcal{S},\pi}^o(Z,X)\}
					\left\{Y - \eta_{\mathcal{S}}^o(Z,X)\right\}
					\middle|X\right]-\psi_{\mathcal{S},\pi}^o\notag\\
					&+\dfrac{1}{\kappa_{\mathcal{S},\pi}(X)}
					\mathbb{E}\left[
					\{\pi(Z,X) - \rho_{\mathcal{S},\pi}^o(Z,X)\}
					\left\{\eta_{\mathcal{S}}^o(Z,X) - \eta_{\mathcal{S}}(Z,X)\right\}
					\middle|X\right]\notag\\
					&+\dfrac{1}{\kappa_{\mathcal{S},\pi}(X)}
					\mathbb{E}\left[
					\{\rho_{\mathcal{S},\pi}^o(Z,X)-\rho_{\mathcal{S},\pi}(Z,X)\}
					\left\{Y - \eta_{\mathcal{S}}^o(Z,X)\right\}
					\middle|X\right]\notag\\
					&+\dfrac{1}{\kappa_{\mathcal{S},\pi}(X)}
					\mathbb{E}\left[
					\{\rho_{\mathcal{S},\pi}^o(Z,X)-\rho_{\mathcal{S},\pi}(Z,X)\}
					\left\{\eta_{\mathcal{S}}^o(Z,X)-\eta_{\mathcal{S}}(Z,X)\right\}
					\middle|X\right]\notag\\
					&+\mu_{\mathcal{S},\pi}(X)\left(1 - \dfrac{1}{\kappa_{\mathcal{S},\pi}(X)}
					\mathbb{E}\left[\left\{\pi(Z,X) -\rho_{\mathcal{S},\pi}(Z,X)\right\}\left\{A-\delta_{\mathcal{S}}(Z,X)\right\}\middle| X\right]\right)\notag\\
					=&\dfrac{1}{\kappa_{\mathcal{S},\pi}(X)}
					\mathbb{E}\left[
					\{\pi(Z,X) - \rho_{\mathcal{S},\pi}^o(Z,X)\}
					\left\{Y - \eta_{\mathcal{S}}^o(Z,X)\right\}
					\middle|X\right]-\psi_{\mathcal{S},\pi}^o\notag\\
					&+\dfrac{1}{\kappa_{\mathcal{S},\pi}(X)}
					\mathbb{E}\left[
					\{\rho_{\mathcal{S},\pi}^o(Z,X)-\rho_{\mathcal{S},\pi}(Z,X)\}
					\left\{\eta_{\mathcal{S}}^o(Z,X)-\eta_{\mathcal{S}}(Z,X)\right\}
					\middle|X\right]\notag\\
					&+\mu_{\mathcal{S},\pi}(X)\left(1 - \dfrac{1}{\kappa_{\mathcal{S},\pi}(X)}
					\mathbb{E}\left[\left\{\pi(Z,X) -\rho_{\mathcal{S},\pi}(Z,X)\right\}\left\{A-\delta_{\mathcal{S}}(Z,X)\right\}\middle| X\right]\right)\label{exp: 5}\\
					=&\dfrac{\kappa_{\mathcal{S},\pi}^o(X)\mu_{\mathcal{S},\pi}^o(X)}{\kappa_{\mathcal{S},\pi}(X)}-\psi_{\mathcal{S},\pi}^o
					+\dfrac{1}{\kappa_{\mathcal{S},\pi}(X)}
					\mathbb{E}\left[
					\{\rho_{\mathcal{S},\pi}^o(Z,X)-\rho_{\mathcal{S},\pi}(Z,X)\}
					\left\{\eta_{\mathcal{S}}^o(Z,X)-\eta_{\mathcal{S}}(Z,X)\right\}
					\middle|X\right]\notag\\
					&+\mu_{\mathcal{S},\pi}(X)\left(1 - \dfrac{1}{\kappa_{\mathcal{S},\pi}(X)}
					\mathbb{E}\left[\left\{\pi(Z,X) -\rho_{\mathcal{S},\pi}(Z,X)\right\}\left\{A-\delta_{\mathcal{S}}(Z,X)\right\}\middle| X\right]\right)\notag\\
					=&\dfrac{\kappa_{\mathcal{S},\pi}^o(X)\mu_{\mathcal{S},\pi}^o(X)}{\kappa_{\mathcal{S},\pi}(X)}-\psi_{\mathcal{S},\pi}^o
					+\dfrac{1}{\kappa_{\mathcal{S},\pi}(X)}
					\mathbb{E}\left[
					\{\rho_{\mathcal{S},\pi}^o(Z,X)-\rho_{\mathcal{S},\pi}(Z,X)\}
					\left\{\eta_{\mathcal{S}}^o(Z,X)-\eta_{\mathcal{S}}(Z,X)\right\}
					\middle|X\right]\notag\\
					&+\mu_{\mathcal{S},\pi}(X) - \dfrac{\mu_{\mathcal{S},\pi}(X)}{\kappa_{\mathcal{S},\pi}(X)}
					\left\{\begin{array}{l}
						\mathbb{E}\left[\left\{\pi(Z,X) -\rho_{\mathcal{S},\pi}^o(Z,X)\right\}\left\{A-\delta_{\mathcal{S}}^o(Z,X)\right\}\middle| X\right]\\
						+\mathbb{E}\left[\left\{ \rho_{\mathcal{S},\pi}^o(Z,X)-\rho_{\mathcal{S},\pi}(Z,X)\right\}\left\{A-\delta_{\mathcal{S}}^o(Z,X)\right\}\middle| X\right]\\
						+\mathbb{E}\left[\left\{\rho_{\mathcal{S},\pi}^o(Z,X) -\rho_{\mathcal{S},\pi}(Z,X)\right\}\left\{\delta_{\mathcal{S}}^o(Z,X)-\delta_{\mathcal{S}}(Z,X)\right\}\middle| X\right]\\
						+\mathbb{E}\left[\left\{\pi(Z,X) -\rho_{\mathcal{S},\pi}^o(Z,X)\right\}\left\{\delta_{\mathcal{S}}^o(Z,X)-\delta_{\mathcal{S}}(Z,X)\right\}\middle| X\right]\\
					\end{array}
					\right\}\notag\\
					=&\dfrac{\kappa_{\mathcal{S},\pi}^o(X)\mu_{\mathcal{S},\pi}^o(X)}{\kappa_{\mathcal{S},\pi}(X)}-\psi_{\mathcal{S},\pi}^o
					+\dfrac{1}{\kappa_{\mathcal{S},\pi}(X)}
					\mathbb{E}\left[
					\{\rho_{\mathcal{S},\pi}^o(Z,X)-\rho_{\mathcal{S},\pi}(Z,X)\}
					\left\{\eta_{\mathcal{S}}^o(Z,X)-\eta_{\mathcal{S}}(Z,X)\right\}
					\middle|X\right]\notag\\
					&+\mu_{\mathcal{S},\pi}(X) - \dfrac{\mu_{\mathcal{S},\pi}(X)}{\kappa_{\mathcal{S},\pi}(X)}
					\left\{\begin{array}{l}
						\mathbb{E}\left[\left\{\pi(Z,X) -\rho_{\mathcal{S},\pi}^o(Z,X)\right\}\left\{A-\delta_{\mathcal{S}}^o(Z,X)\right\}\middle| X\right]\\
						+\mathbb{E}\left[\left\{\rho_{\mathcal{S},\pi}^o(Z,X) -\rho_{\mathcal{S},\pi}(Z,X)\right\}\left\{\delta_{\mathcal{S}}^o(Z,X)-\delta_{\mathcal{S}}(Z,X)\right\}\middle| X\right]
					\end{array}
					\right\}\label{exp: 6}\\
					=&\dfrac{\kappa_{\mathcal{S},\pi}^o(X)\mu_{\mathcal{S},\pi}^o(X)}{\kappa_{\mathcal{S},\pi}(X)}-\psi_{\mathcal{S},\pi}^o+\mu_{\mathcal{S},\pi}(X)\left\{1 - \dfrac{\kappa_{\mathcal{S},\pi}^o(X)}{\kappa_{\mathcal{S},\pi}(X)}\right\}\notag\\
					&+\dfrac{1}{\kappa_{\mathcal{S},\pi}(X)}
					\mathbb{E}\left[
					\{\rho_{\mathcal{S},\pi}^o(Z,X)-\rho_{\mathcal{S},\pi}(Z,X)\}
					\left\{\eta_{\mathcal{S}}^o(Z,X)-\eta_{\mathcal{S}}(Z,X)\right\}
					\middle|X\right]\notag\\
					&-\dfrac{\mu_{\mathcal{S},\pi}(X)}{\kappa_{\mathcal{S},\pi}(X)}
					\mathbb{E}\left[\left\{\rho_{\mathcal{S},\pi}^o(Z,X) -\rho_{\mathcal{S},\pi}(Z,X)\right\}\left\{\delta_{\mathcal{S}}^o(Z,X)-\delta_{\mathcal{S}}(Z,X)\right\}\middle| X\right].\notag
				\end{align}
				Equation~\eqref{exp: 5} follows since $\pi(Z,X)-\rho_{\mathcal{S},\pi}^o(Z,X)\in\mathcal{S}^{\perp}$ and $\eta_{\mathcal{S}}^o(Z,X)-\eta_{\mathcal{S}}(Z,X)\in\mathcal{S}$ are orthogonal. Similarly, $\rho_{\mathcal{S},\pi}^o(Z,X)-\rho_{\mathcal{S},\pi}(Z,X)\in\mathcal{S}$ and $Y-\eta_{\mathcal{S}}^o(Z,X)$ is orthogonal to $\mathcal{S}$. Equation~\eqref{exp: 6} follows since $\rho_{\mathcal{S},\pi}^o(Z,X)-\rho_{\mathcal{S},\pi}(Z,X)\in\mathcal{S}$ and $A-\delta_{\mathcal{S}}^o(Z,X)$ is orthogonal to $\mathcal{S}$. Similarly, $\pi(Z,X)-\rho_{\mathcal{S},\pi}^o(Z,X)\in\mathcal{S}^{\perp}$ and $\delta_{\mathcal{S}}^o(Z,X)-\delta_{\mathcal{S}}(Z,X)\in\mathcal{S}$ are orthogonal.
				
				Consequently, we can integrate $X$ out and deduce that
				\begin{align*}
					&\mathbb{E}[\varphi_{\mathcal{S},\pi}(O;\psi_{\mathcal{S},\pi}^o,\alpha_{\mathcal{S},\pi})]\\
					=&\mathbb{E}\left[\dfrac{\kappa_{\mathcal{S},\pi}^o(X)\mu_{\mathcal{S},\pi}^o(X)}{\kappa_{\mathcal{S},\pi}(X)}-\psi_{\mathcal{S},\pi}^o+\mu_{\mathcal{S},\pi}(X)\left\{1 - \dfrac{\kappa_{\mathcal{S},\pi}^o(X)}{\kappa_{\mathcal{S},\pi}(X)}\right\}\right]\\
					&+
					\mathbb{E}\left[\dfrac{1}{\kappa_{\mathcal{S},\pi}(X)}
					\{\rho_{\mathcal{S},\pi}^o(Z,X)-\rho_{\mathcal{S},\pi}(Z,X)\}
					\left\{\eta_{\mathcal{S}}^o(Z,X)-\eta_{\mathcal{S}}(Z,X)\right\}\right]\\
					&-\mathbb{E}\left[
					\dfrac{\mu_{\mathcal{S},\pi}(X)}{\kappa_{\mathcal{S},\pi}(X)}
					\left\{\rho_{\mathcal{S},\pi}^o(Z,X) -\rho_{\mathcal{S},\pi}(Z,X)\right\}\left\{\delta_{\mathcal{S}}^o(Z,X)-\delta_{\mathcal{S}}(Z,X)\right\}\right]\\
					=&\mathbb{E}\left[\dfrac{1}{\kappa_{\mathcal{S},\pi}(X)}\left\{\mu_{\mathcal{S},\pi}(X)-\mu_{\mathcal{S},\pi}^o(X)\right\}\left\{\kappa_{\mathcal{S},\pi}(X) - \kappa_{\mathcal{S},\pi}^o(X)\right\}\right]\\
					&+
					\mathbb{E}\left[\dfrac{1}{\kappa_{\mathcal{S},\pi}(X)}
					\{\rho_{\mathcal{S},\pi}^o(Z,X)-\rho_{\mathcal{S},\pi}(Z,X)\}
					\left\{\eta_{\mathcal{S}}^o(Z,X)-\eta_{\mathcal{S}}(Z,X)\right\}\right]\\
					&-\mathbb{E}\left[
					\dfrac{\mu_{\mathcal{S},\pi}(X)}{\kappa_{\mathcal{S},\pi}(X)}
					\left\{\rho_{\mathcal{S},\pi}^o(Z,X) -\rho_{\mathcal{S},\pi}(Z,X)\right\}\left\{\delta_{\mathcal{S}}^o(Z,X)-\delta_{\mathcal{S}}(Z,X)\right\}\right].
				\end{align*}
				This completes the proof of the first statement of Proposition~\ref{prop: mix bias property}.	
			\end{proof}
			
			\begin{proof}[Proof of the second equality:]
				\begin{align*}
					&\mathbb{E}\left[\varphi_{\mathcal{S},*}(O;\psi_{\mathcal{S},*}^o,\alpha_{\mathcal{S},*})\middle| X\right]\\
					=&\dfrac{1}{\kappa_{\mathcal{S},*}(X)}
					\mathbb{E}\left[\{\pi_*(Z,X) - \delta_{\mathcal{S}}(Z,X)\}Y\middle|X\right]
					-\psi_{\mathcal{S},*}^o\\
					&+\frac{1}{\kappa_{\mathcal{S},*}(X)}\left\{\mathbb{E}\left[(A-\pi_*(Z,X))\xi_{*}(Z,X) - (A-\delta_{\mathcal{S}}(Z,X))\eta_{\mathcal{S}}(Z,X)\middle|X\right]
					\right\}\\
					&+ \dfrac{\mu_{\mathcal{S},*}(X)}{\kappa_{\mathcal{S},*}(X)}\left\{\kappa_{\mathcal{S},*}(X) + \mathbb{E}\left[(A - \pi_*(Z,X))^2 - (A - \delta_{\mathcal{S}}(Z,X))^2\middle| X\right]  \right\}\\
					=&\dfrac{1}{\kappa_{\mathcal{S},*}(X)}
					\mathbb{E}\left[\{\pi_*^o(Z,X) - \delta_{\mathcal{S}}^o(Z,X)\}Y\middle|X\right]-\psi_{\mathcal{S},*}^o\\
					&+\frac{1}{\kappa_{\mathcal{S},*}(X)}\left\{\begin{array}{l}
						\mathbb{E}\left[\{\pi_*(Z,X) - \pi_*^o(Z,X)\}\mathbb{E}[Y|Z,X]\middle|X\right]\\
						-\mathbb{E}\left[\{\delta_{\mathcal{S}}(Z,X) - \delta_{\mathcal{S}}^o(Z,X)\}Y\middle|X\right]\\
						+\mathbb{E}\left[(A-\pi_*^o(Z,X))\xi_{*}(Z,X) - (A-\delta_{\mathcal{S}}^o(Z,X))\eta_{\mathcal{S}}(Z,X)\middle|X\right]\\
						-\mathbb{E}\left[(\pi_*(Z,X)-\pi_*^o(Z,X))\xi_{*}^o(Z,X) - (\delta_{\mathcal{S}}(Z,X)-\delta_{\mathcal{S}}^o(Z,X))\eta_{\mathcal{S}}^o(Z,X)\middle|X\right]\\
						-\mathbb{E}\left[(\pi_*(Z,X)-\pi_*^o(Z,X))(\xi_{*}(Z,X)-\xi_{*}^o(Z,X))\middle|X\right] \\
						+\mathbb{E}\left[ (\delta_{\mathcal{S}}(Z,X)-\delta_{\mathcal{S}}^o(Z,X))(\eta_{\mathcal{S}}(Z,X)-\eta_{\mathcal{S}}^o(Z,X))\middle|X\right]
					\end{array}\right\}\\
					&+ \dfrac{\mu_{\mathcal{S},*}(X)}{\kappa_{\mathcal{S},*}(X)}\left\{\begin{array}{l}
						\kappa_{\mathcal{S},*}(X) + \mathbb{E}\left[(A - \pi_*^o(Z,X))^2 - (A - \delta_{\mathcal{S}}^o(Z,X))^2\middle| X\right]  \\
						+\mathbb{E}\left[(\pi_*(Z,X) - \pi_*^o(Z,X))^2 - (\delta_{\mathcal{S}}(Z,X) - \delta_{\mathcal{S}}^o(Z,X))^2\middle| X\right]
					\end{array}\right\}\\
					=&\dfrac{\kappa_{\mathcal{S},*}^o(X)}{\kappa_{\mathcal{S},*}(X)}
					\mu_{\mathcal{S},*}^o(X)-\psi_{\mathcal{S},*}^o\\
					&+\frac{1}{\kappa_{\mathcal{S},*}(X)}\left\{\begin{array}{l}
						-\mathbb{E}\left[\{\delta_{\mathcal{S}}(Z,X) - \delta_{\mathcal{S}}^o(Z,X)\}\left\{Y -\eta_{\mathcal{S}}^o(Z,X)\right\} \middle|X\right]\\
						-\mathbb{E}\left[(\pi_*(Z,X)-\pi_*^o(Z,X))(\xi_{*}(Z,X)-\xi_{*}^o(Z,X))\middle|X\right] \\
						+\mathbb{E}\left[ (\delta_{\mathcal{S}}(Z,X)-\delta_{\mathcal{S}}^o(Z,X))(\eta_{\mathcal{S}}(Z,X)-\eta_{\mathcal{S}}^o(Z,X))\middle|X\right]
					\end{array}\right\}\\
					&+ \dfrac{\mu_{\mathcal{S},*}(X)}{\kappa_{\mathcal{S},*}(X)}\left\{
					\kappa_{\mathcal{S},*}(X) - \kappa_{\mathcal{S},*}^o(X) 
					+\mathbb{E}\left[(\pi_*(Z,X) - \pi_*^o(Z,X))^2 - (\delta_{\mathcal{S}}(Z,X) - \delta_{\mathcal{S}}^o(Z,X))^2\middle| X\right]\right\}\\
					=&\frac{1}{\kappa_{\mathcal{S},*}(X)}\left\{\begin{array}{l}
						\left\{\mu_{\mathcal{S},*}(X) - \mu_{\mathcal{S},*}^o(X)\right\}
						\left\{\kappa_{\mathcal{S},*}(X) - \kappa_{\mathcal{S},*}^o(X)\right\}\\
						-\mathbb{E}\left[(\pi_*(Z,X)-\pi_*^o(Z,X))(\xi_{*}(Z,X)-\xi_{*}^o(Z,X))\middle|X\right] \\
						+\mathbb{E}\left[ (\delta_{\mathcal{S}}(Z,X)-\delta_{\mathcal{S}}^o(Z,X))(\eta_{\mathcal{S}}(Z,X)-\eta_{\mathcal{S}}^o(Z,X))\middle|X\right]\\
						+\{\mathbb{E}\left[(\pi_*(Z,X) - \pi_*^o(Z,X))^2 - (\delta_{\mathcal{S}}(Z,X) - \delta_{\mathcal{S}}^o(Z,X))^2\middle| X\right]\}\mu_{\mathcal{S},*}(X)
					\end{array}\right\}\\&
					+\mu_{\mathcal{S},*}^o(X)-\psi_{\mathcal{S},*}^o.
				\end{align*}
				Now we can take an integral with respect to $X$ and deduce that
				\begin{align*}
					&\mathbb{E}\left[\varphi_{\mathcal{S},*}(O;\psi_{\mathcal{S},*}^o,\alpha_{\mathcal{S},*})\right]\\=&\mathbb{E}\left[
					\frac{1}{\kappa_{\mathcal{S},*}(X)}\left\{\begin{array}{l}
						\left\{\mu_{\mathcal{S},*}(X) - \mu_{\mathcal{S},*}^o(X)\right\}
						\left\{\kappa_{\mathcal{S},*}(X) - \kappa_{\mathcal{S},*}^o(X)\right\}\\
						-(\pi_*(Z,X)-\pi_*^o(Z,X))(\xi_{*}(Z,X)-\xi_{*}^o(Z,X)) \\
						+ (\delta_{\mathcal{S}}(Z,X)-\delta_{\mathcal{S}}^o(Z,X))(\eta_{\mathcal{S}}(Z,X)-\eta_{\mathcal{S}}^o(Z,X))\\
						+\{(\pi_*(Z,X) - \pi_*^o(Z,X))^2 - (\delta_{\mathcal{S}}(Z,X) - \delta_{\mathcal{S}}^o(Z,X))^2\}\mu_{\mathcal{S},*}(X)
					\end{array}\right\}
					\right].
				\end{align*}
				This completes the proof of the second statement of Proposition~\ref{prop: mix bias property}.
			\end{proof}

			\subsection{Proof of Proposition~\ref{prop: invalid IV}}
			Under Assumptions~\ref{as: IV independence}--\ref{as: alternative}, the conditions required in Lemma~\ref{lem: connections bewteen PIV and invalid IV}(a) and (b) are satisfied for $\nu_0$. Therefore, by Lemma~\ref{lem: connections bewteen PIV and invalid IV}(a), Definition~\ref{defn: projected IV}(a) is satisfied for all $\gamma\leq |\nu_0|$. By Lemma~\ref{lem: connections bewteen PIV and invalid IV}(b), Definition~\ref{defn: projected IV}(b) is satisfied for all $\gamma\leq \gamma_{\dagger}(\nu_0)$. As a result, $Z$ is an $\mathcal{S}_{\gamma}$-PIV for all $\gamma\leq \min\{|\nu_0|,\gamma_{\dagger}(\nu_0)\}=\gamma_{\dagger}(\nu_0)$, since $\gamma_{\dagger}(\nu_0)\leq |\nu_0|$ by the definition of $\gamma_{\dagger}(\nu_0)$.
			This completes the proof of Proposition~\ref{prop: invalid IV}(a).
			
			If Assumptions~\ref{as: IV independence}, \ref{as: weak latent ignorability}, \ref{as: Z-invariant treatment effect}, and \ref{as: alternative} hold for $\nu_0$, Lemma~\ref{lem: connections bewteen PIV and invalid IV}(b) implies that Definition~\ref{defn: projected IV}(b) holds for $\mathcal{S}_{\gamma}$ with all $\gamma\leq \gamma_\dagger(\nu_0)$. Combining with the fact that Definition~\ref{defn: projected IV}(a) holds for $\gamma_*$, we know that $Z$ is an $\mathcal{S}_{\gamma}$-PIV for all $\gamma\leq \min\{\gamma_*,\gamma_\dagger(\nu_0)\}$.
			This completes the proof of Proposition~\ref{prop: invalid IV}(b).
			
			\subsection{Proof of Proposition~\ref{prop: IVDID}}
			When $Z_{[1]},Z_{[2]}\in\{0,1\}$, it is straightforward to verify that
			\begin{align*}
				&\mathcal{S}_1=\left\{g_0(X)+Z_{[1]}g_1(X)+Z_{[2]}g_2(X):g_0,g_1,g_2\in L_2(X)\right\},\\
				&L_2(Z,X)=\left\{g_0(X)+g_1(X)Z_{[1]}+g_2(X)Z_{[2]}+g_3(X)Z_{[1]}Z_{[2]}:g_0,g_1,g_2,g_3\in L_2(X)\right\}.
			\end{align*} 
			Thus, conditional on \(X\), the orthogonal complement \(\mathcal{S}_1^\perp\) contains a single interaction direction. More precisely, there exists a nonzero generator \(\pi_\dagger\in\mathcal{S}_1^\perp\) such that $\mathcal{S}_1^{\perp}=\overline{\{\pi_\dagger(Z,X)g(X),g\in L_\infty(X)\}}$,
			where the closure is taken in \(L_2(Z,X)\). It therefore suffices to construct one such generator \(\pi_\dagger\). Before giving its explicit form, we establish a general characterization of \(\mathcal{S}_\gamma^\perp\).
			\begin{lemma}\label{lem: orthogonal complement}
				The orthogonal complement of $\mathcal{S}_{\gamma}$ equals
				$$\left\{d \in L_2(Z,X):\mathbb E\{d(Z,X)\mid Z_{[-\nu]}, X\} = 0,\;\text{for all } |\nu|\geq\gamma\right\}.$$
			\end{lemma}
			Now, we verify that 
			\begin{align}
				\pi_\dagger(Z,X) =& (Z_{[1]}-\mathbb{E}[Z_{[1]}| X])(Z_{[2]}-\mathbb{E}[Z_{[2]}| X])\dfrac{p_{Z_{[1]}|X}(Z_{[1]}|X)p_{Z_{[2]}|X}(Z_{[2]}|X)}{p_{Z_{[1]},Z_{[2]}| X}(Z_{[1]},Z_{[2]}| X)}\notag
				\\=& \dfrac{(Z_{[1]}-\mathbb{E}[Z_{[1]}| X])(2Z_{[2]}-1)}{p_{Z_{[2]}|Z_{[1]},X}(Z_{[2]}|Z_{[1]},X)}(1-\mathbb{E}[Z_{[2]}| X])\mathbb{E}[Z_{[2]}| X]\in \mathcal{S}_1^{\perp}.\label{exp: 10}
			\end{align}
			
			\begin{proof}[Proof of Equation~\eqref{exp: 10}:]
				Recall that $\mathbb{S}_0[O](Z_{[1]},Z_{[2]},X)=\mathbb{E}[O \mid Z_{[1]},Z_{[2]},X]$. We have
				\begin{align*}
					&\mathbb{E}[\pi_\dagger(Z,X)\mid Z_{[1]},X]\\
					=&\int \pi_\dagger(Z_{[1]},z_{[2]},X)p(z_{[2]}\mid Z_{[1]},X)\mathrm d z_{[2]}\\
					=&\int (Z_{[1]}-\mathbb{E}[Z_{[1]}| X])(z_{[2]}-\mathbb{E}[Z_{[2]}| X])\dfrac{p(Z_{[1]}|X)p(z_{[2]}|X)}{p(Z_{[1]},z_{[2]}| X)} p(z_{[2]}\mid Z_{[1]},X)\mathrm d z_{[2]}\\
					=&(Z_{[1]}-\mathbb{E}[Z_{[1]}| X])\int (z_{[2]}-\mathbb{E}[Z_{[2]}| X])\dfrac{p(Z_{[1]}|X)p(z_{[2]}\mid Z_{[1]},X)}{p(Z_{[1]},z_{[2]}| X)} p(z_{[2]}|X)\mathrm d z_{[2]}\\
					=&(Z_{[1]}-\mathbb{E}[Z_{[1]}| X])\int (z_{[2]}-\mathbb{E}[Z_{[2]}| X]) p(z_{[2]}|X)\mathrm d z_{[2]}\\
					=&(Z_{[1]}-\mathbb{E}[Z_{[1]}| X])\{\int z_{[2]} p(z_{[2]}|X)\mathrm d z_{[2]}-\mathbb{E}[Z_{[2]}| X]\}=0.
				\end{align*}
				Similarly, one can verify that $\mathbb{E}[\pi_\dagger(Z,X)\mid Z_{[2]},X]=0$.
				Therefore, by Lemma~\ref{lem: orthogonal complement}, $\pi_\dagger\in\mathcal{S}_1^{\perp}$.    
			\end{proof}
			Then, we prove the proposition by verifying the equality in Proposition~\ref{prop: IVDID} for $\pi=\pi_\dagger$.
			By Equation~\eqref{exp: 10}, we can calculate that 
			\begin{align*}
				&\mathbb{E}[\{\pi_\dagger(Z,X)-\mathbb{S}_1[\pi_\dagger](Z,X)\}Y\mid X]
				=\mathbb{E}[\pi_\dagger(Z,X)Y\mid X]\\
				=&\mathbb{E}\left[\dfrac{(Z_{[1]}-\mathbb{E}[Z_{[1]}| X])(2Z_{[2]}-1)}{p_{Z_{[2]}|Z_{[1]},X}(Z_{[2]}|Z_{[1]},X)}(1-\mathbb{E}[Z_{[2]}| X])\mathbb{E}[Z_{[2]}| X]Y\middle| X\right]\\
				=&(1-\mathbb{E}[Z_{[2]}| X])\mathbb{E}[Z_{[2]}| X]\left\{\begin{array}{l}
					\mathbb{E}\left[\dfrac{(Z_{[1]}-\mathbb{E}[Z_{[1]}| X])I\{Z_{[2]}=1\}}{\Pr(Z_{[2]}=1|Z_{[1]},X)}Y\middle| X\right]\\
					-\mathbb{E}\left[\dfrac{(Z_{[1]}-\mathbb{E}[Z_{[1]}| X])I\{Z_{[2]}=0\}}{\Pr(Z_{[2]}=0|Z_{[1]},X)}Y\middle| X\right]
				\end{array}\right\}\\
				=&(1-\mathbb{E}[Z_{[2]}| X])\mathbb{E}[Z_{[2]}| X]\left\{\begin{array}{l}
					\mathbb{E}\left[\dfrac{I\{Z_{[1]}=1,Z_{[2]}=1\}}{\Pr(Z_{[2]}=1|Z_{[1]}=1,X)}Y\middle| X\right]\\
					-\mathbb{E}\left[\dfrac{I\{Z_{[1]}=1,Z_{[2]}=0\}}{\Pr(Z_{[2]}=0|Z_{[1]}=1,X)}Y\middle| X\right]
				\end{array}\right\}\\
				&-(1-\mathbb{E}[Z_{[2]}| X])\mathbb{E}[Z_{[2]}| X]\mathbb{E}[Z_{[1]}| X]\left\{\begin{array}{l}
					\mathbb{E}\left[\dfrac{I\{Z_{[2]}=1\}}{\Pr(Z_{[2]}=1|Z_{[1]},X)}Y\middle| X\right]\\
					-\mathbb{E}\left[\dfrac{I\{Z_{[2]}=0\}}{\Pr(Z_{[2]}=0|Z_{[1]},X)}Y\middle| X\right]
				\end{array}\right\}\\
				=&(1-\mathbb{E}[Z_{[2]}| X])\mathbb{E}[Z_{[2]}| X]\mathbb{E}[Z_{[1]}| X]\left\{\begin{array}{l}
					\mathbb{E}\left[Y\middle| Z_{[1]}=1,Z_{[2]}=1,X\right]\\
					-\mathbb{E}\left[Y\middle| Z_{[1]}=1,Z_{[2]}=0,X\right]
				\end{array}\right\}\\
				&-(1-\mathbb{E}[Z_{[2]}| X])\mathbb{E}[Z_{[2]}| X]\mathbb{E}[Z_{[1]}| X]\left\{\begin{array}{l}
					\mathbb{E}\left[\dfrac{I\{Z_{[2]}=1\}}{\Pr(Z_{[2]}=1|Z_{[1]},X)}Y\middle| Z_{[1]}=1,X\right]\Pr(Z_{[1]}=1|X)\\
					+\mathbb{E}\left[\dfrac{I\{Z_{[2]}=1\}}{\Pr(Z_{[2]}=1|Z_{[1]},X)}Y\middle| Z_{[1]}=0,X\right]\Pr(Z_{[1]}=0|X)\\
					-\mathbb{E}\left[\dfrac{I\{Z_{[2]}=0\}}{\Pr(Z_{[2]}=0|Z_{[1]},X)}Y\middle| Z_{[1]}=1,X\right]\Pr(Z_{[1]}=1|X)\\
					-\mathbb{E}\left[\dfrac{I\{Z_{[2]}=0\}}{\Pr(Z_{[2]}=0|Z_{[1]},X)}Y\middle| Z_{[1]}=0,X\right]\Pr(Z_{[1]}=0|X)
				\end{array}\right\}\\
				=&(1-\mathbb{E}[Z_{[2]}| X])\mathbb{E}[Z_{[2]}| X]\mathbb{E}[Z_{[1]}| X]\left\{\begin{array}{l}
					\mathbb{E}\left[Y\middle| Z_{[1]}=1,Z_{[2]}=1,X\right]\\
					-\mathbb{E}\left[Y\middle| Z_{[1]}=1,Z_{[2]}=0,X\right]
				\end{array}\right\}\\
				&-(1-\mathbb{E}[Z_{[2]}| X])\mathbb{E}[Z_{[2]}| X]\mathbb{E}[Z_{[1]}| X]\left\{\begin{array}{l}
					\mathbb{E}\left[Y\middle| Z_{[1]}=1,Z_{[2]}=1,X\right]\Pr(Z_{[1]}=1|X)\\
					+\mathbb{E}\left[Y\middle| Z_{[1]}=0,Z_{[2]}=1,X\right]\Pr(Z_{[1]}=0|X)\\
					-\mathbb{E}\left[Y\middle| Z_{[1]}=1,Z_{[2]}=0,X\right]\Pr(Z_{[1]}=1|X)\\
					-\mathbb{E}\left[Y\middle| Z_{[1]}=0,Z_{[2]}=0,X\right]\Pr(Z_{[1]}=0|X)
				\end{array}\right\}\\
				=&(1-\mathbb{E}[Z_{[2]}| X])(1-\mathbb{E}[Z_{[1]}| X])\mathbb{E}[Z_{[2]}| X]\mathbb{E}[Z_{[1]}| X]\left\{\begin{array}{l}
					\mathbb{E}\left[Y\middle| Z_{[1]}=1,Z_{[2]}=1,X\right]\\
					-\mathbb{E}\left[Y\middle| Z_{[1]}=0,Z_{[2]}=1,X\right]\\
					-\mathbb{E}\left[Y\middle| Z_{[1]}=1,Z_{[2]}=0,X\right]\\
					+\mathbb{E}\left[Y\middle| Z_{[1]}=0,Z_{[2]}=0,X\right]
				\end{array}\right\}.
			\end{align*}
			Similarly, one can verify that
			\begin{align*}
				&\mathbb{E}[\{\pi_\dagger(Z,X)-\mathbb{S}_1[\pi_\dagger](Z,X)\}A\mid X]
				=(1-\mathbb{E}[Z_{[2]}| X])(1-\mathbb{E}[Z_{[1]}| X])\mathbb{E}[Z_{[2]}| X]\mathbb{E}[Z_{[1]}| X]\\&\times \left\{\begin{array}{l}
					\mathbb{E}\left[A\middle| Z_{[1]}=1,Z_{[2]}=1,X\right]
					-\mathbb{E}\left[A\middle| Z_{[1]}=0,Z_{[2]}=1,X\right]\\
					-\mathbb{E}\left[A\middle| Z_{[1]}=1,Z_{[2]}=0,X\right]
					+\mathbb{E}\left[A\middle| Z_{[1]}=0,Z_{[2]}=0,X\right]
				\end{array}\right\}.
			\end{align*}
			Therefore, we conclude that
			\begin{align*}
				\psi_{\mathcal{S}_1,\pi_\dagger}^o:=&\mathbb{E}\left[\dfrac{\{\pi_\dagger(Z,X)-\mathbb{S}_1[\pi_\dagger](Z,X)\}}{\mathbb{E}[\{\pi_\dagger(Z,X)-\mathbb{S}_1[\pi_\dagger](Z,X)\}A\mid X]}Y\right]\\
				=&\mathbb{E}\left[\dfrac{\left\{\begin{array}{l}
						\mathbb{E}\left[Y\middle| Z_{[1]}=1,Z_{[2]}=1,X\right]
						-\mathbb{E}\left[Y\middle| Z_{[1]}=0,Z_{[2]}=1,X\right]\\
						-\mathbb{E}\left[Y\middle| Z_{[1]}=1,Z_{[2]}=0,X\right]
						+\mathbb{E}\left[Y\middle| Z_{[1]}=0,Z_{[2]}=0,X\right]
					\end{array}\right\}}{\left\{\begin{array}{l}
						\mathbb{E}\left[A\middle| Z_{[1]}=1,Z_{[2]}=1,X\right]
						-\mathbb{E}\left[A\middle| Z_{[1]}=0,Z_{[2]}=1,X\right]\\
						-\mathbb{E}\left[A\middle| Z_{[1]}=1,Z_{[2]}=0,X\right]
						+\mathbb{E}\left[A\middle| Z_{[1]}=0,Z_{[2]}=0,X\right]
					\end{array}\right\}}\right].
			\end{align*}
			This finishes the proof of Proposition~\ref{prop: IVDID} for $\pi_\dagger$.
			The remaining $\mathcal{S}_{1}$-PRWFs all lead to the same quantity, since for all $\beta_0,\ldots,\beta_3\in L_2(X)$, and $\pi_\beta(Z,X)=\beta_0(X) + \beta_1(X)Z_{[1]} + \beta_2(X)Z_{[2]}+ \beta_3(X)Z_{[1]}Z_{[2]}$,
			\begin{align*}
				\mathbb{E}\left[\dfrac{\{\pi_\beta(Z,X)-\mathbb{S}_1[\pi_\beta](Z,X)\}}{\mathbb{E}[\{\pi_\beta(Z,X)-\mathbb{S}_1[\pi_\beta](Z,X)\}A\mid X]}Y\right]=\mathbb{E}\left[\dfrac{\{Z_{[1]}Z_{[2]}-\mathbb{S}_1[Z_{[1]}Z_{[2]}](Z,X)\}}{\mathbb{E}[\{Z_{[1]}Z_{[2]}-\mathbb{S}_1[Z_{[1]}Z_{[2]}](Z,X)\}A\mid X]}Y\right].
			\end{align*}
			This equality follows from $\mathbb{S}_{1}[Z_{[j]}] = Z_{[j]}$ for $j=1,2$. Notably, the right-hand side of the equation above no longer depends on $\beta$.

			\subsection{Proof of Proposition~\ref{prop: uniqueness of largest nu}}
			We prove Proposition~\ref{prop: uniqueness of largest nu} by verifying the following facts in turn:
			\begin{enumerate}[label=(\alph*)]
				\item If Assumption~\ref{as: IV independence} holds for both $\nu_1$ and $\nu_2$, then Assumption~\ref{as: IV independence} holds for $\nu_3:=\nu_1\cup \nu_2$.\label{fact (a)}
				\item Suppose that Assumption~\ref{as: strong latent ignorability} holds for both $\nu_1$ and $\nu_2$, and Assumption~\ref{as: IV independence} holds for $\nu_3=\nu_1\cup \nu_2$. Then, Assumption~\ref{as: strong latent ignorability} holds for $\nu_3$.\label{fact (b)}
				\item If Assumption~\ref{as: weak latent ignorability} (or Assumption~\ref{as: Z-invariant treatment effect}) holds for both $\nu_1$ and $\nu_2$, then Assumption~\ref{as: weak latent ignorability} (or Assumption~\ref{as: Z-invariant treatment effect}) holds for $\nu_3=\nu_1\cup \nu_2$.\label{fact (c)}
				\item If Assumption~\ref{as: alternative}(a) holds for $\nu_1$ at an admissible value of $\gamma$, then it also holds for any $\nu_2\supseteq\nu_1$ at the same value of $\gamma$. \label{fact (d)}
				\item Suppose that Assumption~\ref{as: Z-invariant treatment effect} holds for $\nu_1$ and $\nu_2$ with $\nu_2\supseteq\nu_1$. If Assumption~\ref{as: alternative}(b) holds for $\nu_1$, then Assumption~\ref{as: alternative}(b) also holds for $\nu_2$ with $\gamma_{\dagger}(\nu_2)\geq \gamma_{\dagger}(\nu_1)$. \label{fact (e)}
			\end{enumerate}\vspace{1em}
			
			Combining these results, the first statement in Proposition~\ref{prop: uniqueness of largest nu} is obviously true. To prove the second statement, consider any two regular index sets $\nu_1\subseteq\nu_2$. If Assumption~\ref{as: alternative}(b) holds for $\nu_1$, then Fact~\ref{fact (e)} implies that it also holds for $\nu_2$. Consequently,
			\[
			\gamma_{\dagger}(\nu_1)=|\nu_1|\leq|\nu_2|=\gamma_{\dagger}(\nu_2).
			\]
			Otherwise, Assumption~\ref{as: alternative}(b) fails for $\nu_1$, so the regularity of $\nu_1$ implies that Assumption~\ref{as: alternative}(a) holds for $\nu_1$ at $\gamma=\gamma_{\dagger}(\nu_1)$. By Fact~\ref{fact (d)}, it also holds for $\nu_2$ at the same value of $\gamma$. Hence, $\gamma_{\dagger}(\nu_1)\leq\gamma_{\dagger}(\nu_2)$.
			
			\begin{proof}[Proof of Fact~\ref{fact (a)}:]
				We directly verify that
				\begin{align*}
					&p(Z_{[\nu_3]},U\mid  Z_{[-\nu_3]},X)=p(Z_{[\nu_1]},U\mid Z_{[\nu_3-\nu_1]}, Z_{[-\nu_3]},X)p(Z_{[\nu_3-\nu_1]}\mid Z_{[-\nu_3]},X)\\
					=&p(Z_{[\nu_1]},U\mid Z_{[-\nu_1]},X)p(Z_{[\nu_3-\nu_1]}\mid Z_{[-\nu_3]},X)\\
					=&p(Z_{[\nu_1]}\mid Z_{[-\nu_1]},X)p(U\mid Z_{[-\nu_1]},X)p(Z_{[\nu_3-\nu_1]}\mid Z_{[-\nu_3]},X)\tag{$Z_{[\nu_1]}\indep U\mid Z_{[-\nu_1]},X$}\\
					=&p(Z_{[\nu_1]}\mid Z_{[\nu_3-\nu_1]}, Z_{[-\nu_3]},X)p(U\mid Z_{[-\nu_1]},X)p(Z_{[\nu_3-\nu_1]}\mid Z_{[-\nu_3]},X)\\
					=&p(Z_{[\nu_3]}\mid Z_{[-\nu_3]},X)p(U\mid Z_{[-\nu_1]},X).
				\end{align*}
				Similarly, using the fact that $Z_{[\nu_2]}\indep U\mid Z_{[-\nu_2]},X$, one can verify that 
				$$p(Z_{[\nu_3]},U| Z_{[-\nu_3]}, X)=p(Z_{[\nu_3]}| Z_{[-\nu_3]},X)p(U| Z_{[-\nu_2]}, X).$$ 
				Since $p(Z\mid X)>0$ for all $Z,X$ by Assumption~\ref{as: regular condition}, we know that $p(Z_{[\nu_3]}\mid Z_{[-\nu_3]},X)>0$. Therefore, $p(Z_{[\nu_3]}\mid Z_{[-\nu_3]},X)p(U\mid Z_{[-\nu_1]}, X)=p(Z_{[\nu_3]}\mid Z_{[-\nu_3]},X)p(U\mid Z_{[-\nu_2]}, X)$ implies that
				\begin{align*}
					p(U\mid Z_{[-\nu_1]}, X)=p(U\mid Z_{[-\nu_2]},X)=p(U\mid Z_{[-\nu_3]}, X)\text{ for all $Z$.}
				\end{align*}
				As a result, $p(Z_{[\nu_3]},U\mid Z_{[-\nu_3]},X)=p(Z_{[\nu_3]}\mid Z_{[-\nu_3]}, X)p(U\mid Z_{[-\nu_3]}, X)$. 
			\end{proof}
			\begin{proof}[Proof of Fact~\ref{fact (b)}:]
				Similarly, using the fact that $Y(a)\indep Z_{[\nu_j]}\mid Z_{[-\nu_j]},U,X$ for $j=1,2$, then one can verify that
				\begin{align*}
					&p(Y(a),A,Z_{[\nu_3]}\mid Z_{[-\nu_3]},U,X)=p(A,Z_{[\nu_3]}\mid Z_{[-\nu_3]},U,X)p(Y(a)\mid Z_{[-\nu_1]},U,X)\\
					=&p(A\mid Z,U,X)p(Z_{[\nu_3]}\mid Z_{[-\nu_3]},X)p(Y(a)\mid Z_{[-\nu_1]},U,X).
				\end{align*}
				The final step follows by Assumption~\ref{as: IV independence} that $Z_{[\nu_3]}\indep U\mid Z_{[-\nu_3]},X$. Similarly, one can verify that
				\begin{align*}
					p(Y(a),A,Z_{[\nu_3]}\mid Z_{[-\nu_3]},U,X)=p(A\mid Z,U,X)p(Z_{[\nu_3]}\mid Z_{[-\nu_3]},X)p(Y(a)\mid Z_{[-\nu_2]},U,X).
				\end{align*}
				Therefore, by Assumption~\ref{as: regular condition}, since $p(Z\mid X)> 0$ for all $Z,X$, it holds that
				\begin{align*}
					&p(Y(a)\mid Z_{[-\nu_1]},U,X)p(A\mid Z,U,X)\\
					=&p(Y(a)\mid Z_{[-\nu_2]},U,X)p(A\mid Z,U,X)\\
					=&p(Y(a)\mid Z_{[-\nu_3]},U,X)p(A\mid Z,U,X) \text{ for all $Y(a),Z,U,X$}.
				\end{align*}
				Consequently, $p(Y(a),A,Z_{[\nu_3]}\mid Z_{[-\nu_3]},U,X)=p(Y(a)\mid Z_{[-\nu_3]},U,X)p(A,Z_{[\nu_3]}\mid Z_{[-\nu_3]},U,X).$
			\end{proof}
			
			\begin{proof}[Proof of Fact~\ref{fact (c)}:]
				For $j=1,2$, $\mathbb{E}[Y(1)-Y(0)\mid A,U,Z,X]\perp \{A,Z_{[\nu_j]}\}\mid U,Z_{[-\nu_j]},X$ indicates that $\mathbb{E}[Y(1)-Y(0)\mid A,U,Z,X]$ does not depend on $\{A,Z_{[\nu_1]}\}$ and $\{A,Z_{[\nu_2]}\}$, respectively. Therefore, $\mathbb{E}[Y(1)-Y(0)\mid A,U,Z,X]$ does not depend on $A,Z_{[\nu_3]}$, where $\nu_3=\nu_1\cup\nu_2$. This finishes the proof of Assumption~\ref{as: weak latent ignorability} for $\nu_3$.
				
				The proof for Assumption~\ref{as: Z-invariant treatment effect} can be conducted similarly.
			\end{proof}
			\begin{proof}[Proof of Fact~\ref{fact (d)}]
				In particular, there exist functions
				\(\{b_\nu^{(1)}: \nu\subseteq\nu_1,\ |\nu|=\gamma\}\)
				and \(c_1\) such that
				\[
				\Pr(A=1\mid Z,U,X)=\sum_{\substack{\nu\subseteq\nu_1\\|\nu|=\gamma}}b_\nu^{(1)}(U,Z_{[-\nu]},X)+c_1(Z,X).
				\]
				For every \(\nu\subseteq\nu_2\) with \(|\nu|=\gamma\), define
				\[
				b_\nu^{(2)}(U,Z_{[-\nu]},X)=
				\begin{cases}
					b_\nu^{(1)}(U,Z_{[-\nu]},X), & \nu\subseteq\nu_1,\\
					0,            & \nu\not\subseteq\nu_1.
				\end{cases}
				\]
				It follows that
				\[\Pr(A=1\mid Z,U,X)=\sum_{\substack{\nu\subseteq\nu_2\\|\nu|=\gamma}}b_\nu^{(2)}(U,Z_{[-\nu]},X)+c_1(Z,X).\]
				Hence, Assumption~\ref{as: alternative}(a) holds for \(\nu_2\).
			\end{proof}

			\begin{proof}[Proof of Fact~\ref{fact (e)}:]
				For any $\nu_2\supseteq\nu_1$, we directly verify that
				\begin{align*}
					&\mathbb{E}[Y(1)-Y(0)\mid U,Z_{[-\nu_2]},X]
					=\mathbb{E}[
					\mathbb{E}[Y(1)-Y(0)\mid U,Z_{[-\nu_1]},X]
					\mid U,Z_{[-\nu_2]},X]\\
					=&\mathbb{E}[
					\mathbb{E}[Y(1)-Y(0)\mid Z_{[-\nu_1]},X]
					\mid U,Z_{[-\nu_2]},X]\\
					=&\mathbb{E}[
					\mathbb{E}[Y(1)-Y(0)\mid X]
					\mid U,Z_{[-\nu_2]},X]=\mathbb{E}[Y(1)-Y(0)\mid X].\tag{By Assumption~\ref{as: Z-invariant treatment effect} for $\nu_1$}
				\end{align*}
				Simultaneously, $\mathbb{E}[Y(1)-Y(0)\mid Z_{[-\nu_2]},X]=\mathbb{E}[Y(1)-Y(0)\mid X]$ by Assumption~\ref{as: Z-invariant treatment effect} for $\nu_2$. This completes the proof of $\mathbb{E}[Y(1)-Y(0)\mid U,Z_{[-\nu_2]},X]=\mathbb{E}[Y(1)-Y(0)\mid Z_{[-\nu_2]},X]$.
			\end{proof}
			
			\section{Proof of Theorems}

			\subsection{Proof of Theorem~\ref{thm: identification}}
			We directly verify that
			\begin{align*}
				&\mathbb{E}\left[\left\{\pi(Z,X) - \mathbb{S}[\pi](Z,X)\right\}Y\mid X\right]\\
				=&\mathbb{E}\left[A\left\{\pi(Z,X) - \mathbb{S}[\pi](Z,X)\right\}\{Y(1)-Y(0)\}\mid X\right]\\
				&+\mathbb{E}\left[\left\{\pi(Z,X) - \mathbb{S}[\pi](Z,X)\right\}Y(0)\mid X\right]\tag{By $Y(A)=A\{Y(1)-Y(0)\}+Y(0)$}\\
				=&\mathbb{E}\left[A\left\{\pi(Z,X) - \mathbb{S}[\pi](Z,X)\right\}\mid X\right]
				\times\mathbb{E}\left[Y(1)-Y(0)\mid X\right]\\
				&+\mathbb{E}\left[\left\{\pi(Z,X) - \mathbb{S}[\pi](Z,X)\right\}\mathbb{E}[Y(0)\mid Z,X]\mid X\right]\tag{By Definition~\ref{defn: projected IV}(b)}\\
				=&\mathbb{E}\left[A\left\{\pi(Z,X) - \mathbb{S}[\pi](Z,X)\right\}\mid X\right]
				\times\mathbb{E}\left[Y(1)-Y(0)\mid X\right].
			\end{align*}
			The final step follows from Definition~\ref{defn: projected IV}(a), which states that 
			$\mathbb{E}[Y(0)\mid Z,X]\in\mathcal{S},$
			together with the fact that 
			$\pi - \mathbb{S}[\pi]\in\mathcal{S}^{\perp}$. Consequently,
			\begin{align*}
				\mathbb{E}\left[Y(1)-Y(0)\mid X\right]=&\dfrac{\mathbb{E}\left[\left\{\pi(Z,X) - \mathbb{S}[\pi](Z,X)\right\}Y\mid X\right]}{\mathbb{E}\left[A\left\{\pi(Z,X) - \mathbb{S}[\pi](Z,X)\right\}\mid X\right]}.\\
				\Rightarrow\quad
				\mathbb{E}\left[Y(1)-Y(0)\right]=&\mathbb{E}\left[\dfrac{\left\{\pi(Z,X) - \mathbb{S}[\pi](Z,X)\right\}Y}{\mathbb{E}\left[A\left\{\pi(Z,X) - \mathbb{S}[\pi](Z,X)\right\}\mid X\right]}\right].
			\end{align*}
			This completes the proof of Theorem~\ref{thm: identification}.

			\subsection{Proof of Theorem~\ref{thm: identification treatment bridge}}

			For any function $q(Z,X)$ that satisfies Equation~\eqref{eq: treatment bridge}, if $Z$ is an $\mathcal{S}$-PIV, then $q(Z,X)A = \{q(Z,X)-\mathbb{S}[q](Z,X)\}A$ is conditionally uncorrelated with $Y(1)-Y(0)$ given $X$ by the definition of $\mathcal{S}$-PIV, and 
			\begin{align*}
				&\mathbb{E}\left[q(Z,X)Y\mid X\right]\\
				=&\mathbb{E}\left[q(Z,X)A\{Y(1)-Y(0)\}\mid X\right]
				+\mathbb{E}\left[q(Z,X)\mathbb{E}[Y(0)\mid Z,X]\mid X\right]\\
				=&\mathbb{E}\left[q(Z,X)A\mid X\right]
				\times \mathbb{E}\left[Y(1)-Y(0)\mid X\right]
				\tag{By the definition of $\mathcal{S}$-PIV}\\
				=&\mathbb{E}\left[Y(1)-Y(0)\mid X\right].
			\end{align*}
			By integrating out $X$, we complete the proof of Theorem~\ref{thm: identification treatment bridge}.

			\subsection{Proof of Theorem~\ref{thm: identification outcome bridge}}
			\begin{proof}[Proof of Theorem~\ref{thm: identification outcome bridge}(a):]
				Suppose that $q\in \mathcal{S}^{\perp}$ satisfies Equation~\eqref{eq: treatment bridge}. Then, for any $q_f(Z,X):=f(X)q(Z,X)$ with $f\in L_{\infty}(X)$, $q_f\in\mathcal{S}^{\perp}$, since for any $s\in\mathcal{S}$,
				\begin{align*}
					\mathbb{E}[q_f(Z,X)s(Z,X)]=
					\mathbb{E}[f(X)q(Z,X)s(Z,X)]=\mathbb{E}[f(X)\mathbb{E}[q(Z,X)s(Z,X)\mid X]] = 0.
				\end{align*}
                Let $g\in L_2(X)$ be a function satisfying $\mathbb E[Ad(Z,X)g(X)]=0$ for all $d\in\mathcal{S}^{\perp}$.
				Then, $\mathbb{E}[Aq_f(Z,X)g(X)]=0$ for all $f\in L_{\infty}(X)$ since $q_f\in \mathcal{S}^{\perp}$. Consequently,
				\begin{align*}
					0=\mathbb{E}[Aq_f(Z,X)g(X)]=\mathbb{E}[\mathbb{E}[Aq(Z,X)\mid X]f(X)g(X)]
					=\mathbb{E}[f(X)g(X)],\text{ for all $f\in L_{\infty}(X)$.}
				\end{align*}
				Then $g(X)\equiv 0$, because $L_{\infty}(X)$ is dense in $L_2(X)$. This indicates that the $\mathcal{S}$-completeness condition holds.
			\end{proof}
			
			\begin{proof}[Proof of Theorem~\ref{thm: identification outcome bridge}(b):]
				If there exist two distinct solutions $h_1^o(X)$ and $h_2^o(X)$ to Equation~\eqref{eq: outcome bridge}, then
				\begin{align*}
					&\mathbb E\big[d(Z,X)\{Y - Ah_1^o(X)\}\big] = \mathbb E\big[d(Z,X)\{Y - Ah_2^o(X)\}\big] = 0, \qquad \forall\, d \in \mathcal{S}^{\perp}.\\
					\Longrightarrow\quad&
					\mathbb E\big[Ad(Z,X)\{h_1^o(X) - h_2^o(X)\}\big]=0,\qquad\forall\, d \in \mathcal{S}^{\perp}.
				\end{align*}
				From the $\mathcal{S}$-completeness condition, we know that $h_1^o(X)\equiv h_2^o(X)$. Therefore, the solution to Equation~\eqref{eq: outcome bridge} is unique.
			\end{proof}
			
			\begin{proof}[Proof of Theorem~\ref{thm: identification outcome bridge}(c):]
				We directly show that
				\begin{align*}
					&\mathbb{E}[\{\pi(Z,X)-\mathbb{S}[\pi](Z,X)\}\{Y-A\mathbb{E}[Y(1)-Y(0)\mid X]\}]\\
					=&\mathbb{E}[\{\pi(Z,X)-\mathbb{S}[\pi](Z,X)\}\{A\{Y(1)-Y(0)\}-A\mathbb{E}[Y(1)-Y(0)\mid X]\}]\\
					&+\mathbb{E}[\{\pi(Z,X)-\mathbb{S}[\pi](Z,X)\}Y(0)]\\
					=&\mathbb{E}[A\{\pi(Z,X)-\mathbb{S}[\pi](Z,X)\}\{Y(1)-Y(0)-\mathbb{E}[Y(1)-Y(0)\mid X]\}]\\
					&+\mathbb{E}[\{\pi(Z,X)-\mathbb{S}[\pi](Z,X)\}\mathbb{E}[Y(0)\mid Z,X]]\\
					=&\mathbb{E}[\mathrm{Cov}\{A\{\pi(Z,X)-\mathbb{S}[\pi](Z,X)\},Y(1)-Y(0)\mid X\}] \\&+ 
					\mathbb{E}[\{\pi(Z,X)-\mathbb{S}[\pi](Z,X)\}\mathbb{E}[Y(0)\mid Z,X]]=0.
				\end{align*}
				The final equality follows from \(\mathbb{E}\{Y(0)\mid Z,X\}\in\mathcal{S}\), as stated in Definition~\ref{defn: projected IV}(a), the fact that \(\pi(Z,X)-\mathbb{S}[\pi](Z,X)\in\mathcal{S}^{\perp}\), and Definition~\ref{defn: projected IV}(b), which implies that \[\operatorname{Cov}\!\left[A\{\pi(Z,X)-\mathbb{S}[\pi](Z,X)\},Y(1)-Y(0)\mid X\right]\equiv0.\] Because the collection of residuals \(\pi(Z,X)-\mathbb{S}[\pi](Z,X)\) spans \(\mathcal{S}^{\perp}\), this is equivalent to Equation~\eqref{eq: outcome bridge} with \(h^o(X)=\mathbb{E}\{Y(1)-Y(0)\mid X\}\).
			\end{proof}

			\subsection{Proof of Theorem~\ref{thm: EIF ATE}}
			We parametrize the marginal density $p_O(o)$ by a differentiable parametric path $\{p_O^{(\theta)}(o):\theta\in[-c,c]\}$, for some sufficiently small constant $c>0$, such that $p_O^{(0)}(o)=p_O(o)$. This construction defines a regular parametric submodel along which pathwise derivatives are taken.
			We define the score function $s_{\theta}(o):=\nabla_{\theta}\log\{p_O^{(\theta)}(o)\}\eval_{\theta=0}$, and denote expectations under $p_O^{(\theta)}$ by
			$\mathbb{E}^{\theta}[f(O)]:=\int f(o)p_O^{(\theta)}(o)\mathrm d o$. We further let $\mathbb{S}^{(\theta)}: L_2(O)\to \mathcal{S}$ denote a projection operator defined under the model $p_O^{(\theta)}$:
			\begin{align*}
				\mathbb{S}^{(\theta)}[f]:=\underset{s\in\mathcal{S}}{\arg\min}\mathbb{E}^{(\theta)}[\left\{f(O)-s(Z,X)\right\}^2],
			\end{align*}
			which will be used to characterize pathwise derivatives along this parametric submodel.
			
			Next, we introduce a useful lemma for conducting semiparametric analysis.
			\begin{lemma}\label{lem: semiparametric calculation}
				Let $f\in L_2(O)$ be fixed, and consider a regular parametric path $\{p_O^{(\theta)}(o):\theta\in[-c,c]\}$ along which $\mathcal{S}$ is the same closed linear space and $\mathbb{S}^{(\theta)}[f]$ is differentiable in $L_2$ at zero with derivative in $\mathcal{S}$. Assume that differentiation may be interchanged with expectation and that $(f-\mathbb{S}[f])s_{\theta}\in L_2(O)$. Then
				\begin{align*}
					\nabla_{\theta}\mathbb{S}^{(\theta)}[f](Z,X)\eval_{\theta=0}
					=\mathbb{S}[\{f - \mathbb{S}[f]\}s_{\theta}](Z,X).
				\end{align*}    
				In particular, if we set $\mathcal{S} = L_2(X)$, so that $\mathbb{S}^{(\theta)}[f](Z,X) = \mathbb{E}^{(\theta)}[f(O)\mid X]$, then
				\begin{align*}
					\nabla_{\theta}\mathbb{E}^{(\theta)}[f(O)\mid X]\eval_{\theta=0}
					=\mathbb{E}[\{f(O) - \mathbb{E}[f(O)\mid X]\}s_{\theta}(O)\mid X].
				\end{align*}
			\end{lemma}

			We then define the perturbed functional $\psi_{\mathcal{S},\pi}^{(\theta)}$ of interest as
			\begin{align*}
				\psi_{\mathcal{S},\pi}^{(\theta)}:=\mathbb{E}^{(\theta)}\left[\dfrac{\pi(Z,X) - \mathbb{S}^{(\theta)}[\pi](Z,X)}{\mathbb{E}^{(\theta)}\left[\left\{\pi(Z,X) - \mathbb{S}^{(\theta)}[\pi](Z,X)\right\}A|X\right]}Y\right].
			\end{align*}
			Then by repeatedly applying Lemma~\ref{lem: semiparametric calculation}, we derive that
			\begin{align*}
				&\nabla_{\theta}\psi_{\mathcal{S},\pi}^{(\theta)}\eval_{\theta=0}
				=\nabla_{\theta}\mathbb{E}^{(\theta)}\left[\dfrac{\pi(Z,X) - \mathbb{S}^{(\theta)}[\pi](Z,X)}{\mathbb{E}^{(\theta)}\left[\left\{\pi(Z,X) - \mathbb{S}^{(\theta)}[\pi](Z,X)\right\}A|X\right]}Y\right]\eval_{\theta=0}
				\\
				=&\mathbb{E}\left[\left\{\dfrac{\pi(Z,X) - \mathbb{S}[\pi](Z,X)}{\mathbb{E}\left[\left\{\pi(Z,X) - \mathbb{S}[\pi](Z,X)\right\}A|X\right]}Y-\psi_{\mathcal{S},\pi}^o\right\} s_{\theta}(O)\right]\\
				&-\mathbb{E}\left[\dfrac{ \nabla_{\theta}\mathbb{S}^{(\theta)}[\pi](Z,X)\eval_{\theta=0}}{\mathbb{E}\left[\left\{\pi(Z,X) - \mathbb{S}[\pi](Z,X)\right\}A|X\right]}Y\right]\\
				&-\mathbb{E}\left[ \dfrac{\pi(Z,X) - \mathbb{S}[\pi](Z,X)}
				{\kappa_{\mathcal{S},\pi}^o(X)^2}Y\nabla_{\theta}\mathbb{E}^{(\theta)}\left[\left\{\pi(Z,X) - \mathbb{S}^{(\theta)}[\pi](Z,X)\right\}A|X\right]\eval_{\theta=0}\right]\\
				=&\mathbb{E}\left[\left\{\dfrac{\pi(Z,X) - \mathbb{S}[\pi](Z,X)}{\mathbb{E}\left[\left\{\pi(Z,X) - \mathbb{S}[\pi](Z,X)\right\}A|X\right]}Y-\psi_{\mathcal{S},\pi}^o\right\} s_{\theta}(O)\right]\\
				&-\mathbb{E}\left[\dfrac{ \nabla_{\theta}\mathbb{S}^{(\theta)}[\pi](Z,X)\eval_{\theta=0}}{\mathbb{E}\left[\left\{\pi(Z,X) - \mathbb{S}[\pi](Z,X)\right\}A|X\right]}Y\right]\\
				&-\mathbb{E}\left[ \dfrac{\pi(Z,X) - \mathbb{S}[\pi](Z,X)}
				{\kappa_{\mathcal{S},\pi}^o(X)^2}Y\mathbb{E}\left[(\left\{\pi(Z,X) - \mathbb{S}[\pi](Z,X)\right\}A-\kappa_{\mathcal{S},\pi}^o(X))s_{\theta}(O)|X\right]\right]\\
				&+\mathbb{E}\left[ \dfrac{\pi(Z,X) - \mathbb{S}[\pi](Z,X)}
				{\kappa_{\mathcal{S},\pi}^o(X)^2}Y\mathbb{E}\left[\nabla_{\theta} \mathbb{S}^{(\theta)}[\pi](Z,X)\eval_{\theta=0} A|X\right]\right]\\
				=&\mathbb{E}\left[\left\{\dfrac{\pi(Z,X) - \mathbb{S}[\pi](Z,X)}{\kappa_{\mathcal{S},\pi}^o(X)}Y-\psi_{\mathcal{S},\pi}^o\right\} s_{\theta}(O)\right]\\
				&-\mathbb{E}\left[\dfrac{\mathbb{S}[\{\pi -\mathbb{S}[\pi]\}s_{\theta} ](Z,X)}{\kappa_{\mathcal{S},\pi}^o(X)}Y\right]\\
				&-\mathbb{E}\left[ \dfrac{\pi(Z,X) - \mathbb{S}[\pi](Z,X)}
				{\kappa_{\mathcal{S},\pi}^o(X)^2}Y\mathbb{E}\left[(\left\{\pi(Z,X) - \mathbb{S}[\pi](Z,X)\right\}A-\kappa_{\mathcal{S},\pi}^o(X))s_{\theta}(O)|X\right]\right]\\
				&+\mathbb{E}\left[ \dfrac{\pi(Z,X) - \mathbb{S}[\pi](Z,X)}
				{\kappa_{\mathcal{S},\pi}^o(X)^2}Y\mathbb{E}\left[\mathbb{S}[\{\pi -\mathbb{S}[\pi]\}s_{\theta} ](Z,X)A|X\right]\right]\\
				=&\mathbb{E}\left[\left\{\dfrac{\pi(Z,X) - \mathbb{S}[\pi](Z,X)}{\kappa_{\mathcal{S},\pi}^o(X)}Y-\psi_{\mathcal{S},\pi}^o\right\} s_{\theta}(O)\right]\\
				&-\mathbb{E}\left[\dfrac{\{\pi -\mathbb{S}[\pi]\}s_{\theta}(O)}{\kappa_{\mathcal{S},\pi}^o(X)}\mathbb{S}[Y](Z,X)\right]\\
				&-\mathbb{E}\left[ 
				\mathbb{E}\left[\dfrac{\pi(Z,X) - \mathbb{S}[\pi](Z,X)}
				{\kappa_{\mathcal{S},\pi}^o(X)^2}Y\middle| X\right]
				(\left\{\pi(Z,X) - \mathbb{S}[\pi](Z,X)\right\}A-\kappa_{\mathcal{S},\pi}^o(X))s_{\theta}(O)\right]\\
				&+\mathbb{E}\left[ \mathbb{E}\left[\dfrac{\pi(Z,X) - \mathbb{S}[\pi](Z,X)}
				{\kappa_{\mathcal{S},\pi}^o(X)^2}Y\middle| X\right]\mathbb{S}[\{\pi -\mathbb{S}[\pi]\}s_{\theta} ](Z,X)A\right]\\
				=&\mathbb{E}\left[\left\{\dfrac{\pi(Z,X) - \mathbb{S}[\pi](Z,X)}{\kappa_{\mathcal{S},\pi}^o(X)}Y-\psi_{\mathcal{S},\pi}^o\right\} s_{\theta}(O)\right]
				-\mathbb{E}\left[\dfrac{\{\pi -\mathbb{S}[\pi]\}s_{\theta}(O)}{\kappa_{\mathcal{S},\pi}^o(X)}\mathbb{S}[Y](Z,X)\right]\\
				&-\mathbb{E}\left[ 
				\dfrac{\mu_{\mathcal{S},\pi}^o(X)}{\kappa_{\mathcal{S},\pi}^o(X)}
				(\left\{\pi(Z,X) - \mathbb{S}[\pi](Z,X)\right\}A-\kappa_{\mathcal{S},\pi}^o(X))s_{\theta}(O)\right]\\
				&+\mathbb{E}\left[ \dfrac{\mu_{\mathcal{S},\pi}^o(X)}{\kappa_{\mathcal{S},\pi}^o(X)}\mathbb{S}[\{\pi -\mathbb{S}[\pi]\}s_{\theta} ](Z,X)A\right]\\
				=&\mathbb{E}\left[\left\{\dfrac{\pi(Z,X) - \mathbb{S}[\pi](Z,X)}{\kappa_{\mathcal{S},\pi}^o(X)}Y-\psi_{\mathcal{S},\pi}^o\right\} s_{\theta}(O)\right]\\
				&-\mathbb{E}\left[\dfrac{\{\pi(Z,X) -\mathbb{S}[\pi](Z,X)\}s_{\theta}(O)}{\kappa_{\mathcal{S},\pi}^o(X)}\mathbb{S}[Y](Z,X)\right]\\
				&-\mathbb{E}\left[ 
				\dfrac{\mu_{\mathcal{S},\pi}^o(X)}{\kappa_{\mathcal{S},\pi}^o(X)}
				(\left\{\pi(Z,X) - \mathbb{S}[\pi](Z,X)\right\}A-\kappa_{\mathcal{S},\pi}^o(X))s_{\theta}(O)\right]\\
				&+\mathbb{E}\left[ \dfrac{\mu_{\mathcal{S},\pi}^o(X)}{\kappa_{\mathcal{S},\pi}^o(X)}\{\pi(Z,X) -\mathbb{S}[\pi](Z,X)\}s_{\theta}(O) \mathbb{S}[A](Z,X)\right].
			\end{align*}
			Finally, we find an influence function for $\psi_{\mathcal{S},\pi}^o$, defined as 
			\begin{align*}
				&\varphi_{\mathcal{S},\pi}(O;\psi_{\mathcal{S},\pi}^o,\alpha_{\mathcal{S},\pi}^o)\\:=&\dfrac{\pi(Z,X) - \mathbb{S}[\pi](Z,X)}{\kappa_{\mathcal{S},\pi}^o(X)}Y-\psi_{\mathcal{S},\pi}^o\\
				&-\dfrac{\{\pi(Z,X) -\mathbb{S}[\pi](Z,X)\}}{\kappa_{\mathcal{S},\pi}^o(X)}\mathbb{S}[Y](Z,X)\\
				&+\dfrac{\mu_{\mathcal{S},\pi}^o(X)}{\kappa_{\mathcal{S},\pi}^o(X)}
				(\left\{-\pi(Z,X) + \mathbb{S}[\pi](Z,X)\right\}A+\kappa_{\mathcal{S},\pi}^o(X))\\
				&+\dfrac{\mu_{\mathcal{S},\pi}^o(X)}{\kappa_{\mathcal{S},\pi}^o(X)}\{\pi(Z,X) -\mathbb{S}[\pi](Z,X)\} \mathbb{S}[A](Z,X)\\
				=&\dfrac{\pi(Z,X) - \mathbb{S}[\pi](Z,X)}{\kappa_{\mathcal{S},\pi}^o(X)}\left\{Y - \mathbb{S}[Y](Z,X)\right\}-\psi_{\mathcal{S},\pi}^o\\
				&+\mu_{\mathcal{S},\pi}^o(X)\left(1 - \dfrac{\left\{\pi(Z,X) -\mathbb{S}[\pi](Z,X)\right\}\left\{A-\mathbb{S}[A](Z,X)\right\}}{\kappa_{\mathcal{S},\pi}^o(X)}\right).
			\end{align*}
			The final step follows from the definition of nuisance functions:
			\begin{align*}
				&\delta_{\mathcal{S}}^o(Z,X) := \mathbb{S}[A](Z,X), \qquad
				\eta_{\mathcal{S}}^o(Z,X) := \mathbb{S}[Y](Z,X),\qquad
				\rho_{\mathcal{S},\pi}^o(Z,X) := \mathbb{S}[\pi](Z,X), \\
				&\kappa_{\mathcal{S},\pi}^o(X)=\mathbb{E}[\{\pi(Z,X)-\mathbb{S}[\pi](Z,X)\}\{A-\mathbb{S}[A](Z,X)\}\mid X],\\
				&\mu_{\mathcal{S},\pi}^o(X) := \dfrac{\mathbb{E}[\{\pi(Z,X)-\mathbb{S}[\pi](Z,X)\}\{Y-\mathbb{S}[Y](Z,X)\}\mid X]}{\mathbb{E}[\{\pi(Z,X)-\mathbb{S}[\pi](Z,X)\}\{A-\mathbb{S}[A](Z,X)\}\mid X]}.
			\end{align*}
			Since we conduct the semiparametric analysis in the nonparametric space, the influence function is the EIF, which concludes the proof of Theorem~\ref{thm: EIF ATE}.

			\subsection{Proof of Theorem~\ref{thm: ATE EIF adaptive}}
			We use a similar method as in Theorem~\ref{thm: EIF ATE} to calculate the EIF for $\psi_{\mathcal{S},*}^o$. First, we define 
			\begin{align*}
				\psi_{\mathcal{S},*}^{(\theta)}:=\mathbb{E}^{(\theta)}\left[\dfrac{\pi_*^{(\theta)}(Z,X) - \mathbb{S}^{(\theta)}[A](Z,X)}{\mathbb{E}^{(\theta)}\left[\{\pi_*^{(\theta)}(Z,X) - \mathbb{S}^{(\theta)}[A](Z,X)\}^2\middle|X\right]}Y\right].
			\end{align*}
			By repeatedly applying Lemma~\ref{lem: semiparametric calculation} and writing $s=s_{\theta}$ for brevity, we obtain
			\begin{align*}
				\nabla_{\theta}\mathbb{E}^{(\theta)}[A|Z,X]\eval_{\theta=0}
				=&\mathbb{E}[\{A-\pi_*^o(Z,X)\}s(O)|Z,X],\\
				\nabla_{\theta}\mathbb{S}^{(\theta)}[A](Z,X)\eval_{\theta=0}
				=&\mathbb{S}[\{A - \mathbb{S}[A]\}s](Z,X).
			\end{align*}
			Moreover, we can derive that
			\begin{align*}
				&\nabla_{\theta}\mathbb{E}^{(\theta)}\left[\{\pi_*^{(\theta)}(Z,X) - \mathbb{S}^{(\theta)}[A](Z,X)\}^2\middle|X\right]\eval_{\theta=0}\\
				=&\mathbb{E}\left[\left\{
				(\pi_*^o(Z,X) - \mathbb{S}[A](Z,X))^2 - \kappa_{\mathcal{S},*}^o(X)
				\right\} s(O)\middle|X\right]\\
				&+\mathbb{E}\left[2\{\pi_*^o(Z,X) - \mathbb{S}[A](Z,X)\}\left\{
				\nabla_{\theta}\pi_*^{(\theta)}(Z,X)\eval_{\theta=0}
				-\nabla_{\theta}\mathbb{S}^{(\theta)}[A](Z,X)\eval_{\theta=0}
				\right\}\middle|X\right]\\
				=&\mathbb{E}\left[\left\{
				(\pi_*^o(Z,X) - \mathbb{S}[A](Z,X))^2 - \kappa_{\mathcal{S},*}^o(X)
				\right\} s(O)\middle|X\right]\\
				&+\mathbb{E}\left[2\{\pi_*^o(Z,X) - \mathbb{S}[A](Z,X)\}
				\mathbb{E}[\{A-\pi_*^o(Z,X)\}s(O)|Z,X]
				\middle|X\right]\\
				&-\mathbb{E}\left[2\{\pi_*^o(Z,X) - \mathbb{S}[A](Z,X)\}\left\{
				\mathbb{S}[\{A - \mathbb{S}[A]\}s](Z,X)
				\right\}\middle|X\right]\\
				=&\mathbb{E}\left[\left\{
				(\pi_*^o(Z,X) - \mathbb{S}[A](Z,X))^2 - \kappa_{\mathcal{S},*}^o(X)
				\right\} s(O)\middle|X\right]\\
				&+\mathbb{E}\left[2\{\pi_*^o(Z,X) - \mathbb{S}[A](Z,X)\}
				\{A-\pi_*^o(Z,X)\}s(O)
				\middle|X\right]\\
				=&\mathbb{E}\left[\left\{
				(A - \mathbb{S}[A](Z,X))^2 -(A - \pi_*^o(Z,X))^2 - \kappa_{\mathcal{S},*}^o(X)
				\right\} s(O)\middle|X\right].
			\end{align*}
			Then, we calculate
			\begin{align*}
				\nabla_{\theta}\psi_{\mathcal{S},*}^{(\theta)}\eval_{\theta=0}
				=&\nabla_{\theta}\mathbb{E}^{(\theta)}\left[
				\frac{\pi_*^{(\theta)}(Z,X) - \mathbb{S}^{(\theta)}[A](Z,X)}{\mathbb{E}^{(\theta)}\left[\{\pi_*^{(\theta)}(Z,X) - \mathbb{S}^{(\theta)}[A](Z,X)\}^2|X\right]}Y
				\right]\eval_{\theta=0}\\
				=&\mathbb{E}\left[\left\{
				\frac{\pi_*^o(Z,X) - \mathbb{S}[A](Z,X)}{\kappa_{\mathcal{S},*}^o(X)}Y
				-\psi_{\mathcal{S},*}^o
				\right\}s(O)\right]\\
				&+\mathbb{E}\left[
				\frac{\nabla_{\theta}\{\pi_*^{(\theta)}(Z,X) - \mathbb{S}^{(\theta)}[A](Z,X)\}\eval_{\theta=0}}{\kappa_{\mathcal{S},*}^o(X)}Y
				\right]\\
				&-\mathbb{E}\left[\begin{array}{l}
					\dfrac{\pi_*^o(Z,X) - \mathbb{S}[A](Z,X)}{\kappa_{\mathcal{S},*}^o(X)^2}Y
					\nabla_{\theta}\mathbb{E}^{(\theta)}\left[\{\pi_*^{(\theta)}(Z,X) - \mathbb{S}^{(\theta)}[A](Z,X)\}^2\middle|X\right]\eval_{\theta=0}
				\end{array}\right]\\
				=&\mathbb{E}\left[\left\{
				\frac{\pi_*^o(Z,X) - \mathbb{S}[A](Z,X)}{\kappa_{\mathcal{S},*}^o(X)}Y
				-\psi_{\mathcal{S},*}^o
				\right\}s(O)\right]\\
				&+\mathbb{E}\left[
				\frac{\mathbb{E}\left[\{A-\pi_*^o(Z,X)\}s(O)|Z,X\right]
					-\mathbb{S}[\{A - \mathbb{S}[A]\}s](Z,X)
				}{\kappa_{\mathcal{S},*}^o(X)}Y
				\right]\\
				&-\mathbb{E}\left[\begin{array}{l}
					\dfrac{\pi_*^o(Z,X) - \mathbb{S}[A](Z,X)}{\kappa_{\mathcal{S},*}^o(X)^2}Y\\
					\mathbb{E}\left[\left\{
					(A - \mathbb{S}[A](Z,X))^2 -(A - \pi_*^o(Z,X))^2 - \kappa_{\mathcal{S},*}^o(X)
					\right\} s(O)\middle|X\right]
				\end{array}\right]\\
				=&\mathbb{E}\left[\left\{
				\frac{\pi_*^o(Z,X) - \mathbb{S}[A](Z,X)}{\kappa_{\mathcal{S},*}^o(X)}Y
				-\psi_{\mathcal{S},*}^o
				\right\}s(O)\right]\\
				&+\mathbb{E}\left[
				\frac{\{A-\pi_*^o(Z,X)\}\mathbb{E}[Y|Z,X] - \{A-\mathbb{S}[A](Z,X)\}\mathbb{S}[Y](Z,X)
				}{\kappa_{\mathcal{S},*}^o(X)}s(O)
				\right]\\
				&-\mathbb{E}\left[\begin{array}{l}
					\mathbb{E}\left[\dfrac{\pi_*^o(Z,X) - \mathbb{S}[A](Z,X)}{\kappa_{\mathcal{S},*}^o(X)^2}Y\middle|X\right]\\
					\left\{(A - \mathbb{S}[A](Z,X))^2 -(A - \pi_*^o(Z,X))^2 - \kappa_{\mathcal{S},*}^o(X)\right\} s(O)
				\end{array}\right]\\
				=&\mathbb{E}\left[\left\{
				\frac{\pi_*^o(Z,X) - \mathbb{S}[A](Z,X)}{\kappa_{\mathcal{S},*}^o(X)}Y
				-\psi_{\mathcal{S},*}^o
				\right\}s(O)\right]\\
				&+\mathbb{E}\left[
				\frac{\{A-\pi_*^o(Z,X)\}\mathbb{E}[Y|Z,X] - \{A-\mathbb{S}[A](Z,X)\}\mathbb{S}[Y](Z,X)
				}{\kappa_{\mathcal{S},*}^o(X)}s(O)
				\right]\\
				&-\mathbb{E}\left[\begin{array}{l}
					\dfrac{\mu_{\mathcal{S},*}^o(X)}{\kappa_{\mathcal{S},*}^o(X)}
					\left\{(A - \mathbb{S}[A](Z,X))^2 -(A - \pi_*^o(Z,X))^2 - \kappa_{\mathcal{S},*}^o(X)\right\} s(O)
				\end{array}\right].
			\end{align*}
			
			As a result, the EIF for $\psi_{\mathcal{S},*}^o$ is given by
			\begin{align*}
				&\varphi_{\mathcal{S},*}(O;\psi_{\mathcal{S},*}^o,\alpha_{\mathcal{S},*}^o)\\
				=&\frac{\pi_*^o(Z,X) - \mathbb{S}[A](Z,X)}{\kappa_{\mathcal{S},*}^o(X)}Y
				-\psi_{\mathcal{S},*}^o\\
				&+\frac{\{A-\pi_*^o(Z,X)\}\mathbb{E}[Y|Z,X] - \{A-\mathbb{S}[A](Z,X)\}\mathbb{S}[Y](Z,X)
				}{\kappa_{\mathcal{S},*}^o(X)}\\
				&+\dfrac{\mu_{\mathcal{S},*}^o(X)}{\kappa_{\mathcal{S},*}^o(X)}
				\left\{\kappa_{\mathcal{S},*}^o(X) + (A - \pi_*^o(Z,X))^2 - (A - \mathbb{S}[A](Z,X))^2\right\}\\
				=&\frac{\pi_*^o(Z,X) - \delta_{\mathcal{S}}^o(Z,X)}{\kappa_{\mathcal{S},*}^o(X)}Y
				-\psi_{\mathcal{S},*}^o
				+ \dfrac{\mu_{\mathcal{S},*}^o(X)}{\kappa_{\mathcal{S},*}^o(X)}
				\left\{\kappa_{\mathcal{S},*}^o(X) + (A - \pi_*^o(Z,X))^2 - (A - \delta_{\mathcal{S}}^o(Z,X))^2  \right\}\\
				&+\frac{1}{\kappa_{\mathcal{S},*}^o(X)}\left\{(A-\pi_*^o(Z,X))\xi_{*}^o(Z,X) - (A-\delta_{\mathcal{S}}^o(Z,X))\eta_{\mathcal{S}}^o(Z,X)
				\right\}.
			\end{align*}
			This completes the proof of Theorem~\ref{thm: ATE EIF adaptive}.

			\subsection{Proof of Theorem~\ref{thm: asymptotic normality}}
			Define the conditional expectation on the $k$-th fold as $\mathbb{E}_k[O]=\mathbb{E}[O\mid I_{-k}]$. By the definition of $\hat \psi_{\mathcal{S},\pi}$,
			\begin{align}
				0=&\dfrac{\sqrt{n}}{K}\sum_{k=1}^K\mathbb{E}_{nk}\left[\varphi_{\mathcal{S},\pi}(O;\hat\psi_{\mathcal{S},\pi},\hat\alpha_{\mathcal{S},\pi}^{(n,-k)})\right]\notag\\
				=&\dfrac{\sqrt{n}}{K}\sum_{k=1}^K\mathbb{E}_{nk}\left[\varphi_{\mathcal{S},\pi}(O;\psi_{\mathcal{S},\pi}^{o},\hat\alpha_{\mathcal{S},\pi}^{(n,-k)})\right]-\sqrt{n}\{\hat\psi_{\mathcal{S},\pi}-\psi_{\mathcal{S},\pi}^{o}\}\notag\\
				=&\dfrac{\sqrt{n}}{K}\sum_{k=1}^K\mathbb{E}_{nk}\left[\varphi_{\mathcal{S},\pi}(O;\psi_{\mathcal{S},\pi}^{o},\alpha_{\mathcal{S},\pi}^{o})\right]-\sqrt{n}\{\hat\psi_{\mathcal{S},\pi}-\psi_{\mathcal{S},\pi}^{o}\}\notag\\
				&+\dfrac{\sqrt{n}}{K}\sum_{k=1}^K\mathbb{E}_{k}\left[\varphi_{\mathcal{S},\pi}(O;\psi_{\mathcal{S},\pi}^{o},\hat\alpha_{\mathcal{S},\pi}^{(n,-k)})-\varphi_{\mathcal{S},\pi}(O;\psi_{\mathcal{S},\pi}^{o},\alpha_{\mathcal{S},\pi}^{o})\right]\label{exp:10}\\
				&+\dfrac{\sqrt{n}}{K}\sum_{k=1}^K\{\mathbb{E}_{nk}-\mathbb{E}_{k}\}\left[\varphi_{\mathcal{S},\pi}(O;\psi_{\mathcal{S},\pi}^{o},\hat\alpha_{\mathcal{S},\pi}^{(n,-k)})-\varphi_{\mathcal{S},\pi}(O;\psi_{\mathcal{S},\pi}^{o},\alpha_{\mathcal{S},\pi}^{o})\right].\label{exp:9}
			\end{align}
			From Proposition~\ref{prop: mix bias property}, we can deduce that the quantity in Equation~\eqref{exp:10} is $o_p(1)$, since
			\begin{align*}
				&\left|\mathbb{E}_{k}\left[\varphi_{\mathcal{S},\pi}(O;\psi_{\mathcal{S},\pi}^{o},\hat\alpha_{\mathcal{S},\pi}^{(n,-k)})-\varphi_{\mathcal{S},\pi}(O;\psi_{\mathcal{S},\pi}^{o},\alpha_{\mathcal{S},\pi}^{o})\right]\right|\\
				\lesssim&
				\left\{\begin{array}{l}
					\|\hat\kappa_{\mathcal{S},\pi}^{(n,-k)}(X)-\kappa_{\mathcal{S},\pi}^o(X)\|_2\times \|\hat\mu_{\mathcal{S},\pi}^{(n,-k)}(X)-\mu_{\mathcal{S},\pi}^o(X)\|_2\\
					+\|\hat\rho_{\mathcal{S},\pi}^{(n,-k)}(Z,X)-\rho_{\mathcal{S},\pi}^o(Z,X)\|_2\times
					\|\hat\delta_{\mathcal{S}}^{(n,-k)}(Z,X)-\delta_{\mathcal{S}}^o(Z,X)\|_2\\
					+\|\hat\rho_{\mathcal{S},\pi}^{(n,-k)}(Z,X)-\rho_{\mathcal{S},\pi}^o(Z,X)\|_2\times
					\|\hat\eta_{\mathcal{S}}^{(n,-k)}(Z,X)-\eta_{\mathcal{S}}^o(Z,X)\|_2
				\end{array}\right\}
				=o_p(n^{-1/2}).
			\end{align*}
			Define the empirical process as $\mathbb{G}_{nk}[f(O)]:=\sqrt{n/K}\{\mathbb{E}_{nk}-\mathbb{E}_{k}\}[f(O)],$ and we have
			\begin{align*}
				&\Pr\left(\left|\mathbb{G}_{nk}\left[\varphi_{\mathcal{S},\pi}(O;\psi_{\mathcal{S},\pi}^{o},\hat\alpha_{\mathcal{S},\pi}^{(n,-k)})-\varphi_{\mathcal{S},\pi}(O;\psi_{\mathcal{S},\pi}^{o},\alpha_{\mathcal{S},\pi}^{o})\right]\right|\geq \epsilon_0\middle| I_{-k}\right)\\
				\lesssim&\dfrac{1}{\epsilon_0^2}\mathrm{Var}\left[\mathbb{G}_{nk}\left\{\varphi_{\mathcal{S},\pi}(O;\psi_{\mathcal{S},\pi}^{o},\hat\alpha_{\mathcal{S},\pi}^{(n,-k)})-\varphi_{\mathcal{S},\pi}(O;\psi_{\mathcal{S},\pi}^{o},\alpha_{\mathcal{S},\pi}^{o})\right\}\middle| I_{-k}\right]\\
				\lesssim&\dfrac{1}{\epsilon_0^2}\mathrm{Var}\left[\varphi_{\mathcal{S},\pi}(O;\psi_{\mathcal{S},\pi}^{o},\hat\alpha_{\mathcal{S},\pi}^{(n,-k)})-\varphi_{\mathcal{S},\pi}(O;\psi_{\mathcal{S},\pi}^{o},\alpha_{\mathcal{S},\pi}^{o})\middle| I_{-k}\right]\\
				\lesssim&\dfrac{1}{\epsilon_0^2}\mathbb{E}\left[\left\{\varphi_{\mathcal{S},\pi}(O;\psi_{\mathcal{S},\pi}^{o},\hat\alpha_{\mathcal{S},\pi}^{(n,-k)})-\varphi_{\mathcal{S},\pi}(O;\psi_{\mathcal{S},\pi}^{o},\alpha_{\mathcal{S},\pi}^{o})\right\}^2\middle| I_{-k}\right]\\
				\lesssim&\|\hat\rho_{\mathcal{S},\pi}^{(n,-k)}(Z,X)-\rho_{\mathcal{S},\pi}^o(Z,X)\|_2^2+
				\|\hat\delta_{\mathcal{S}}^{(n,-k)}(Z,X)-\delta_{\mathcal{S}}^o(Z,X)\|_2^2+
				\|\hat\eta_{\mathcal{S}}^{(n,-k)}(Z,X)-\eta_{\mathcal{S}}^o(Z,X)\|_2^2\\&+
				\|\hat\kappa_{\mathcal{S},\pi}^{(n,-k)}(X)-\kappa_{\mathcal{S},\pi}^o(X)\|_2^2+
				\|\hat\mu_{\mathcal{S},\pi}^{(n,-k)}(X)-\mu_{\mathcal{S},\pi}^o(X)\|_2^2\\\lesssim&
				\|\hat\alpha_{\mathcal{S},\pi}^{(n,-k)}-\alpha_{\mathcal{S},\pi}^o\|_2^2.
			\end{align*}
			We can now integrate $I_{-k}$ out to deduce that
			\begin{align*}
				\Pr\left(\left|\mathbb{G}_{nk}\left[\varphi_{\mathcal{S},\pi}(O;\psi_{\mathcal{S},\pi}^{o},\hat\alpha_{\mathcal{S},\pi}^{(n,-k)})-\varphi_{\mathcal{S},\pi}(O;\psi_{\mathcal{S},\pi}^{o},\alpha_{\mathcal{S},\pi}^{o})\right]\right|\geq \epsilon_0\right)\lesssim 
				\mathbb{E}[\|\hat\alpha_{\mathcal{S},\pi}^{(n,-k)}-\alpha_{\mathcal{S},\pi}^o\|_2^2]=o(1).
			\end{align*}
			This is equivalent to saying that the quantity in Equation~\eqref{exp:9} is $o_p(1)$. Next, we can see that
			\begin{align*}
				\sqrt{n}\{\hat\psi_{\mathcal{S},\pi}-\psi_{\mathcal{S},\pi}^{o}\}=&\dfrac{\sqrt{n}}{K}\sum_{k=1}^K\mathbb{E}_{nk}\left[\varphi_{\mathcal{S},\pi}(O;\psi_{\mathcal{S},\pi}^{o},\alpha_{\mathcal{S},\pi}^{o})\right]+o_p(1)\\
				=&\dfrac{1}{\sqrt{n}}\sum_{i = 1}^n\varphi_{\mathcal{S},\pi}(O_i;\psi_{\mathcal{S},\pi}^{o},\alpha_{\mathcal{S},\pi}^{o})+o_p(1).
			\end{align*}
			From the central limit theorem and Slutsky's lemma, we have
			$\sqrt{n}\{\hat\psi_{\mathcal{S},\pi}-\psi_{\mathcal{S},\pi}^{o}\}$ converges in distribution to $\mathcal{N}(0,(\sigma_{\mathcal{S},\pi}^o)^2)$.

            Next, we prove the consistency of the variance estimator $$\left(\hat\sigma_{\mathcal{S},\pi}^{(n,-k)}\right)^2:=\mathbb{E}_{nk}\left[\varphi_{\mathcal{S},\pi}(O;\hat\psi_{\mathcal{S},\pi},\hat\alpha_{\mathcal{S},\pi}^{(n,-k)})^2\right].$$
            First, we define
            $
            \left(\tilde\sigma_{\mathcal{S},\pi}^{(n,-k)}\right)^2:=\mathbb{E}_{nk}\left[\varphi_{\mathcal{S},\pi}(O;\psi_{\mathcal{S},\pi}^{o},\alpha_{\mathcal{S},\pi}^{o})^2\right].
            $
            By the law of large numbers, $\left(\tilde\sigma_{\mathcal{S},\pi}^{(n,-k)}\right)^2$ converges to $\left(\sigma_{\mathcal{S},\pi}^{o}\right)^2$ in probability. 
            Moreover, $\varphi_{\mathcal{S},\pi}(O;\psi_{\mathcal{S},\pi}^o,\hat\alpha_{\mathcal{S},\pi}^{(n,-k)})$ and the corresponding true influence function are uniformly bounded. For notational simplicity, define $d_n:=\hat\psi_{\mathcal{S},\pi}-\psi_{\mathcal{S},\pi}^o$. This gives
            \[
            \varphi_{\mathcal{S},\pi}(O;\hat\psi_{\mathcal{S},\pi},\hat\alpha_{\mathcal{S},\pi}^{(n,-k)})
            =\varphi_{\mathcal{S},\pi}(O;\psi_{\mathcal{S},\pi}^o,\hat\alpha_{\mathcal{S},\pi}^{(n,-k)})-d_n.
            \]
            Expanding the square and using the uniform bound established above, we obtain
            \begin{align*}
            	&\left|\left(\hat\sigma_{\mathcal{S},\pi}^{(n,-k)}\right)^2-
            	\mathbb{E}_{nk}\left[\varphi_{\mathcal{S},\pi}(O;\psi_{\mathcal{S},\pi}^o,\hat\alpha_{\mathcal{S},\pi}^{(n,-k)})^2\right]\right|\\
            	&\quad=\left|-2d_n\mathbb{E}_{nk}\left[\varphi_{\mathcal{S},\pi}(O;\psi_{\mathcal{S},\pi}^o,\hat\alpha_{\mathcal{S},\pi}^{(n,-k)})\right]+d_n^2\right|\\
            	&\quad\leq 2|d_n|\mathbb{E}_{nk}\left[\left|\varphi_{\mathcal{S},\pi}(O;\psi_{\mathcal{S},\pi}^o,\hat\alpha_{\mathcal{S},\pi}^{(n,-k)})\right|\right]+d_n^2
            	\lesssim |d_n|+d_n^2=o_p(1).
            \end{align*}
            The final equality follows from $d_n=o_p(1)$, as established above. By  Markov's inequality, sample splitting, and the Cauchy-Schwarz inequality, for any $\epsilon>0$,
            \begin{align*}
            	&\Pr\left(\left|\mathbb{E}_{nk}\left[\varphi_{\mathcal{S},\pi}(O;\psi_{\mathcal{S},\pi}^o,\hat\alpha_{\mathcal{S},\pi}^{(n,-k)})^2-\varphi_{\mathcal{S},\pi}(O;\psi_{\mathcal{S},\pi}^{o},\alpha_{\mathcal{S},\pi}^{o})^2\right]\right|\geq\epsilon\middle| I_{-k}\right)\\
            	\leq&\dfrac{1}{\epsilon}\mathbb{E}_{k}\left[\left|\mathbb{E}_{nk}\left[\varphi_{\mathcal{S},\pi}(O;\psi_{\mathcal{S},\pi}^o,\hat\alpha_{\mathcal{S},\pi}^{(n,-k)})^2-\varphi_{\mathcal{S},\pi}(O;\psi_{\mathcal{S},\pi}^{o},\alpha_{\mathcal{S},\pi}^{o})^2\right]\right|\right]\\
            	\lesssim&\dfrac{1}{\epsilon}\mathbb{E}_{k}\left[\left|\varphi_{\mathcal{S},\pi}(O;\psi_{\mathcal{S},\pi}^o,\hat\alpha_{\mathcal{S},\pi}^{(n,-k)})^2-\varphi_{\mathcal{S},\pi}(O;\psi_{\mathcal{S},\pi}^{o},\alpha_{\mathcal{S},\pi}^{o})^2\right|\right]\\
            	\lesssim&\dfrac{1}{\epsilon}\left(\mathbb{E}_{k}\left[\left|\varphi_{\mathcal{S},\pi}(O;\psi_{\mathcal{S},\pi}^o,\hat\alpha_{\mathcal{S},\pi}^{(n,-k)})-\varphi_{\mathcal{S},\pi}(O;\psi_{\mathcal{S},\pi}^{o},\alpha_{\mathcal{S},\pi}^{o})\right|^2\right]\right)^{1/2}\\
                &\times \left(\mathbb{E}_{k}\left[\left|\varphi_{\mathcal{S},\pi}(O;\psi_{\mathcal{S},\pi}^o,\hat\alpha_{\mathcal{S},\pi}^{(n,-k)})+\varphi_{\mathcal{S},\pi}(O;\psi_{\mathcal{S},\pi}^{o},\alpha_{\mathcal{S},\pi}^{o})\right|^2\right]\right)^{1/2}\\
            	\lesssim&\dfrac{1}{\epsilon}\|\hat\alpha_{\mathcal{S},\pi}^{(n,-k)}-\alpha_{\mathcal{S},\pi}^{o}\|_2.
            \end{align*}
            Here, conditioning on $I_{-k}$ means conditioning on the training observations (and any independent fitting randomization), and $O$ in the single-observation conditional expectations denotes an independent observation. The last inequality follows directly from the expression for $\varphi_{\mathcal{S},\pi}$ and the stated uniform bounds.
            Now we can integrate the training data out to deduce that
            \begin{align*}
            	&\Pr\left(\left|\mathbb{E}_{nk}\left[\varphi_{\mathcal{S},\pi}(O;\psi_{\mathcal{S},\pi}^o,\hat\alpha_{\mathcal{S},\pi}^{(n,-k)})^2-\varphi_{\mathcal{S},\pi}(O;\psi_{\mathcal{S},\pi}^{o},\alpha_{\mathcal{S},\pi}^{o})^2\right]\right|\geq\epsilon\right)\\
            	\lesssim&\dfrac{1}{\epsilon}\mathbb{E}\left[\|\hat\alpha_{\mathcal{S},\pi}^{(n,-k)}-\alpha_{\mathcal{S},\pi}^{o}\|_2\right]
            	\leq\dfrac{1}{\epsilon}\left\{\mathbb{E}\left[\|\hat\alpha_{\mathcal{S},\pi}^{(n,-k)}-\alpha_{\mathcal{S},\pi}^{o}\|_2^2\right]\right\}^{1/2}=o(1).
            \end{align*}
            Consequently, $\left(\hat\sigma_{\mathcal{S},\pi}^{(n,-k)}\right)^2=\left(\tilde\sigma_{\mathcal{S},\pi}^{(n,-k)}\right)^2+o_p(1)=\left(\sigma_{\mathcal{S},\pi}^{o}\right)^2+o_p(1)$ for every $k$. Since $K$ is fixed, averaging over the folds gives $(\hat\sigma_{\mathcal{S},\pi})^2\overset{p}{\to}(\sigma_{\mathcal{S},\pi}^o)^2$, and the continuous mapping theorem yields $\hat\sigma_{\mathcal{S},\pi}\overset{p}{\to}\sigma_{\mathcal{S},\pi}^o$, finishing the proof of Theorem~\ref{thm: asymptotic normality}.
            Finally, since the proof of Theorem~\ref{thm: asymptotic normality adaptive} is similar to that of Theorem~\ref{thm: asymptotic normality}, we omit it for brevity.

			\subsection{Proof of Theorem~\ref{thm: convergence rate of nuisance estimator}}
			We first review the notion of the critical radius. Let $\mathcal{F}_{\kappa}$ and $\mathcal{F}_{\Gamma}$ be two function classes satisfying $\sup_{f \in \mathcal{F}_{\kappa} \cup \mathcal{F}_{\Gamma}} \|f\|_{\infty} < \infty$. For the function classes $\mathcal{F}_{\kappa}$ and $\mathcal{F}_{\Gamma}$, define their localized empirical process complexity
			\begin{align*}
				&\mathcal{G}_n(\delta; \mathcal{F}_{\kappa}):= \mathbb{E}\left[\sup_{\substack{f_{\kappa} \in \mathcal{F}_{\kappa},\;\|f_{\kappa} - f_{\kappa}^*\|_2 \le \delta}}\left| \frac{1}{n}\sum_{i=1}^n \epsilon_i \{f_{\kappa}(X_i)-f_{\kappa}^*(X_i)\} \right| \right],\\
				&\mathcal{G}_n(\delta; \mathcal{F}_{\Gamma}):= \mathbb{E}\left[\sup_{\substack{f_{\Gamma} \in \mathcal{F}_{\Gamma},\;\|f_{\Gamma} - f_{\Gamma}^*\|_2 \le \delta}}\left| \frac{1}{n}\sum_{i=1}^n \epsilon_i \{f_{\Gamma}(X_i)-f_{\Gamma}^*(X_i)\} \right| \right],
			\end{align*}
			where $f_{\kappa}^*=\kappa_{\mathcal{S},*}^o \in \mathcal{F}_{\kappa}$, $f_{\Gamma}^*=\Gamma_{\mathcal{S},*}^o \in \mathcal{F}_{\Gamma}$, and $\{\epsilon_i\}_{i=1}^n$ are independent Rademacher variables. For a function class $\mathcal{F}$, the critical radius $\delta_n(\mathcal{F})$ is defined as the smallest positive solution to
			\[
			\mathcal{G}_n(\delta; \mathcal{F}) \le c \delta^2,
			\]
			for some universal constant $c > 0$. We denote the corresponding critical radii of $\mathcal{F}_{\kappa}$ and $\mathcal{F}_{\Gamma}$ by $c_{n,\kappa}$ and $c_{n,\Gamma}$, respectively.

			We verify Assumptions~1--4 of \citet{foster2023orthogonalstatisticallearning} in turn. 
			First, we verify that the estimating equation has the functional mixed bias property
			\begin{align*}
				&\nabla_{t} \mathbb{E}\left[\mathcal{L}_{\kappa}(O;\kappa_{\mathcal{S},*}^o+t\kappa_{v},\alpha_{\mathcal{S},*} )\mid X\right]\eval_{t=0}\\
				=&-\mathbb{E}\left[(A-\delta_{\mathcal{S}}(Z,X))^2 - (A-\pi_*(Z,X))^2 - \kappa_{\mathcal{S},*}^o(X)\mid X\right]\kappa_{v}(X)\\
				=&-\mathbb{E}\left[\left\{\begin{array}{l}
					(A-\delta_{\mathcal{S}}(Z,X))^2 - (A-\pi_*(Z,X))^2\\
					-(A-\delta_{\mathcal{S}}^o(Z,X))^2 + (A-\pi_*^o(Z,X))^2
				\end{array}\right\}\middle| X\right]\kappa_{v}(X)\\
				=&-\mathbb{E}\left[(\delta_{\mathcal{S}}^o(Z,X)-\delta_{\mathcal{S}}(Z,X))^2 - (\pi_*^o(Z,X)-\pi_*(Z,X))^2 \mid X\right]\kappa_{v}(X).
			\end{align*}
			Furthermore, $\nabla_{t}^2 \mathbb{E}\left[\mathcal{L}_{\kappa}(O;\kappa_{\mathcal{S},*}^o+t\kappa_{v},\alpha_{\mathcal{S},*} )\mid X\right]\eval_{t=0}\equiv \kappa_{v}(X)^2$, and
			\begin{align*}
				&\nabla_{s}\nabla_{t} \mathbb{E}\left[\mathcal{L}_{\kappa}(O;\kappa_{\mathcal{S},*}^o+t\kappa_{v},\alpha_{\mathcal{S},*}^o+s\alpha_{v} )\middle| X\right]\eval_{t=0}\\
				=&-\nabla_{s}\mathbb{E}\left[s^2\delta_{v}(Z,X)^2 - s^2\pi_{v}(Z,X)^2 \mid X\right]\kappa_{v}(X)\\
				=&-2s\mathbb{E}\left[\delta_{v}(Z,X)^2 - \pi_{v}(Z,X)^2 \mid X\right]\kappa_{v}(X).
			\end{align*}
			This indicates that $\nabla_{s}\nabla_{t} \mathbb{E}\left[\mathcal{L}_{\kappa}(O;\kappa_{\mathcal{S},*}^o+t\kappa_{v},\alpha_{\mathcal{S},*}^o+s\alpha_{v} )\middle| X\right]\eval_{t=0,s=0}=0$, and
			\begin{align*}
				&\left|\nabla_{s}^2\nabla_{t} \mathbb{E}\left[\mathcal{L}_{\kappa}(O;\kappa_{\mathcal{S},*}^o+t\kappa_{v},\alpha_{\mathcal{S},*}^o+s\alpha_{v} )\right]\eval_{t=0}\right|\\
				=&2\left|\mathbb{E}\left[\{\delta_{v}(Z,X)^2 - \pi_{v}(Z,X)^2\} \kappa_{v}(X)\right]\right|
				\lesssim \|\alpha_{v}\|_4^2\times \|\kappa_{v}\|_2.
			\end{align*}
			Consequently, we have checked that
			\begin{align*}
				\begin{cases}
					\nabla_{s}\nabla_{t} \mathbb{E}\left[\mathcal{L}_{\kappa}(O;\kappa_{\mathcal{S},*}^o+t\kappa_{v},\alpha_{\mathcal{S},*}^o+s\alpha_{v} )\right]\eval_{t=0,s=0}=0;\\
					\nabla_{t} \mathbb{E}\left[\mathcal{L}_{\kappa}(O;\kappa_{\mathcal{S},*}^o+t\kappa_{v},\alpha_{\mathcal{S},*}^o )\right]\eval_{t=0}= 0;\\
					\nabla_{t}^2 \mathbb{E}\left[\mathcal{L}_{\kappa}(O;\kappa_{\mathcal{S},*}^o+t\kappa_{v},\alpha_{\mathcal{S},*} )\right]\eval_{t=0}=\|\kappa_{v}\|_2^2;\\
					\left|\nabla_{s}^2\nabla_{t} \mathbb{E}\left[\mathcal{L}_{\kappa}(O;\kappa_{\mathcal{S},*}^o+t\kappa_{v},\alpha_{\mathcal{S},*}^o+s\alpha_{v} )\right]\eval_{t=0}\right|\lesssim \|\alpha_{v}\|_4^2 \|\kappa_{v}\|_2.
				\end{cases}
			\end{align*}
			By Theorem~3 of \citet{foster2023orthogonalstatisticallearning}, with probability at least $1-\epsilon$,
			\begin{align*}
				\|\hat\kappa_{\mathcal{S},*}^{(n,-k)}-\kappa_{\mathcal{S},*}^o\|_2^2 \lesssim  c_{n,\kappa}^2 +\frac{\log(1/\epsilon)}{n}+ \sum_{l=1}^L\|\hat\alpha_{\mathcal{S},*}^{(n,-kl)}-\alpha_{\mathcal{S},*}^o\|_4^4.
			\end{align*}

			Next, we verify the corresponding conditions for $\Gamma_{\mathcal{S},*}^o$:
			\begin{align*}
				&\nabla_{t} \mathbb{E}\left[\mathcal{L}_{\Gamma}(O;\Gamma_{\mathcal{S},*}^o+t\Gamma_{v},\alpha_{\mathcal{S},*} )\mid X\right]\eval_{t=0}\\
				=&-\mathbb{E}\left[\begin{array}{l}
					(A-\pi_*(Z,X))\xi_*(Z,X) - (A-\delta_{\mathcal{S}}(Z,X))\eta_{\mathcal{S}}(Z,X)\\
					+(\pi_*(Z,X)-\delta_{\mathcal{S}}(Z,X))Y -\Gamma_{\mathcal{S},*}^o(X)
				\end{array}\middle| X\right]\Gamma_{v}(X)\\
				=&-\mathbb{E}\left[\begin{array}{l}
					(\pi_*^o(Z,X)-\pi_*(Z,X))\xi_*(Z,X) - (\delta_{\mathcal{S}}^o(Z,X)-\delta_{\mathcal{S}}(Z,X))\eta_{\mathcal{S}}(Z,X)\\
					+\pi_*(Z,X)\xi_*^o(Z,X)-\delta_{\mathcal{S}}(Z,X)\eta_{\mathcal{S}}^o(Z,X)-\Gamma_{\mathcal{S},*}^o(X)
				\end{array}\middle| X\right]\Gamma_{v}(X)\\
				=&-\mathbb{E}\left[\begin{array}{l}
					(\pi_*^o(Z,X)-\pi_*(Z,X))(\xi_*(Z,X)- \xi_*^o(Z,X))\\
					-(\delta_{\mathcal{S}}^o(Z,X)-\delta_{\mathcal{S}}(Z,X))(\eta_{\mathcal{S}}(Z,X)-\eta_{\mathcal{S}}^o(Z,X))\\
					+\pi_*^o(Z,X)\xi_*^o(Z,X)-\delta_{\mathcal{S}}^o(Z,X)\eta_{\mathcal{S}}^o(Z,X)-\Gamma_{\mathcal{S},*}^o(X)
				\end{array}\middle| X\right]\Gamma_{v}(X)\\
				=&-\mathbb{E}\left[\begin{array}{l}
					-(\pi_*(Z,X)-\pi_*^o(Z,X))(\xi_*(Z,X)- \xi_*^o(Z,X))\\
					+(\delta_{\mathcal{S}}(Z,X)-\delta_{\mathcal{S}}^o(Z,X))(\eta_{\mathcal{S}}(Z,X)-\eta_{\mathcal{S}}^o(Z,X))
				\end{array}\middle| X\right]\Gamma_{v}(X).
			\end{align*}
			In addition, we can verify that 
			\begin{align*}
				&\nabla_{t}^2 \mathbb{E}\left[\mathcal{L}_{\Gamma}(O;\Gamma_{\mathcal{S},*}^o+t\Gamma_{v},\alpha_{\mathcal{S},*})\mid X\right]\eval_{t=0}
				=\Gamma_{v}^2(X);\\
				&\nabla_{s}\nabla_{t} \mathbb{E}\left[\mathcal{L}_{\Gamma}(O;\Gamma_{\mathcal{S},*}^o+t\Gamma_{v},\alpha_{\mathcal{S},*}^o+s\alpha_{v})\mid X\right]\eval_{t=0}\\
				=&-2s\mathbb{E}\left[-\pi_{v}(Z,X)\xi_{v}(Z,X)+\delta_{v}(Z,X)\eta_{v}(Z,X)\mid X\right]\Gamma_{v}(X).
			\end{align*}
			This indicates that 
			\begin{align*}
				\begin{cases}
					\nabla_{s}\nabla_{t} \mathbb{E}\left[\mathcal{L}_{\Gamma}(O;\Gamma_{\mathcal{S},*}^o+t\Gamma_{v},\alpha_{\mathcal{S},*}^o+s\alpha_{v})\right]\eval_{t=0}\eval_{s=0}\equiv 0;\\
					\nabla_{t} \mathbb{E}\left[\mathcal{L}_{\Gamma}(O;\Gamma_{\mathcal{S},*}^o+t\Gamma_{v},\alpha_{\mathcal{S},*}^o )\right]\eval_{t=0}= 0;\\
					\nabla_{t}^2 \mathbb{E}\left[\mathcal{L}_{\Gamma}(O;\Gamma_{\mathcal{S},*}^o+t\Gamma_{v},\alpha_{\mathcal{S},*} )\right]\eval_{t=0}=\|\Gamma_{v}\|_2^2;\\
					|\nabla_{s}^2\nabla_{t} \mathbb{E}\left[\mathcal{L}_{\Gamma}(O;\Gamma_{\mathcal{S},*}^o+t\Gamma_{v},\alpha_{\mathcal{S},*}^o+s\alpha_{v})\right]\eval_{t=0}\eval_{s=0}|\\
					=2|\mathbb{E}\left[\left\{
					-\pi_{v}(Z,X)\xi_{v}(Z,X)
					+\delta_{v}(Z,X)\eta_{v}(Z,X)\right\}
					\Gamma_{v}(X)\right]|
					\lesssim \|\alpha_{v}\|_4^2\|\Gamma_{v}\|_2.
				\end{cases}
			\end{align*}
			Applying Theorem~3 of \citet{foster2023orthogonalstatisticallearning} once more, we obtain, with probability at least $1-\epsilon$,
			\begin{align*}
				\|\hat\Gamma_{\mathcal{S},*}^{(n,-k)}-\Gamma_{\mathcal{S},*}^o\|_2^2 \lesssim  c_{n,\Gamma}^2 +\frac{\log(1/\epsilon)}{n}+ \sum_{l=1}^L\|\hat\alpha_{\mathcal{S},*}^{(n,-kl)}-\alpha_{\mathcal{S},*}^o\|_4^4.
			\end{align*}
			This completes the proof of Theorem~\ref{thm: convergence rate of nuisance estimator}.

			\subsection{Proof of Theorem~\ref{thm: identification partially linear effect}}
			By Equation~\eqref{eq: partially linear model},
			\begin{align*}
				&Y = \beta^oA + s^o(Z,X) + \epsilon_y \text{,\qquad where $\mathbb{E}[\epsilon_y|Z,X]=0$,\; and  $s^o\in\mathcal{S}$}.
			\end{align*}
			Therefore,
			\begin{align*}
				&\mathbb{E}[\{\pi(Z,X)-\mathbb{S}[\pi](Z,X)\}\{Y-\mathbb{S}[Y](Z,X)\}\mid X]\\
				=&\mathbb{E}[\{\pi(Z,X)-\mathbb{S}[\pi](Z,X)\}\{\beta^oA + s^o(Z,X)+\epsilon_y-\mathbb{S}[\beta^oA + s^o+ \epsilon_y](Z,X)\}\mid X]\\
				=&\mathbb{E}[\{\pi(Z,X)-\mathbb{S}[\pi](Z,X)\}\{\beta^oA-\mathbb{S}[\beta^oA](Z,X)\}\mid X]\\
				=&\beta^o\mathbb{E}[\{\pi(Z,X)-\mathbb{S}[\pi](Z,X)\}\{A-\mathbb{S}[A](Z,X)\}\mid X].
			\end{align*}
			Now, we can integrate $X$ to deduce that
			\begin{align*}
				&\mathbb{E}[\{\pi(Z,X)-\mathbb{S}[\pi](Z,X)\}\{Y-\mathbb{S}[Y](Z,X)\}]\\=&
				\beta^o\mathbb{E}[\{\pi(Z,X)-\mathbb{S}[\pi](Z,X)\}\{A-\mathbb{S}[A](Z,X)\}].
			\end{align*}
			Since the expectation on the right-hand side of the equation above is nonzero, 
			\begin{align*}
				\beta^o=\dfrac{\mathbb{E}[\{\pi(Z,X)-\mathbb{S}[\pi](Z,X)\}\{Y-\mathbb{S}[Y](Z,X)\}]}{\mathbb{E}[\{\pi(Z,X)-\mathbb{S}[\pi](Z,X)\}\{A-\mathbb{S}[A](Z,X)\}]},
			\end{align*}
			finishing the proof of $\beta_{\mathcal{S},\pi}^o=\beta^o$ in Theorem~\ref{thm: identification partially linear effect}. 
			Applying the preceding argument with \(\pi=\pi_*^o\) yields
			$\beta_{\mathcal{S},*}^o=\beta^o$.

			\subsection{Proof of Theorem~\ref{thm: EIF partially linear effect}}
			Analogous to the proof of Theorem~\ref{thm: EIF ATE}, define
			\begin{align*}
				&\beta_{\mathcal{S},\pi}^{(\theta)}= 
				\dfrac{\mathbb{E}^{(\theta)}[\{\pi(Z,X)-\mathbb{S}^{(\theta)}[\pi](Z,X)\}\{Y-\mathbb{S}^{(\theta)}[Y](Z,X)\}]}{\mathbb{E}^{(\theta)}[\{\pi(Z,X)-\mathbb{S}^{(\theta)}[\pi](Z,X)\}\{A-\mathbb{S}^{(\theta)}[A](Z,X)\}]};\\
				&\beta_{\mathcal{S},*}^{(\theta)}= 
				\dfrac{\mathbb{E}^{(\theta)}[\{\mathbb{E}^{(\theta)}[A|Z,X]-\mathbb{S}^{(\theta)}[A](Z,X)\}\{Y-\mathbb{S}^{(\theta)}[Y](Z,X)\}]}{\mathbb{E}^{(\theta)}[\{\mathbb{E}^{(\theta)}[A|Z,X]-\mathbb{S}^{(\theta)}[A](Z,X)\}^2]}.
			\end{align*}
			By repeatedly using Lemma~\ref{lem: semiparametric calculation}, we can derive that
			\begin{align*}
				\nabla_{\theta}\beta_{\mathcal{S},\pi}^{(\theta)}\eval_{\theta=0}
				=&\mathbb{E}\left[\left\{\dfrac
				{\{\pi(Z,X)-\mathbb{S}[\pi](Z,X)\}\{Y-\mathbb{S}[Y](Z,X)\}}
				{\mathbb{E}[\{\pi(Z,X)-\mathbb{S}[\pi](Z,X)\}\{A-\mathbb{S}[A](Z,X)\}]}-\beta_{\mathcal{S},\pi}^o\right\}s_{\theta}(O)\right]\\
				&-\mathbb{E}\left[\dfrac
				{\mathbb{S}[(\pi-\mathbb{S}[\pi])s_{\theta}](Z,X)\{Y-\mathbb{S}[Y](Z,X)\}}
				{\mathbb{E}[\{\pi(Z,X)-\mathbb{S}[\pi](Z,X)\}\{A-\mathbb{S}[A](Z,X)\}]}\right]\tag{\text{vanish}}\\
				&-\mathbb{E}\left[\dfrac
				{\{\pi(Z,X)-\mathbb{S}[\pi](Z,X)\}\mathbb{S}[(Y-\mathbb{S}[Y])s_{\theta}](Z,X)}
				{\mathbb{E}[\{\pi(Z,X)-\mathbb{S}[\pi](Z,X)\}\{A-\mathbb{S}[A](Z,X)\}]}\right]\tag{\text{vanish}}\\
				&-\dfrac
				{\mathbb{E}\left[\{\pi(Z,X)-\mathbb{S}[\pi](Z,X)\}\{Y-\mathbb{S}[Y](Z,X)\}\right]}
				{\mathbb{E}[\{\pi(Z,X)-\mathbb{S}[\pi](Z,X)\}\{A-\mathbb{S}[A](Z,X)\}]^2}\\
				&\times\nabla_{\theta}
				\mathbb{E}^{(\theta)}[\{\pi(Z,X)-\mathbb{S}^{(\theta)}[\pi](Z,X)\}\{A-\mathbb{S}^{(\theta)}[A](Z,X)\}]\eval_{\theta=0}\\
				=&\mathbb{E}\left[\left\{\dfrac
				{\{\pi(Z,X)-\mathbb{S}[\pi](Z,X)\}\{Y-\mathbb{S}[Y](Z,X)\}}
				{\mathbb{E}[\{\pi(Z,X)-\mathbb{S}[\pi](Z,X)\}\{A-\mathbb{S}[A](Z,X)\}]}-\beta_{\mathcal{S},\pi}^o\right\}s_{\theta}(O)\right]\\
				&-\beta_{\mathcal{S},\pi}^o
				\mathbb{E}\left[\left\{\dfrac{
					\{\pi(Z,X)-\mathbb{S}[\pi](Z,X)\}\{A-\mathbb{S}[A](Z,X)\}
				}{\mathbb{E}[\{\pi(Z,X)-\mathbb{S}[\pi](Z,X)\}\{A-\mathbb{S}[A](Z,X)\}]}-1\right\}s_{\theta}(O)\right]\\
				=&\mathbb{E}\left[\left\{\begin{array}{l}
					\dfrac
					{\{\pi(Z,X)-\mathbb{S}[\pi](Z,X)\}\{Y-\mathbb{S}[Y](Z,X)\}}
					{\mathbb{E}[\{\pi(Z,X)-\mathbb{S}[\pi](Z,X)\}\{A-\mathbb{S}[A](Z,X)\}]}\\
					-\dfrac{
						\{\pi(Z,X)-\mathbb{S}[\pi](Z,X)\}\{A-\mathbb{S}[A](Z,X)\}
					}{\mathbb{E}[\{\pi(Z,X)-\mathbb{S}[\pi](Z,X)\}\{A-\mathbb{S}[A](Z,X)\}]}\beta_{\mathcal{S},\pi}^o
				\end{array}\right\}s_{\theta}(O)\right].
			\end{align*}
			Thus, the EIF of $\beta_{\mathcal{S},\pi}^o$ is 
			\begin{align*}
				\phi_{\mathcal{S},\pi}(O;\beta_{\mathcal{S},\pi}^o,\alpha_{\mathcal{S},\pi}^o,\tau_{\mathcal{S},\pi}^o)
				=&\dfrac
				{\{\pi(Z,X)-\rho_{\mathcal{S},\pi}^o(Z,X)\}\{Y-\eta_{\mathcal{S}}^o(Z,X)-(A-\delta_{\mathcal{S}}^o(Z,X))\beta_{\mathcal{S},\pi}^o\}}
				{\tau_{\mathcal{S},\pi}^o},\text{ where }\\
                \tau_{\mathcal{S},\pi}^o=&\mathbb{E}[\{\pi(Z,X)-\rho_{\mathcal{S},\pi}^o(Z,X)\}\{A-\delta_{\mathcal{S}}^o(Z,X)\}].
			\end{align*}
			Next, by repeatedly using Lemma~\ref{lem: semiparametric calculation}, we calculate the EIF of $\beta_{\mathcal{S},*}^o$ as
			\begin{align*}
				\nabla_{\theta}\beta_{\mathcal{S},*}^{(\theta)}\eval_{\theta=0}
				=&\nabla_{\theta}\left(\dfrac{\mathbb{E}^{(\theta)}[\{\mathbb{E}^{(\theta)}[A|Z,X]-\mathbb{S}^{(\theta)}[A](Z,X)\}\{Y-\mathbb{S}^{(\theta)}[Y](Z,X)\}]}{\mathbb{E}^{(\theta)}[\{\mathbb{E}^{(\theta)}[A|Z,X]-\mathbb{S}^{(\theta)}[A](Z,X)\}^2]}\right)\eval_{\theta=0}\\
				=&\mathbb{E}\left[\left\{\dfrac
				{\{\pi_*^o(Z,X)-\mathbb{S}[A](Z,X)\}\{Y-\mathbb{S}[Y](Z,X)\}}
				{\mathbb{E}[\{\pi_*^o(Z,X)-\mathbb{S}[A](Z,X)\}^2]}-\beta_{\mathcal{S},*}^o\right\}s_{\theta}(O)\right]\\
				&+\mathbb{E}\left[\dfrac
				{\mathbb{E}[\{A-\pi_*^o(Z,X)\}s_{\theta}(O)\mid Z,X]\{Y-\mathbb{S}[Y](Z,X)\}}
				{\mathbb{E}[\{\pi_*^o(Z,X)-\mathbb{S}[A](Z,X)\}^2]}\right]\\
				&-\dfrac{\beta_{\mathcal{S},*}^o}{\mathbb{E}[\{\pi_*^o(Z,X)-\mathbb{S}[A](Z,X)\}^2]}\nabla_{\theta}\mathbb{E}^{(\theta)}\left[\{\mathbb{E}^{(\theta)}[A|Z,X]-\mathbb{S}^{(\theta)}[A](Z,X)\}^2\right]\eval_{\theta=0}\\
				=&\mathbb{E}\left[\left\{\dfrac
				{\{\pi_*^o(Z,X)-\mathbb{S}[A](Z,X)\}\{Y-\mathbb{S}[Y](Z,X)\}}
				{\mathbb{E}[\{\pi_*^o(Z,X)-\mathbb{S}[A](Z,X)\}^2]}-\beta_{\mathcal{S},*}^o\right\}s_{\theta}(O)\right]\\
				&+\mathbb{E}\left[\dfrac
				{\{A-\pi_*^o(Z,X)\}\{\mathbb{E}[Y\mid Z,X]-\mathbb{S}[Y](Z,X)\}}
				{\mathbb{E}[\{\pi_*^o(Z,X)-\mathbb{S}[A](Z,X)\}^2]}s_{\theta}(O)\right]\\
				&-\beta_{\mathcal{S},*}^o\left\{\begin{array}{l}
					\mathbb{E}\left[\left\{\dfrac
					{\{\pi_*^o(Z,X)-\mathbb{S}[A](Z,X)\}\{A-\mathbb{S}[A](Z,X)\}}
					{\mathbb{E}[\{\pi_*^o(Z,X)-\mathbb{S}[A](Z,X)\}^2]}-1\right\}s_{\theta}(O)\right]\\
					+\mathbb{E}\left[\dfrac
					{\{A-\pi_*^o(Z,X)\}\{\pi_*^o(Z,X)-\mathbb{S}[A](Z,X)\}}
					{\mathbb{E}[\{\pi_*^o(Z,X)-\mathbb{S}[A](Z,X)\}^2]}s_{\theta}(O)\right]
				\end{array}\right\}\\
				=&\mathbb{E}\left[\dfrac
				{\{\pi_*^o(Z,X)-\mathbb{S}[A](Z,X)\}\{Y-\mathbb{S}[Y](Z,X)\}}
				{\mathbb{E}[\{\pi_*^o(Z,X)-\mathbb{S}[A](Z,X)\}^2]}s_{\theta}(O)\right]\\
				&+\mathbb{E}\left[\dfrac
				{\{A-\pi_*^o(Z,X)\}\{\xi_*^o(Z,X)-\mathbb{S}[Y](Z,X)\}}
				{\mathbb{E}[\{\pi_*^o(Z,X)-\mathbb{S}[A](Z,X)\}^2]}s_{\theta}(O)\right]\\
				&-\beta_{\mathcal{S},*}^o\mathbb{E}\left[\left\{\begin{array}{l}
					\dfrac
					{\{\pi_*^o(Z,X)-\mathbb{S}[A](Z,X)\}\{A-\mathbb{S}[A](Z,X)\}}
					{\mathbb{E}[\{\pi_*^o(Z,X)-\mathbb{S}[A](Z,X)\}^2]}\\
					+\dfrac
					{\{A-\pi_*^o(Z,X)\}\{\pi_*^o(Z,X)-\mathbb{S}[A](Z,X)\}}
					{\mathbb{E}[\{\pi_*^o(Z,X)-\mathbb{S}[A](Z,X)\}^2]}
				\end{array}\right\}s_{\theta}(O)\right].
			\end{align*}
			Thus, define $\tau_{\mathcal{S},*}^o:=\mathbb{E}[\{\pi_*^o(Z,X)-\mathbb{S}[A](Z,X)\}^2]$, and the EIF of $\beta_{\mathcal{S},*}^o$ is 
			\begin{align*}
				&\phi_{\mathcal{S},*}(O;\beta_{\mathcal{S},*}^o,\alpha_{\mathcal{S},*}^o,\tau_{\mathcal{S},*}^o)\\
				=&\dfrac{1}{\tau_{\mathcal{S},*}^o}\{\pi_*^o(Z,X)-\mathbb{S}[A](Z,X)\}\{Y-\mathbb{S}[Y](Z,X)-\beta_{\mathcal{S},*}^o(A-\mathbb{S}[A](Z,X))\}\\
				&+\dfrac{1}{\tau_{\mathcal{S},*}^o}\{A-\pi_*^o(Z,X)\}\{\xi_*^o(Z,X)-\mathbb{S}[Y](Z,X)-\beta_{\mathcal{S},*}^o(\pi_*^o(Z,X)-\mathbb{S}[A](Z,X))\}\\
				=&\dfrac{1}{\tau_{\mathcal{S},*}^o}\{\pi_*^o(Z,X)-\delta_{\mathcal{S}}^o(Z,X)\}\{Y-\eta_{\mathcal{S}}^o(Z,X)-\beta_{\mathcal{S},*}^o(A-\delta_{\mathcal{S}}^o(Z,X))\}\\
				&+\dfrac{1}{\tau_{\mathcal{S},*}^o}\{A-\pi_*^o(Z,X)\}\{\xi_*^o(Z,X)-\eta_{\mathcal{S}}^o(Z,X)-\beta_{\mathcal{S},*}^o(\pi_*^o(Z,X)-\delta_{\mathcal{S}}^o(Z,X))\}.
			\end{align*}
			This completes the proof of Theorem~\ref{thm: EIF partially linear effect}.

			\subsection{Proof of Theorem~\ref{thm: identification of MPO}}
			First, we prove the results under Assumptions~\ref{as: IV independence}, \ref{as: strong latent ignorability}, \ref{as: Z-invariant treatment effect'}, and \ref{as: alternative'}(a). Specifically, for any $\gamma\leq \gamma_{\dagger,a}(\nu_0)$,
			\begin{align*}
				&\mathbb{E}\left[\{\pi(Z,X) - \mathbb{S}_{\gamma}[\pi](Z,X)\}I(A=a)Y\mid U,Z_{[-\nu_0]},X\right]\\
				=&\mathbb{E}\left[\{\pi(Z,X) - \mathbb{S}_{\gamma}[\pi](Z,X)\}\Pr(A=a\mid Z,U,X)Y(a)\mid U,Z_{[-\nu_0]},X\right]\\
				=&\mathbb{E}\left[\{\pi(Z,X) - \mathbb{S}_{\gamma}[\pi](Z,X)\}\left\{\sum_{\nu: |\nu|=\gamma,\nu\subseteq \nu_0}b_{\nu,a}(U,Z_{[-\nu]},X) + c_a(Z,X)\right\}Y(a)\mid U,Z_{[-\nu_0]},X\right]\tag{By Assumption~\ref{as: alternative'}(a)}\\
				=&\sum_{\nu: |\nu|=\gamma,\nu\subseteq \nu_0}\mathbb{E}\left[b_{\nu,a}(U,Z_{[-\nu]},X)\{\pi(Z,X) - \mathbb{S}_{\gamma}[\pi](Z,X)\}Y(a)\mid U,Z_{[-\nu_0]},X\right]\\
				&+\mathbb{E}\left[\{\pi(Z,X) - \mathbb{S}_{\gamma}[\pi](Z,X)\}c_a(Z,X)Y(a)\mid U,Z_{[-\nu_0]},X\right]\\
				=&\sum_{\nu: |\nu|=\gamma,\nu\subseteq \nu_0}\mathbb{E}\left[b_{\nu,a}(U,Z_{[-\nu]},X)\{\pi(Z,X) - \mathbb{S}_{\gamma}[\pi](Z,X)\}\mid U,Z_{[-\nu_0]},X\right]\mathbb{E}\left[Y(a)\mid U,Z_{[-\nu_0]},X\right]\\
				&+\mathbb{E}\left[\{\pi(Z,X) - \mathbb{S}_{\gamma}[\pi](Z,X)\}c_a(Z,X)\mid U,Z_{[-\nu_0]},X\right]\mathbb{E}\left[Y(a)\mid U,Z_{[-\nu_0]},X\right]\tag{By Assumption~\ref{as: strong latent ignorability}}\\
				=&\sum_{\nu: |\nu|=\gamma,\nu\subseteq \nu_0}\mathbb{E}\left[b_{\nu,a}(U,Z_{[-\nu]},X)\mathbb{E}\left[\pi(Z,X) - \mathbb{S}_{\gamma}[\pi](Z,X)\middle| U,Z_{[-\nu]},X\right]\mid U,Z_{[-\nu_0]},X\right]\mathbb{E}\left[Y(a)\mid U,Z_{[-\nu_0]},X\right]\\
				&+\mathbb{E}\left[\{\pi(Z,X) - \mathbb{S}_{\gamma}[\pi](Z,X)\}c_a(Z,X)\mid Z_{[-\nu_0]},X\right]\mathbb{E}\left[Y(a)\mid U,Z_{[-\nu_0]},X\right]\\
				=&\sum_{\nu: |\nu|=\gamma,\nu\subseteq \nu_0}\mathbb{E}\left[b_{\nu,a}(U,Z_{[-\nu]},X)\mathbb{E}\left[\pi(Z,X) - \mathbb{S}_{\gamma}[\pi](Z,X)\middle| Z_{[-\nu]},X\right]\mid U,Z_{[-\nu_0]},X\right]\mathbb{E}\left[Y(a)\mid U,Z_{[-\nu_0]},X\right]\\
				&+\mathbb{E}\left[\{\pi(Z,X) - \mathbb{S}_{\gamma}[\pi](Z,X)\}c_a(Z,X)\mid Z_{[-\nu_0]},X\right]\mathbb{E}\left[Y(a)\mid U,Z_{[-\nu_0]},X\right]\tag{By Assumption~\ref{as: IV independence} that $U\indep Z_{[\nu_0]}\mid Z_{[-\nu_0]},X\Longrightarrow U\indep Z_{[\nu]}\mid Z_{[-\nu]},X$ for $\nu\subseteq \nu_0$}\\
				=&\mathbb{E}\left[\{\pi(Z,X) - \mathbb{S}_{\gamma}[\pi](Z,X)\}c_a(Z,X)\mid Z_{[-\nu_0]},X\right]\mathbb{E}\left[Y(a)\mid U,Z_{[-\nu_0]},X\right].
			\end{align*}
			The final step follows from $\pi-\mathbb{S}_{\gamma}[\pi] \in\mathcal{S}_{\gamma}^{\perp}$, so that $\mathbb{E}\left[\{\pi(Z,X) - \mathbb{S}_{\gamma}[\pi](Z,X)\}\mid Z_{[-\nu]},X\right]\equiv 0$ with $\gamma=|\nu|$.
			Next, we integrate $U$ out to deduce that
			\begin{align*}
				&\mathbb{E}\left[\{\pi(Z,X) - \mathbb{S}_{\gamma}[\pi](Z,X)\}I(A=a)Y\mid Z_{[-\nu_0]},X\right]\\
				=&\mathbb{E}\left[\{\pi(Z,X) - \mathbb{S}_{\gamma}[\pi](Z,X)\}c_a(Z,X)\mid Z_{[-\nu_0]},X\right]\mathbb{E}\left[Y(a)\mid Z_{[-\nu_0]},X\right]\\
				=&\mathbb{E}\left[\{\pi(Z,X) - \mathbb{S}_{\gamma}[\pi](Z,X)\}c_a(Z,X)\mid Z_{[-\nu_0]},X\right]\mathbb{E}\left[Y(a)\mid X\right]\tag{By Assumption~\ref{as: Z-invariant treatment effect'}}\\
				=&\mathbb{E}\left[\{\pi(Z,X) - \mathbb{S}_{\gamma}[\pi](Z,X)\}I\{A=a\}\mid Z_{[-\nu_0]},X\right]\mathbb{E}\left[Y(a)\mid X\right].
			\end{align*}
			The final step follows from the same reason as in the proof of Proposition~\ref{prop: invalid IV}. Now, we integrate $Z_{[-\nu_0]}$ out to deduce that
			\begin{align*}
				&\mathbb{E}\left[\{\pi(Z,X) - \mathbb{S}_{\gamma}[\pi](Z,X)\}I\{A=a\}Y\mid X\right]\\
				=&\mathbb{E}\left[\{\pi(Z,X) - \mathbb{S}_{\gamma}[\pi](Z,X)\}I\{A=a\}\mid X\right]\mathbb{E}\left[Y(a)\mid X\right].
			\end{align*}
			This indicates that
			\begin{align*}
				\mathbb{E}\left[Y(a)\right]=\mathbb{E}\left[\dfrac{
					\{\pi(Z,X) - \mathbb{S}_{\gamma}[\pi](Z,X)\}I\{A=a\}Y
				}{\mathbb{E}\left[\{\pi(Z,X) - \mathbb{S}_{\gamma}[\pi](Z,X)\}I\{A=a\}\mid X\right]
				}\right].
			\end{align*}
			Alternatively, under Assumptions~\ref{as: IV independence}, \ref{as: strong latent ignorability}, \ref{as: Z-invariant treatment effect'} and \ref{as: alternative'}(b),
			\begin{align*}
				&\mathbb{E}\left[\{\pi(Z,X) - \mathbb{S}_{\gamma}[\pi](Z,X)\}I(A=a)Y\mid U,Z_{[-\nu_0]},X\right]\\
				=&\mathbb{E}\left[\{\pi(Z,X) - \mathbb{S}_{\gamma}[\pi](Z,X)\}I(A=a)\mid U,Z_{[-\nu_0]},X\right]\mathbb{E}\left[Y(a)\mid U,Z_{[-\nu_0]},X\right]\tag{By Assumption~\ref{as: strong latent ignorability}}\\
				=&\mathbb{E}\left[\{\pi(Z,X) - \mathbb{S}_{\gamma}[\pi](Z,X)\}I(A=a)\mid U,Z_{[-\nu_0]},X\right]\mathbb{E}\left[Y(a)\mid Z_{[-\nu_0]},X\right]\tag{By Assumption \ref{as: alternative'}(b)}\\
				=&\mathbb{E}\left[\{\pi(Z,X) - \mathbb{S}_{\gamma}[\pi](Z,X)\}I(A=a)\mid U,Z_{[-\nu_0]},X\right]\mathbb{E}\left[Y(a)\mid X\right].\tag{By Assumption~\ref{as: Z-invariant treatment effect'}}
			\end{align*} 
			This completes the proof of Theorem~\ref{thm: identification of MPO}.

			\subsection{Proof of Theorem~\ref{thm: identification ATT}}
			Under Assumptions~\ref{as: IV independence}, \ref{as: strong latent ignorability}, \ref{as: Z-invariant treatment effect''}, and \ref{as: alternative''}(a), 
			\begin{align*}
				&\mathbb{E}\left[\{\pi(Z,X) - \mathbb{S}_{\gamma}[\pi](Z,X)\}(A-1)Y\middle| U,Z_{[-\nu_0]},X\right]\notag\\
				=&\mathbb{E}\left[\{\pi(Z,X) - \mathbb{S}_{\gamma}[\pi](Z,X)\}(A-1)Y(0)\middle| U,Z_{[-\nu_0]},X\right]\notag\\
				=&\mathbb{E}\left[\{\pi(Z,X) - \mathbb{S}_{\gamma}[\pi](Z,X)\}(A-1)\middle| U,Z_{[-\nu_0]},X\right]\times
				\mathbb{E}\left[Y(0)\middle| U,Z_{[-\nu_0]},X\right]
				\tag{\text{By Assumption~\ref{as: strong latent ignorability}}}\\
				=&\mathbb{E}\left[\{\pi(Z,X) - \mathbb{S}_{\gamma}[\pi](Z,X)\}A\middle| U,Z_{[-\nu_0]},X\right]\times
				\mathbb{E}\left[Y(0)\middle| U,Z_{[-\nu_0]},X\right]\\
				&-\mathbb{E}\left[\{\pi(Z,X) - \mathbb{S}_{\gamma}[\pi](Z,X)\}\middle| Z_{[-\nu_0]},X\right]\times
				\mathbb{E}\left[Y(0)\middle| U,Z_{[-\nu_0]},X\right]
				\tag{\text{By Assumption~\ref{as: IV independence}}}\\
				=&\mathbb{E}\left[\{\pi(Z,X) - \mathbb{S}_{\gamma}[\pi](Z,X)\}A\middle| U,Z_{[-\nu_0]},X\right]\times
				\mathbb{E}\left[Y(0)\middle| U,Z_{[-\nu_0]},X\right].
				\tag{By $\pi - \mathbb{S}_{\gamma}[\pi]\in \mathcal{S}_{\gamma}^{\perp}$.}
			\end{align*}
			By Assumption~\ref{as: alternative''}(a), there exist functions $b(U,Z_{[-\nu_0]},X)$ and $c(Z,X)$ such that
			\begin{align*}
				&\Pr(A=1\mid Z, U, X) = b(U,Z_{[-\nu_0]}, X) \times c(Z, X).\\
				\Rightarrow\qquad&
				\mathbb{E}[A\mid Z,X] = 
				\mathbb{E}[b(U,Z_{[-\nu_0]}, X)\mid Z,X]\times c(Z, X)\\
				=&\mathbb{E}[b(U,Z_{[-\nu_0]}, X)\mid Z_{[-\nu_0]},X]\times c(Z, X).
				\tag{By Assumption~\ref{as: IV independence}}\\
				\Rightarrow\qquad&
				\dfrac{\mathbb{E}[A\mid U,Z_{[-\nu_0]},X]}{\mathbb{E}[A\mid Z_{[-\nu_0]},X]}
				=\dfrac{b(U,Z_{[-\nu_0]}, X)}{\mathbb{E}[b(U,Z_{[-\nu_0]}, X)\mid Z_{[-\nu_0]},X]}.
			\end{align*}
			Consequently,
			\begin{align*}
				&\mathbb{E}\left[\{\pi(Z,X) - \mathbb{S}_{\gamma}[\pi](Z,X)\}(A-1)Y\middle| U,Z_{[-\nu_0]},X\right]\\
				=&b(U,Z_{[-\nu_0]}, X)\mathbb{E}\left[\{\pi(Z,X) - \mathbb{S}_{\gamma}[\pi](Z,X)\}c(Z, X)\middle| U,Z_{[-\nu_0]},X\right]\times
				\mathbb{E}\left[Y(0)\middle| U,Z_{[-\nu_0]},X\right]\\
				=&b(U,Z_{[-\nu_0]}, X)\mathbb{E}\left[\{\pi(Z,X) - \mathbb{S}_{\gamma}[\pi](Z,X)\}c(Z, X)\middle| Z_{[-\nu_0]},X\right]\times
				\mathbb{E}\left[Y(0)\middle| U,Z_{[-\nu_0]},X\right]\tag{By Assumption~\ref{as: IV independence}}\\
				=&\mathbb{E}\left[\{\pi(Z,X) - \mathbb{S}_{\gamma}[\pi](Z,X)\}A\middle| Z_{[-\nu_0]},X\right]\times\dfrac{b(U,Z_{[-\nu_0]}, X)\mathbb{E}[Y(0)\mid U,Z_{[-\nu_0]}, X]}{\mathbb{E}[b(U,Z_{[-\nu_0]}, X)\mid Z_{[-\nu_0]},X]}\\
				=&\mathbb{E}\left[\{\pi(Z,X) - \mathbb{S}_{\gamma}[\pi](Z,X)\}A\middle| Z_{[-\nu_0]},X\right]\times
				\dfrac{\mathbb{E}[A\mid U,Z_{[-\nu_0]},X]}{\mathbb{E}[A\mid Z_{[-\nu_0]},X]}\mathbb{E}[Y(0)\mid U,Z_{[-\nu_0]}, X]\\
				=&\mathbb{E}\left[\{\pi(Z,X) - \mathbb{S}_{\gamma}[\pi](Z,X)\}A\middle| Z_{[-\nu_0]},X\right]\times
				\dfrac{\mathbb{E}[AY(0)\mid U,Z_{[-\nu_0]},X]}{\mathbb{E}[A\mid Z_{[-\nu_0]},X]}.\tag{By Assumption~\ref{as: strong latent ignorability}}
			\end{align*}
			Therefore, we can integrate $U$ out to deduce that
			\begin{align*}
				&\mathbb{E}\left[\{\pi(Z,X) - \mathbb{S}_{\gamma}[\pi](Z,X)\}(A-1)Y\middle| Z_{[-\nu_0]},X\right]\\
				=&\mathbb{E}\left[\{\pi(Z,X) - \mathbb{S}_{\gamma}[\pi](Z,X)\}A\middle| Z_{[-\nu_0]},X\right]\times
				\dfrac{\mathbb{E}[AY(0)\mid Z_{[-\nu_0]},X]}{\mathbb{E}[A\mid Z_{[-\nu_0]},X]}\\
				=&\mathbb{E}\left[\{\pi(Z,X) - \mathbb{S}_{\gamma}[\pi](Z,X)\}A\middle| Z_{[-\nu_0]},X\right]\times
				\mathbb{E}[Y(0)\mid A=1,Z_{[-\nu_0]},X]\\
				=&\mathbb{E}\left[\{\pi(Z,X) - \mathbb{S}_{\gamma}[\pi](Z,X)\}A\middle| Z_{[-\nu_0]},X\right]\times
				\mathbb{E}[Y(0)\mid A=1,X].\tag{By Assumption~\ref{as: Z-invariant treatment effect''}}
			\end{align*}
			Next, we can integrate $Z_{[-\nu_0]}$ out to deduce that
			\begin{align*}
				&\mathbb{E}\left[\{\pi(Z,X) - \mathbb{S}_{\gamma}[\pi](Z,X)\}(A-1)Y\middle| X\right]\\
				=&\mathbb{E}\left[\{\pi(Z,X) - \mathbb{S}_{\gamma}[\pi](Z,X)\}A\middle| X\right]\times
				\mathbb{E}[Y(0)\mid A=1,X]\\
				\Longrightarrow\qquad&
				\mathbb{E}[Y(0)\mid A=1,X]=\mathbb{E}\left[\dfrac{\{\pi(Z,X) - \mathbb{S}_{\gamma}[\pi](Z,X)\}(A-1)}{\mathbb{E}\left[\{\pi(Z,X) - \mathbb{S}_{\gamma}[\pi](Z,X)\}A\middle| X\right]}Y\middle| X\right].
			\end{align*}
			Alternatively, under Assumption~\ref{as: alternative''}(b), 
			\begin{align*}
				&\mathbb{E}\left[\{\pi(Z,X) - \mathbb{S}_{\gamma}[\pi](Z,X)\}(A-1)Y\middle| U,Z_{[-\nu_0]},X\right]\\
				=&\mathbb{E}\left[\{\pi(Z,X) - \mathbb{S}_{\gamma}[\pi](Z,X)\}A\middle| U,Z_{[-\nu_0]},X\right]\times
				\mathbb{E}\left[Y(0)\middle| U,Z_{[-\nu_0]},X\right]\\
				=&\mathbb{E}\left[\{\pi(Z,X) - \mathbb{S}_{\gamma}[\pi](Z,X)\}A\middle| U,Z_{[-\nu_0]},X\right]\times
				\mathbb{E}\left[Y(0)\middle| A=1,U,Z_{[-\nu_0]},X\right]\tag{By Assumption~\ref{as: strong latent ignorability}}\\
				=&\mathbb{E}\left[\{\pi(Z,X) - \mathbb{S}_{\gamma}[\pi](Z,X)\}A\middle| U,Z_{[-\nu_0]},X\right]\times
				\mathbb{E}\left[Y(0)\middle| A=1,Z_{[-\nu_0]},X\right]\tag{By Assumption~\ref{as: alternative''}(b)}\\
				=&\mathbb{E}\left[\{\pi(Z,X) - \mathbb{S}_{\gamma}[\pi](Z,X)\}A\middle| U,Z_{[-\nu_0]},X\right]\times
				\mathbb{E}\left[Y(0)\middle| A=1,X\right].\tag{By Assumption~\ref{as: Z-invariant treatment effect''}}
			\end{align*}
			Now we can integrate $(U,Z_{[-\nu_0]})$ out to deduce that under Assumptions~\ref{as: IV independence}--\ref{as: strong latent ignorability}, \ref{as: Z-invariant treatment effect''} and \ref{as: alternative''},
			\begin{align*}
				&\mathbb{E}\left[\{\pi(Z,X) - \mathbb{S}_{\gamma}[\pi](Z,X)\}(A-1)Y\middle|X\right]\\
				=&\mathbb{E}\left[\{\pi(Z,X) - \mathbb{S}_{\gamma}[\pi](Z,X)\}A\middle| X\right]\times
				\mathbb{E}\left[Y(0)\middle| A=1,X\right].
			\end{align*}
			
			As a result, we deduce that
			\begin{align*}
				&\mathbb{E}[Y(0)\mid A=1]
				=\mathbb{E}\left[\mathbb{E}\left[Y(0)\middle| A=1,X\right]\middle| A=1\right]\\
				=&\mathbb{E}\left[\mathbb{E}\left[\dfrac{\{\pi(Z,X) - \mathbb{S}_{\gamma}[\pi](Z,X)\}(A-1)}{\mathbb{E}\left[\{\pi(Z,X) - \mathbb{S}_{\gamma}[\pi](Z,X)\}A\middle| X\right]}Y\middle| X\right]\middle| A=1\right].
			\end{align*}
			This completes the proof of Theorem~\ref{thm: identification ATT}.

			\subsection{Proof of Theorem~\ref{thm: DGMM}}
			By the definition of $\hat{\beta}_{\mathcal{S}}$, we know that
			\begin{align*}
				\left\{\nabla_{\beta}\frac{1}{K}\sum_{k=1}^K\mathbb{E}_{nk}[\hat\varphi_{\mathcal{S},\Pi}^{(n,-k)}(\hat{\beta}_{\mathcal{S}})] \right\}^\top\hat\Sigma_{\mathcal{S}}\left\{\frac{1}{K}\sum_{k=1}^K\mathbb{E}_{nk}[\hat\varphi_{\mathcal{S},\Pi}^{(n,-k)}(\hat{\beta}_{\mathcal{S}})] \right\}
				=0.
			\end{align*}
			Since $\nabla_{\beta}\hat\varphi_{\mathcal{S},\Pi}^{(n,-k)}(\hat{\beta}_{\mathcal{S}})=-\left\{\Pi(Z,X)-\hat\rho_{\mathcal{S},\Pi}^{(n,-k)}(Z,X)\right\}\left\{A - \hat\delta_{\mathcal{S}}^{(n,-k)}(Z,X)\right\}$, this indicates that
			\begin{align*}
				&(\hat q_{\mathcal{S}})^\top\hat\Sigma_{\mathcal{S}}\left\{\frac{1}{K}\sum_{k=1}^K\mathbb{E}_{nk}[\hat\varphi_{\mathcal{S},\Pi}^{(n,-k)}(\hat{\beta}_{\mathcal{S}})] \right\}=0,\text{ where }\\
				&\hat q_{\mathcal{S}}:=\frac{1}{K}\sum_{k=1}^K\mathbb{E}_{nk}\left[\left\{\Pi(Z,X)-\hat\rho_{\mathcal{S},\Pi}^{(n,-k)}(Z,X)\right\}\left\{A - \hat\delta_{\mathcal{S}}^{(n,-k)}(Z,X)\right\}\right].
			\end{align*}
			Therefore, by expanding $\mathbb{E}_{nk}[\hat\varphi_{\mathcal{S},\Pi}^{(n,-k)}(\hat{\beta}_{\mathcal{S}})]$, we know that
			\begin{align*}
				&\hat\beta_{\mathcal{S}}=\left\{(\hat q_{\mathcal{S}})^\top\hat\Sigma_{\mathcal{S}}\hat q_{\mathcal{S}}\right\}^{-1}(\hat q_{\mathcal{S}})^\top\hat\Sigma_{\mathcal{S}}\hat r_{\mathcal{S}}
				,\text{ where }\\
				&\hat r_{\mathcal{S}}:=\frac{1}{K}\sum_{k=1}^K\mathbb{E}_{nk}\left[\left\{\Pi(Z,X)-\hat\rho_{\mathcal{S},\Pi}^{(n,-k)}(Z,X)\right\}\left\{Y-\hat\eta_{\mathcal{S}}^{(n,-k)}(Z,X)\right\}\right].
			\end{align*}
			Now we only need to prove that
			\begin{equation}\label{exp: 11}
				\begin{aligned}
					&\hat q_{\mathcal{S}} = \mathbb{E}_{n}\left[\left\{\Pi(Z,X)-\rho_{\mathcal{S},\Pi}^o(Z,X)\right\}\left\{A - \delta_{\mathcal{S}}^o(Z,X)\right\}\right]+o_p(n^{-1/2}),\\
					&\hat\Sigma_{\mathcal{S}}=\Sigma_{\mathcal{S}}+o_p(1),\\
					&\hat r_{\mathcal{S}}=\mathbb{E}_n\left[\left\{\Pi(Z,X)-\rho_{\mathcal{S},\Pi}^o(Z,X)\right\}\left\{Y-\eta_{\mathcal{S}}^o(Z,X)\right\}\right]+o_p(n^{-1/2}).
				\end{aligned}
			\end{equation}
			By Equation~\eqref{exp: 11}, one may use Slutsky's theorem to deduce that
			\begin{align*}
				&\hat\beta_{\mathcal{S}}-\beta^o=\left\{(\hat q_{\mathcal{S}})^\top\hat\Sigma_{\mathcal{S}}\hat q_{\mathcal{S}}\right\}^{-1}(\hat q_{\mathcal{S}})^\top\hat\Sigma_{\mathcal{S}}\left\{\hat r_{\mathcal{S}}-\hat q_{\mathcal{S}}\beta^o\right\}\\
				=&\left\{(\hat q_{\mathcal{S}})^\top\hat\Sigma_{\mathcal{S}}\hat q_{\mathcal{S}}\right\}^{-1}(\hat q_{\mathcal{S}})^\top\hat\Sigma_{\mathcal{S}}\left\{\begin{array}{l}
					\mathbb{E}_n\left[\left\{\Pi(Z,X)-\rho_{\mathcal{S},\Pi}^o(Z,X)\right\}\left\{Y-\eta_{\mathcal{S}}^o(Z,X)\right\}\right]\\
					-\mathbb{E}_{n}\left[\left\{\Pi(Z,X)-\rho_{\mathcal{S},\Pi}^o(Z,X)\right\}\left\{A - \delta_{\mathcal{S}}^o(Z,X)\right\}\right]\beta^o + o_p(n^{-1/2})
				\end{array}\right\}\\
				=&\left\{(\hat q_{\mathcal{S}})^\top\hat\Sigma_{\mathcal{S}}\hat q_{\mathcal{S}}\right\}^{-1}(\hat q_{\mathcal{S}})^\top\hat\Sigma_{\mathcal{S}}
				\left\{\mathbb{E}_n\left[\begin{array}{l}
					\left\{\Pi(Z,X)-\rho_{\mathcal{S},\Pi}^o(Z,X)\right\}\\
					\times \left\{Y-\eta_{\mathcal{S}}^o(Z,X)-\{A-\delta_{\mathcal{S}}^o(Z,X)\}\beta^o\right\}
				\end{array}\right]+o_p(n^{-1/2})\right\}\\
				=&\left\{q_{\mathcal{S}}^\top\Sigma_{\mathcal{S}}q_{\mathcal{S}} + o_p(1)\right\}^{-1}\{q_{\mathcal{S}}^\top\Sigma_{\mathcal{S}}+o_p(1)\}
				\left\{\mathbb{E}_n\left[\begin{array}{l}
					\left\{\Pi(Z,X)-\rho_{\mathcal{S},\Pi}^o(Z,X)\right\}\\
					\times \left\{Y-\eta_{\mathcal{S}}^o(Z,X)-\{A-\delta_{\mathcal{S}}^o(Z,X)\}\beta^o\right\}
				\end{array}\right]+o_p(n^{-1/2})\right\}.
			\end{align*}
			This indicates that 
			\begin{align*}
				&\sqrt{n}\{\hat\beta_{\mathcal{S}}-\beta^o\}=\sqrt{n}\left\{(\hat q_{\mathcal{S}})^\top\hat\Sigma_{\mathcal{S}}\hat q_{\mathcal{S}}\right\}^{-1}(\hat q_{\mathcal{S}})^\top\hat\Sigma_{\mathcal{S}}\left\{\hat r_{\mathcal{S}}-\hat q_{\mathcal{S}}\beta^o\right\}\\
				=&\sqrt{n}\left\{q_{\mathcal{S}}^\top\Sigma_{\mathcal{S}}q_{\mathcal{S}}\right\}^{-1}q_{\mathcal{S}}^\top\Sigma_{\mathcal{S}}
				\mathbb{E}_n\left[\begin{array}{l}
					\left\{\Pi(Z,X)-\rho_{\mathcal{S},\Pi}^o(Z,X)\right\}\\
					\times \left\{Y-\eta_{\mathcal{S}}^o(Z,X)-\{A-\delta_{\mathcal{S}}^o(Z,X)\}\beta^o\right\}
				\end{array}\right]+o_p(1).
			\end{align*}
			The results then follow from the central limit theorem.
			
			\begin{proof}[Proof of Equation~\eqref{exp: 11}:]
				We directly calculate that
				\begin{align*}
					&\mathbb{E}_{nk}\left[\left\{\Pi(Z,X)-\hat\rho_{\mathcal{S},\Pi}^{(n,-k)}(Z,X)\right\}\left\{A - \hat\delta_{\mathcal{S}}^{(n,-k)}(Z,X)\right\}\right]-q_{\mathcal{S}}\\
					=&\{\mathbb{E}_{nk}-\mathbb{E}_k\}\left[\left\{\Pi(Z,X)-\hat\rho_{\mathcal{S},\Pi}^{(n,-k)}(Z,X)\right\}\left\{A - \hat\delta_{\mathcal{S}}^{(n,-k)}(Z,X)\right\}\right]\\
					&+\mathbb{E}_k\left[\left\{\Pi(Z,X)-\hat\rho_{\mathcal{S},\Pi}^{(n,-k)}(Z,X)\right\}\left\{A - \hat\delta_{\mathcal{S}}^{(n,-k)}(Z,X)\right\}\right]\\
					&-\mathbb{E}_k\left[\left\{\Pi(Z,X)-\rho_{\mathcal{S},\Pi}^o(Z,X)\right\}\left\{A - \delta_{\mathcal{S}}^o(Z,X)\right\}\right]\\
					=&\{\mathbb{E}_{nk}-\mathbb{E}_k\}\left[\left\{\Pi(Z,X)-\rho_{\mathcal{S},\Pi}^o(Z,X)\right\}\left\{A - \delta_{\mathcal{S}}^o(Z,X)\right\}\right]+o_p(n^{-1/2})\\
					&+\mathbb{E}_k\left[\left\{\Pi(Z,X)-\hat\rho_{\mathcal{S},\Pi}^{(n,-k)}(Z,X)\right\}\left\{A - \hat\delta_{\mathcal{S}}^{(n,-k)}(Z,X)\right\}\right]\\
					&-\mathbb{E}_k\left[\left\{\Pi(Z,X)-\rho_{\mathcal{S},\Pi}^o(Z,X)\right\}\left\{A - \delta_{\mathcal{S}}^o(Z,X)\right\}\right]\\
					=&\{\mathbb{E}_{nk}-\mathbb{E}_k\}\left[\left\{\Pi(Z,X)-\rho_{\mathcal{S},\Pi}^o(Z,X)\right\}\left\{A - \delta_{\mathcal{S}}^o(Z,X)\right\}\right]+o_p(n^{-1/2})\\
					&+\mathbb{E}_k\left[\left\{\rho_{\mathcal{S},\Pi}^o(Z,X)-\hat\rho_{\mathcal{S},\Pi}^{(n,-k)}(Z,X)\right\}\left\{\delta_{\mathcal{S}}^o(Z,X) - \hat\delta_{\mathcal{S}}^{(n,-k)}(Z,X)\right\}\right]\\
					=&\{\mathbb{E}_{nk}-\mathbb{E}_k\}\left[\left\{\Pi(Z,X)-\rho_{\mathcal{S},\Pi}^o(Z,X)\right\}\left\{A - \delta_{\mathcal{S}}^o(Z,X)\right\}\right]+o_p(n^{-1/2}).
				\end{align*}
				The second equality follows from the same argument used to establish Equation~\eqref{exp:9} by using the fact that
				$\mathbb{E}[\sum\limits_{l=1}^{L}\|\hat\rho_{\mathcal{S},\pi_l}^{(n,-k)}-\rho_{\mathcal{S},\pi_l}^o\|_2^2] + \mathbb{E}[\|\hat\delta_{\mathcal{S}}^{(n,-k)}-\delta_{\mathcal{S}}^o\|_2^2] + \mathbb{E}[\|\hat\eta_{\mathcal{S}}^{(n,-k)}-\eta_{\mathcal{S}}^o\|_2^2]=o(1)$.
				The final equality follows from the fact that $\|\hat\rho_{\mathcal{S},\Pi}^{(n,-k)}-\rho_{\mathcal{S},\Pi}^o\|_2
				\|\hat\delta_{\mathcal{S}}^{(n,-k)}-\delta_{\mathcal{S}}^o\|_2=o_p(n^{-1/2})$. 
				This indicates that
				\begin{align*}
					\hat q_{\mathcal{S}}=&\dfrac{1}{K}\sum_{k=1}^K\mathbb{E}_{nk}\left[\left\{\Pi(Z,X)-\hat\rho_{\mathcal{S},\Pi}^{(n,-k)}(Z,X)\right\}\left\{A - \hat\delta_{\mathcal{S}}^{(n,-k)}(Z,X)\right\}\right]\\
					=&\mathbb{E}_{n}\left[\left\{\Pi(Z,X)-\rho_{\mathcal{S},\Pi}^o(Z,X)\right\}\left\{A - \delta_{\mathcal{S}}^o(Z,X)\right\}\right]+o_p(n^{-1/2}).
				\end{align*}
				The proof of the remaining facts follows similarly by repeatedly using the required convergence rate of nuisance functions.
			\end{proof}
			
			\section{Proof of Lemmas}
			\subsection{Proof of Lemma~\ref{lem: connections bewteen PIV and invalid IV}}
			\begin{proof}[Proof of Lemma~\ref{lem: connections bewteen PIV and invalid IV}(a):]
				Under Assumptions~\ref{as: IV independence} and \ref{as: strong latent ignorability}, we have:
				\begin{align*}
					\mathbb{E}[Y(0) \mid Z,X] 
					&= \mathbb{E}[\mathbb{E}[Y(0) \mid U,Z,X] \mid Z,X] \\
					&= \mathbb{E}[\mathbb{E}[Y(0) \mid U,Z_{[-\nu_0]},X] \mid Z,X] \tag{by Assumption~\ref{as: strong latent ignorability}} \\
					&= \mathbb{E}[\mathbb{E}[Y(0) \mid U,Z_{[-\nu_0]},X] \mid Z_{[-\nu_0]},X] \tag{by Assumption~\ref{as: IV independence}} \\
					&= \mathbb{E}[Y(0) \mid Z_{[-\nu_0]},X].
				\end{align*}
				This finishes the proof of Equation~\eqref{eq: relation to sun2023}. Since $\mathbb{E}[Y(0)\mid Z,X]$ does not depend on $Z_{[\nu_0]}$, we know that $\mathbb{E}[Y(0)\mid Z,X]\in\mathcal{S}_{\gamma}$ for all $\gamma\leq |\nu_0|$, completing the proof of Lemma~\ref{lem: connections bewteen PIV and invalid IV}(a).    
			\end{proof}
			\begin{proof}[Proof of Lemma~\ref{lem: connections bewteen PIV and invalid IV}(b) under Assumption~\ref{as: alternative}(b):]
				In this setting, $\gamma_{\dagger}(\nu_0)=|\nu_0|$. Therefore, for all $\pi\in L_2(Z,X)$ and $\gamma\leq|\nu_0|$, 
				\begin{align*}
					&\mathbb{E}\left[A\{\pi(Z,X) - \mathbb{S}_{\gamma}[\pi](Z,X)\}\{Y(1)-Y(0)\}\mid U,Z_{[-\nu_0]},X\right]\\
					=&\mathbb{E}\left[A\{\pi(Z,X) - \mathbb{S}_{\gamma}[\pi](Z,X)\}\mid U,Z_{[-\nu_0]},X\right] \mathbb{E}\left[Y(1)-Y(0)\mid U,Z_{[-\nu_0]},X\right]\tag{By Assumption~\ref{as: weak latent ignorability}}\\
					=&\mathbb{E}\left[A\{\pi(Z,X) - \mathbb{S}_{\gamma}[\pi](Z,X)\}\mid U,Z_{[-\nu_0]},X\right]\mathbb{E}\left[Y(1)-Y(0)\mid Z_{[-\nu_0]},X\right]\tag{By Assumption~\ref{as: alternative}(b)}\\
					=&\mathbb{E}\left[A\{\pi(Z,X) - \mathbb{S}_{\gamma}[\pi](Z,X)\}\mid U,Z_{[-\nu_0]},X\right]\times \mathbb{E}\left[Y(1)-Y(0)\mid X\right]\tag{By Assumption~\ref{as: Z-invariant treatment effect}}.
				\end{align*}
				Now we can integrate $U$ and $Z_{[-\nu_0]}$ to deduce that
				\begin{align*}
					&\mathbb{E}\left[A\{\pi(Z,X) - \mathbb{S}_{\gamma}[\pi](Z,X)\}\{Y(1)-Y(0)\}\mid X\right]\\
					=&\mathbb{E}\left[A\{\pi(Z,X) - \mathbb{S}_{\gamma}[\pi](Z,X)\}\mid X\right]\times \mathbb{E}\left[Y(1)-Y(0)\mid X\right]\\
					\Longrightarrow\qquad&
					\mathrm{Cov}\left\{A\{\pi(Z,X) - \mathbb{S}_{\gamma}[\pi](Z,X)\},Y(1)-Y(0)\mid X\right\}=0.
				\end{align*}
				Thus, Definition~\ref{defn: projected IV}(b) is verified for $\mathcal{S}_{\gamma}$ with $\gamma\leq |\nu_0|=\gamma_{\dagger}(\nu_0)$.
			\end{proof}
			
			\begin{proof}[Proof of Lemma~\ref{lem: connections bewteen PIV and invalid IV}(b) under Assumption~\ref{as: alternative}(a):]
				Under Assumptions~\ref{as: IV independence}, \ref{as: weak latent ignorability}, \ref{as: Z-invariant treatment effect}, and \ref{as: alternative}(a), for any weighting functions $\pi$ and $\gamma\leq \gamma_\dagger(\nu_0)$,
				\begin{align*}
					&\mathbb{E}\left[A\{\pi(Z,X) - \mathbb{S}_{\gamma}[\pi](Z,X)\}\{Y(1)-Y(0)\}\mid U,Z_{[-\nu_0]},X\right]\\
					=&\mathbb{E}\left[A\{\pi(Z,X) - \mathbb{S}_{\gamma}[\pi](Z,X)\}\mid U,Z_{[-\nu_0]},X\right] \mathbb{E}\left[Y(1)-Y(0)\mid U,Z_{[-\nu_0]},X\right]\tag{By Assumption~\ref{as: weak latent ignorability}}\\
					=&\sum_{\nu:|\nu|=\gamma,\nu\subseteq \nu_0}\mathbb{E}\left[b_{\nu}(U,Z_{[-\nu]},X)\{\pi(Z,X) - \mathbb{S}_{\gamma}[\pi](Z,X)\}\mid U,Z_{[-\nu_0]},X\right] \mathbb{E}\left[Y(1)-Y(0)\mid U,Z_{[-\nu_0]},X\right]\\
					&+\mathbb{E}\left[c(Z,X)\{\pi(Z,X) - \mathbb{S}_{\gamma}[\pi](Z,X)\}\mid U,Z_{[-\nu_0]},X\right] \mathbb{E}\left[Y(1)-Y(0)\mid U,Z_{[-\nu_0]},X\right]\tag{By Assumption~\ref{as: alternative}(a)}\\
					=&\sum_{\nu:|\nu|=\gamma,\nu\subseteq \nu_0}\left\{\begin{array}{l}
						\mathbb{E}\left[b_{\nu}(U,Z_{[-\nu]},X)\mathbb{E}[\{\pi(Z,X) - \mathbb{S}_{\gamma}[\pi](Z,X)\}\mid U,Z_{[-\nu]},X]\mid U,Z_{[-\nu_0]},X\right] \\\times\mathbb{E}\left[Y(1)-Y(0)\mid U,Z_{[-\nu_0]},X\right]
					\end{array}\right\}\\
					&+\mathbb{E}\left[c(Z,X)\{\pi(Z,X) - \mathbb{S}_{\gamma}[\pi](Z,X)\}\mid U,Z_{[-\nu_0]},X\right] \mathbb{E}\left[Y(1)-Y(0)\mid U,Z_{[-\nu_0]},X\right]\\
					=&\sum_{\nu:|\nu|=\gamma,\nu\subseteq \nu_0}\left\{\begin{array}{l}
						\mathbb{E}\left[b_{\nu}(U,Z_{[-\nu]},X)\mathbb{E}\left[\{\pi(Z,X) - \mathbb{S}_{\gamma}[\pi](Z,X)\}\mid Z_{[-\nu]},X\right]\mid U,Z_{[-\nu_0]},X\right] \\
						\times\mathbb{E}\left[Y(1)-Y(0)\mid U,Z_{[-\nu_0]},X\right]
					\end{array}\right\}\\
					&+\mathbb{E}\left[c(Z,X)\{\pi(Z,X) - \mathbb{S}_{\gamma}[\pi](Z,X)\}\mid Z_{[-\nu_0]},X\right] \mathbb{E}\left[Y(1)-Y(0)\mid U,Z_{[-\nu_0]},X\right]\tag{By Assumption~\ref{as: IV independence} that $Z_{[\nu_0]}\indep U\mid Z_{[-\nu_0]},X$}\\
					=&\mathbb{E}\left[c(Z,X)\{\pi(Z,X) - \mathbb{S}_{\gamma}[\pi](Z,X)\}\mid Z_{[-\nu_0]},X\right] \mathbb{E}\left[Y(1)-Y(0)\mid U,Z_{[-\nu_0]},X\right].
				\end{align*}
				The final step uses the fact that $\mathbb{E}[\{\pi(Z,X) - \mathbb{S}_{\gamma}[\pi](Z,X)\}\mid Z_{[-\nu]},X]\equiv 0$, since $\gamma=|\nu|$.
				We can now integrate $U$ out to deduce that
				\begin{align*}
					&\mathbb{E}\left[A\{\pi(Z,X) - \mathbb{S}_{\gamma}[\pi](Z,X)\}\{Y(1)-Y(0)\}\middle| Z_{[-\nu_0]},X\right]\\
					=&\mathbb{E}\left[A\{\pi(Z,X) - \mathbb{S}_{\gamma}[\pi](Z,X)\}\middle| Z_{[-\nu_0]},X\right]\times
					\mathbb{E}\left[Y(1)-Y(0)\middle| Z_{[-\nu_0]},X\right]\\
					=&\mathbb{E}\left[A\{\pi(Z,X) - \mathbb{S}_{\gamma}[\pi](Z,X)\}\middle| Z_{[-\nu_0]},X\right]\times
					\mathbb{E}\left[Y(1)-Y(0)\middle| X\right].
					\tag{\text{By Assumption~\ref{as: Z-invariant treatment effect}}}
				\end{align*}
				Next, we can integrate $Z_{[-\nu_0]}$ out to deduce that
				\begin{align*}
					&\mathbb{E}\left[A\{\pi(Z,X) - \mathbb{S}_{\gamma}[\pi](Z,X)\}\{Y(1)-Y(0)\}\middle| X\right]\\
					=&\mathbb{E}\left[A\{\pi(Z,X) - \mathbb{S}_{\gamma}[\pi](Z,X)\}\middle| X\right]\times
					\mathbb{E}\left[Y(1)-Y(0)\middle| X\right].
				\end{align*}
				This completes the proof of Definition~\ref{defn: projected IV}(b) for $\mathcal{S}_{\gamma}$ with $\gamma\leq \gamma_{\dagger}(\nu_0)$. 
			\end{proof}
			
			\subsection{Proof of Lemma~\ref{lem: orthogonal complement}}
			We define
			$$
				\mathcal{D}_{\gamma}
				:=
				\left\{
				d\in L_2(Z,X):
				\mathbb{E}\left[d(Z,X)\mid Z_{[-\nu]},X\right]=0,\;
				\text{for all }|\nu|\geq\gamma
				\right\}.
			$$
			We directly verify that \(L_2(Z,X)\) admits an orthogonal decomposition into \(\mathcal{S}_{\gamma}\) and \(\mathcal{D}_{\gamma}\).
			
			For any \(\nu\subseteq\langle d_Z\rangle\), define the map
			\(\mathcal{F}_{[\nu]}:L_2(Z,X)\to L_2(Z,X)\) by $\mathcal{F}_{[\nu]}[d](Z,X)=\mathbb{E}\left[d(Z,X)\mid Z_{[-\nu]},X\right].$
			Define the kernel and range of \(\mathcal{F}_{[\nu]}\) as
			\begin{align*}
				\operatorname{Ker}\{\mathcal{F}_{[\nu]}\}
				&:=\left\{d\in L_2(Z,X):\mathcal{F}_{[\nu]}[d]=0\right\},\\
				\operatorname{Range}\{\mathcal{F}_{[\nu]}\}
				&:=\left\{\mathcal{F}_{[\nu]}[d]:d\in L_2(Z,X)\right\}=L_2(Z_{[-\nu]},X).
			\end{align*}
			Because \(\mathcal{F}_{[\nu]}\) is the orthogonal projection onto
			\(L_2(Z_{[-\nu]},X)\), its range is closed and
			$\operatorname{Ker}\{\mathcal{F}_{[\nu]}\}^{\perp}
			=\operatorname{Range}\{\mathcal{F}_{[\nu]}\}.$
			Since $\mathcal{D}_{\gamma}=\bigcap_{\nu:|\nu|\geq\gamma}\operatorname{Ker}\{\mathcal{F}_{[\nu]}\},$ we obtain
			\begin{align*}
				\mathcal{D}_{\gamma}^{\perp}
				&=\left\{\bigcap_{\nu:|\nu|\geq\gamma}\operatorname{Ker}\{\mathcal{F}_{[\nu]}\}\right\}^{\perp}
				=\overline{\sum_{\nu:|\nu|\geq\gamma}\operatorname{Ker}\{\mathcal{F}_{[\nu]}\}^{\perp}}\\
				&=\overline{\sum_{\nu:|\nu|\geq\gamma}\operatorname{Range}\{\mathcal{F}_{[\nu]}\}}
				=\overline{\sum_{\nu:|\nu|\geq\gamma}L_2(Z_{[-\nu]},X)}=\mathcal{S}_{\gamma}.
			\end{align*}
			Because \(\mathcal{D}_{\gamma}\) is an intersection of closed subspaces, it is closed. Taking orthogonal complements therefore gives $\mathcal{S}_{\gamma}^{\perp}=\mathcal{D}_{\gamma}.$ Hence, $L_2(Z,X)=\mathcal{S}_{\gamma}\oplus\mathcal{D}_{\gamma},$ which completes the proof of Lemma~\ref{lem: orthogonal complement}.

			\subsection{Proof of Lemma~\ref{lem: semiparametric calculation}}
			We define $\langle f,g\rangle_{\theta} := \mathbb{E}^{(\theta)}[f(O)g(O)]$, which is an inner product in $L_2^{(\theta)}(O):=\{f: \mathbb{E}^{(\theta)}[f(O)^2]<\infty\}$.
			We start from the definition of a projection operator. That is, for all $g\in\mathcal{S}$, 
			\begin{align*}
				\mathbb{E}^{(\theta)}[\{f(O)-\mathbb{S}^{(\theta)}[f](Z,X)\}g(Z,X)]=\langle f-\mathbb{S}^{(\theta)}[f], \; g\rangle_{\theta} \equiv 0.
			\end{align*}
			We then take the derivative with respect to $\theta$ and derive that
			\begin{align}
				0=&\nabla_{\theta}\mathbb{E}^{(\theta)}[\{f(O)-\mathbb{S}^{(\theta)}[f](Z,X)\}g(Z,X)]\eval_{\theta=0}\notag\\
				=&\mathbb{E}[\{f(O)-\mathbb{S}[f](Z,X)\}g(Z,X)s_{\theta}(O)]
				+\mathbb{E}[-\nabla_{\theta}\mathbb{S}^{(\theta)}[f](Z,X)\eval_{\theta=0}g(Z,X)]\notag\\
				=&\mathbb{E}[\mathbb{S}[\{f-\mathbb{S}[f]\}s_{\theta}](Z,X)g(Z,X)]
				+\mathbb{E}[-\nabla_{\theta}\mathbb{S}^{(\theta)}[f](Z,X)\eval_{\theta=0}g(Z,X)]\notag\\
				=&\mathbb{E}\left[\left\{\mathbb{S}[\{f-\mathbb{S}[f]\}s_{\theta}](Z,X) - \nabla_{\theta}\mathbb{S}^{(\theta)}[f](Z,X)\eval_{\theta=0}\right\}g(Z,X)\right].\label{exp: 3}
			\end{align}
			The third step follows since $g\in\mathcal{S}$ and
			$\mathbb{E}[f(O)g(Z,X)] = \mathbb{E}[\mathbb{S}[f](Z,X)g(Z,X)]$ by the property of projection operator.
			Now since Equation~\eqref{exp: 3} holds for all $g\in\mathcal{S}$, it follows that
			\begin{align*}
				\mathbb{S}[\{f-\mathbb{S}[f]\}s_{\theta}](Z,X) - \nabla_{\theta}\mathbb{S}^{(\theta)}[f](Z,X)\eval_{\theta=0}\in\mathcal{S}^{\perp}.
			\end{align*}
			Simultaneously, $\mathbb{S}[\{f-\mathbb{S}[f]\}s_{\theta}](Z,X) - \nabla_{\theta}\mathbb{S}^{(\theta)}[f](Z,X)\eval_{\theta=0}\in\mathcal{S}$ by definition of $\mathbb{S}$ and the assumption that the derivative belongs to $\mathcal S$. Therefore,
			\begin{align*}
				\mathbb{S}[\{f-\mathbb{S}[f]\}s_{\theta}](Z,X) - \nabla_{\theta}\mathbb{S}^{(\theta)}[f](Z,X)\eval_{\theta=0}\in\mathcal{S}^{\perp}\bigcap\mathcal{S} = \{0\}.
			\end{align*}
			This completes the proof of Lemma~\ref{lem: semiparametric calculation}.

			{\setlength{\baselineskip}{0.63\baselineskip}
				\let\newline\space
				\putbib[ref]
			}
		\end{bibunit}
	\end{document}